%% file: SchU18-ppt.tex
\documentclass[11pt]{article}
\usepackage[utf8]{inputenc}
\usepackage[T1]{fontenc}
\usepackage[english]{babel}
\usepackage[margin=3cm]{geometry}

\usepackage{listings}
\usepackage{amsmath}
\usepackage{amsfonts}
\usepackage{amssymb}
\usepackage{parskip}
\usepackage{amsthm}
\usepackage{mathtools}
\usepackage{thmtools}
\usepackage{etoolbox}
\usepackage{cite}
\usepackage{color}
\usepackage{subcaption}
\usepackage{todonotes}
\usepackage{booktabs}
\usepackage{siunitx}

\begingroup
    \makeatletter
    \@for\theoremstyle:=definition,remark,plain\do{%
        \expandafter\g@addto@macro\csname th@\theoremstyle\endcsname{%
            \addtolength\thm@preskip\parskip
            }%
        }
\endgroup

\usepackage{tikz}
\usetikzlibrary{circuits}
\usetikzlibrary{circuits.ee.IEC}
\usetikzlibrary{patterns}
\usetikzlibrary{arrows}
\usepackage{pgfplots}
\pgfplotsset{compat=1.10}

\usepackage{cite}

\usepackage{microtype}	
\usepackage[colorlinks,citecolor=red]{hyperref}
\usepackage[nameinlink]{cleveref}
\newtheorem{theorem}{Theorem}[section]
\crefname{theorem}{theorem}{theorems}
\Crefname{theorem}{Theorem}{Theorems}

\newtheorem{assumption}[theorem]{Assumption}
\crefname{assumption}{assumption}{assumptions}
\Crefname{assumption}{Assumption}{Assumptions}

\newtheorem{prop}[theorem]{Proposition}
\crefname{prop}{proposition}{propositions}
\Crefname{prop}{Proposition}{Propositions}

\crefname{lemma}{lemma}{lemmas}
\Crefname{lemma}{Lemma}{Lemmas}

\crefname{corollary}{corollary}{corollaries}
\Crefname{corollary}{Corollary}{Corollaries}

\theoremstyle{definition}
\newtheorem{remark}[theorem]{Remark}
\AtEndEnvironment{remark}{\hfill\ensuremath{\diamondsuit}}
\crefname{remark}{remark}{remarks}
\Crefname{remark}{Remark}{Remarks}

\theoremstyle{definition}
\newtheorem{defn}[theorem]{Definition}
\crefname{defn}{definition}{definitions}
\Crefname{defn}{Definition}{Definitions}

\newtheorem{example}[theorem]{Example}
\AtEndEnvironment{example}{\hfill\ensuremath{\bigcirc}}
\crefname{example}{example}{example}
\Crefname{example}{Example}{Examples}

\tikzstyle{Resistor} = [draw, rectangle, 
    minimum height=1cm, minimum width=2cm]
\tikzstyle{Inductance} = [draw, rectangle, fill=black,
    minimum height=1cm, minimum width=2cm]
\tikzstyle{input} = [coordinate]
\tikzset{circuit declare symbol = ac source}
\tikzset{set ac source graphic = ac source IEC graphic}
\tikzset{
         ac source IEC graphic/.style=
          {
           transform shape,
           circuit symbol lines,
           circuit symbol size = width 3 height 3,
           shape=generic circle IEC,
           /pgf/generic circle IEC/before background=
            {
             \pgfpathmoveto{\pgfpoint{-0.8pt}{0pt}}
             \pgfpathsine{\pgfpoint{0.4pt}{0.4pt}}
             \pgfpathcosine{\pgfpoint{0.4pt}{-0.4pt}}
             \pgfpathsine{\pgfpoint{0.4pt}{-0.4pt}}
             \pgfpathcosine{\pgfpoint{0.4pt}{0.4pt}}
             \pgfusepathqstroke
            }
          }
        }

\DeclareMathOperator{\rank}{rank}
\DeclareMathOperator{\blkdiag}{blkdiag}

\newcommand{\bs}[1]{\boldsymbol{#1}}
\newcommand{\state}{\bs{x}}
\newcommand{\inVar}{\bs{u}}
\newcommand{\outVar}{\bs{y}}

\newcommand{\switchVar}{\bs{z}}

\newcommand{\stateDim}{n}
\newcommand{\inDim}{m}
\newcommand{\outDim}{p}

\newcommand{\perturbationDimA}{\beta}

\newcommand{\E}{E}
\newcommand{\A}{A}
\newcommand{\B}{B}
\newcommand{\C}{C}
\newcommand{\D}{D}
\newcommand{\J}{J}
\newcommand{\R}{R}
\newcommand{\Q}{Q}
\newcommand{\Ham}{\mathcal{H}}
\renewcommand{\H}{H}
\newcommand{\system}{\Sigma}
\newcommand{\stateTrans}{K}

\newcommand{\switch}{{\sigma}}
\newcommand{\switchNr}{\ell}

\newcommand{\lowRankLeft}{S}
\newcommand{\lowRankRight}{T}
\newcommand{\eSystem}{\system_\mathrm{E}}

\newcommand{\eA}{\A_\mathrm{E}}
\newcommand{\eB}{\B_\mathrm{E}}
\newcommand{\eC}{\C_\mathrm{E}}
\newcommand{\eD}{\D_\mathrm{E}}

\newcommand{\stateDimE}{\stateDim_\mathrm{E}}
\newcommand{\inDimE}{\inDim_\mathrm{E}}
\newcommand{\outDimE}{\outDim_\mathrm{E}}
\newcommand{\eState}{\state_\mathrm{E}}
\newcommand{\deState}{\dot{\state}_\mathrm{E}}
\newcommand{\eInVar}{\inVar_\mathrm{E}}
\newcommand{\eOutVar}{\outVar_\mathrm{E}}
\newcommand{\rhoc}{\zeta}

\newcommand{\feedback}{K}

\newcommand{\midMatr}{M}

\newcommand{\svdLeft}{U}
\newcommand{\svdMid}{\Sigma}
\newcommand{\svdRight}{V}

\newcommand{\leftProj}{W}
\newcommand{\rightProj}{V}

\newcommand{\stateRed}{\widetilde{\state}}
\newcommand{\outVarRed}{\widetilde{\outVar}}

\newcommand{\stateDimRed}{r}

\newcommand{\Ared}{\widetilde{\A}}
\newcommand{\Bred}{\widetilde{\B}}
\newcommand{\Cred}{\widetilde{\C}}
\newcommand{\Dred}{\widetilde{\D}}
\newcommand{\Hred}{\widetilde{H}}
\newcommand{\systemRed}{\widetilde{\system}}
\newcommand{\stateTransRed}{\widetilde{\stateTrans}}

\newcommand{\eSystemRed}{\widetilde{\system}_\mathrm{E}}

\newcommand{\eAred}{\Ared_\mathrm{E}}
\newcommand{\eBred}{\Bred_\mathrm{E}}
\newcommand{\eCred}{\Cred_\mathrm{E}}
\newcommand{\eDred}{\Dred_\mathrm{E}}
\newcommand{\eOutVarRed}{\widetilde{\outVar}_\mathrm{E}}
\newcommand{\eInVarRed}{\widetilde{\inVar}_\mathrm{E}}

\newcommand{\LtNorm}[1]{\|#1\|_{\mathcal{L}_2}}
\newcommand{\LinfNorm}[1]{\|#1\|_{\mathcal{L}_\infty}}

\newcommand{\HinfNorm}[1]{\|#1\|_{\mathcal{H}_\infty}}

\newcommand{\lyap}{Q}

\date{\today}
\title{Model reduction for linear systems with low-rank switching}
\author{
Philipp Schulze\footnotemark[1]~\footnotemark[2]
\and
Benjamin Unger\footnotemark[1]~\footnotemark[3]
}

\begin{document}

\maketitle
\renewcommand{\thefootnote}{\fnsymbol{footnote}}
\footnotetext[1]{
Institut f\"ur Mathematik,
Technische Universität Berlin, Str.\ des 17.~Juni~136,
10623~Berlin,
Federal Republic of Germany,
{\tt \{pschulze,unger\}@math.tu-berlin.de}.}
\footnotetext[2]{The author was supported by the DFG Collaborative Research Center 1029 \emph{Substantial efficiency increase in gas turbines through direct use of coupled unsteady combustion and flow dynamics}, project A02.}
\footnotetext[3]{The author was supported by the DFG Collaborative Research Center 910 \emph{Control of self-organizing nonlinear systems: Theoretical methods and concepts of application}, project A2.}
%
\begin{abstract}
We introduce a novel model order reduction method for large-scale linear switched systems (LSS) where the coefficient matrices are affected by a low-rank switching. 
The key idea is to replace the LSS by a non-switched system with extended input and output vectors -- called the envelope system -- which is able to reproduce the dynamical behavior of the original LSS by applying a certain feedback law. 
The envelope system can be reduced using standard model order reduction schemes and then transformed back to an LSS. 
Furthermore, we present an upper bound for the output error of the reduced-order LSS and show how to preserve quadratic Lyapunov stability. 
The approach is tested by means of various numerical examples demonstrating the efficacy of the presented method.
\vskip .3truecm

{\bf Keywords:} switched system, model reduction, balanced truncation, rational interpolation, stability, port-Hamiltonian System
\vskip .3truecm

{\bf AMS(MOS) subject classification:} 93C30, 93A15, 30E05, 93B52
\end{abstract}

\section{Introduction}
Many real world applications are a combination of continuous and discrete dynamics, which are mathematically described by \emph{hybrid systems}. Popular examples are manufacturing processes \cite{PepC00}, automotive engine control \cite{BalBBPS00}, and transportation systems \cite{LeeS03}. In the context of control design the discrete dynamics may be used to switch between operating modes of the dynamical system. Thus the focus is on the continuous dynamics only. This results in the concept of  \emph{switched systems}.

In this paper we study linear switched systems (LSS) of the form
\begin{equation}
	\label{eq:switchedSystem}
	\system = \left\{
	\begin{aligned}
		\dot{\state}(t) &= \A_\switch\state(t) + \B_\switch\inVar(t),\\
		\outVar(t) &= \C_\switch\state(t) + \D_\switch\inVar(t),\\
		\state(0) &= \state_0,
	\end{aligned}
	\right.\qquad\qquad t\geq0
\end{equation}
with matrices $\A_i\in\mathbb{R}^{\stateDim\times\stateDim}$, $\B_i\in\mathbb{R}^{\stateDim\times\inDim}$, $\C_i\in\mathbb{R}^{\outDim\times\stateDim}$, $\D_i\in\mathbb{R}^{\outDim\times\inDim}$ for $i=1,\ldots,\switchNr$, and \emph{switching signal} $\sigma:[0,\infty)\times\mathbb{R}^\outDim\to\{1,\ldots,\switchNr\}$. 
We refer to $\state$, $\inVar$, and $\outVar$ as the \emph{state}, \emph{input}, and \emph{output} of the system, respectively. 
Without loss of generality we assume a zero initial condition, i.\,e. $\state_0 = 0$, since otherwise we can simply append the initial condition to the input matrices as in \cite{HeiRA11} or solve two independent problems as in \cite{BeaGM17}.
We include that case that the switching signal depends on the output of the system \eqref{eq:switchedSystem}, thus allowing feedback-based switching laws. 
Although each subsystem in \eqref{eq:switchedSystem} is linear, the overall dynamics are nonlinear with a discontinuous vector field and thus even linear time-invariant (LTI) switched systems exhibit complex phenomena.

If \eqref{eq:switchedSystem} results from semi-discretization of a switched partial differential equation then $\stateDim$ is large and controller design, optimization, or system verification of such a large-scale system is computationally demanding or even intractable. 
For LTI systems without switching the computational complexity is nowadays often avoided by replacing the system with a cheap-to-evaluate surrogate model. 
A standard approach to build reliable surrogate models is model order reduction (MOR), which usually constructs a low-dimensional approximation via Petrov-Galerkin projection -- see \cite{Ant05,BauBF14,AntBG10,BenCOW17} for an overview. 
Popular reduction methods are balanced truncation \cite{MulR76,Moo81}, rational interpolation (formerly known as moment matching) \cite{GugAB08}, and proper orthogonal decomposition \cite{Vol01,HinV05}. 

A straigthforward way to generalize these methods to switched systems is to apply the reduction technique of choice to each subsystem $\system_i = (\A_i,\B_i,\C_i,\D_i)$ separately, see for example \cite{MazVBB08,PapP14,PapP16}. 
However, even if during this reduction each input-output mapping is preserved, the approximation error for the switched system may be arbitrarily large (see \Cref{ex:independentReduction}). 
Instead, MOR methods for switched systems should be tailored to switching specific phenomena. 
A generalization of balanced truncation to switched systems is developed in \cite{ShaW09} and analyzed in \cite{PetWL13}. 
In these papers the authors assume that all subsystems in \eqref{eq:switchedSystem} can be balanced simultaneously via generalized Gramians, which is in general not feasible \cite{Lib03}. 
To compute the generalized Gramians, $2\switchNr$ coupled Lyapunov inequalities need to be solved, which may become computationally intractable in a large-scale setting. 
A modified version with similar computational complexity is presented in \cite{ShaW12} and the authors claim (without proof or numerical example) that this method performs better in terms of the approximation error. 
Model reduction of switched systems via interpolation is considered in \cite{BasPWL14b,BasPWL16,ScaA16}. 
In \cite{BasPWL14b,BasPWL16} the notion of moment is defined for LSS based on the realization theory derived in \cite{Pet11}. Then a certain selection of moments -- called \emph{nice selections} -- is interpolated. In contrast, the authors of \cite{ScaA16} base their work on the signal generator approach \cite{Ast10}. Let us also mention the recent technical report \cite{GosPAF17}, where model reduction is performed by introducing Gramians for the switched system that are similar to those of bilinear systems.

As is standard in MOR theory, all methods assume that the solution of the dynamical system evolves (approximately) in a low-dimensional subspace. 
The existence of such a low-dimensional subspace is based on the smoothness of the input-output mapping \cite{OhlR16} -- a feature that switched systems do not possess in general if arbitrary switching is allowed. 
In order to expect a good approximation quality with low-dimensional surrogates we either need to assume that the switching occurs very rarely in the desired time horizon or that the switching affects only a low-dimensional part of the system. 

\begin{example}
	\label{ex:twoRooms}
	Similar to \cite{PapP14,PapP16} we consider the heating of two adjacent rooms that are connected via a door that is either open or closed. For simplicity we restrict ourselves to a one-dimensional model. The situation is depicted in \Cref{fig:ExampleHeatingRoom}.
	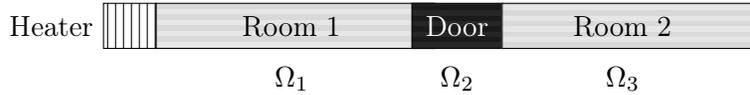
\begin{figure}[ht]
	\centering
	\begin{tikzpicture}		
		\draw[pattern=vertical lines, pattern color=black] (0.3,0.7) rectangle (1,1.3);
		\draw[pattern=horizontal lines light gray] (1,0.7) rectangle (4.7,1.3);
		\draw[pattern=horizontal lines dark gray] (4.4,0.7) rectangle (5.6,1.3);
		\draw[pattern=horizontal lines light gray] (5.6,0.7) rectangle (9,1.3);
		
		\node at (2.8,1) {Room 1};
		\node at (7.2,1) {Room 2};
		\node at (5,1) {\color{white}Door};
		\node[anchor=east] at (0.3,1) {Heater};

		\node at (2.8,0.3) {$\Omega_1$};
		\node at (5,0.3) {$\Omega_2$};
		\node at (7.2,0.3) {$\Omega_3$};		
	\end{tikzpicture}
	\caption{Heating of two adjacent rooms}
	\label{fig:ExampleHeatingRoom}
	\end{figure}
	Opening or closing the door results in an abrupt change of the thermal conductivity and the heat capacity in the domain $\Omega_2$. 
	Since the domain of the door is small compared to the overall domain $\Omega\vcentcolon= \Omega_1\cup\Omega_2\cup\Omega_3$, the switching influences only a small part of the dynamical system. Thus after a semi-discretization (cf. \Cref{ex:twoRooms2}), only a low-dimensional structure of the overall system is affected.
\end{example}

In this paper we allow arbitrary switching sequences and thus assume that the switching affects only a low-dimensional part of the system. We exploit this to compute a multiple-input multiple-output (MIMO) LTI system without switching -- called the \emph{envelope system} -- that covers all the dynamics of the original switched system. More precisely, using a certain feedback control law for the envelope system, the two systems are equivalent. Model reduction is achieved by reducing the envelope system by a standard LTI MOR method. The resulting reduced-order model (ROM) can be transformed back to a switched system of the same form as \eqref{eq:switchedSystem}. The advantage of this approach is that a comparison of models and hence an evaluation of approximation properties can be made for the envelope system using standard tools. Using this abstraction, the difficulties associated with switched systems and different switching signals can be circumvented. Of course, one cannot expect similar approximation qualities as for a standard LTI system -- this is reflected by a slower decay of the Hankel singular values of the envelope system.

The main contributions of this paper are the following:
\begin{enumerate}
	\item We show (cf. \Cref{ex:independentReduction}) that independent reduction of the subsystems in \eqref{eq:switchedSystem} may result in arbitrarily large errors and should thus be avoided.
	\item \Cref{ex:outputDependentSwitching} demonstrates that if the switching signal depends on the state or the output of the system itself, then any approximation of the state or the output that effects the switching law may result in arbitrarily large errors and thus no error bounds for such switching signals can be derived.
	\item We propose an abstraction of the switched system \eqref{eq:switchedSystem} in form of an LTI system without switching (cf. \Cref{lem:envelopeFeedback}). This abstraction allows to use standard MOR methods for switched systems and to derive an error bound for the approximation error in the time domain (\Cref{thm:errorBound}) provided that the switching signal depends only on time.
	\item \Cref{prop:stableDH} shows that quadratic Lyapunov stability of a switched system implies that all subsystems are dissipative Hamiltonian systems and hence MOR techniques for port-Hamiltonian systems can be used to preserve quadratic stability in the reduced model (\Cref{thm:quadraticStability}).
\end{enumerate}

\section*{Notation}
The symbol $A^T$ denotes the transpose of the real matrix $A$. 
If $A$ is quadratic, we write $A>0$ ($A\geq 0$) to indicate that the matrix is positive definite (semidefinite). 
The matrix $I_n$ is the $n\times n$ identity matrix and $0_{n,m}$ describes the $n\times m$ zero matrix. We use $\blkdiag(A_1,\ldots,A_k)$ to denote the block diagonal matrix with matrices $A_1,\ldots,A_k$ on the diagonal.
For a square-integrable function $\boldsymbol{f}:[0,T]\to\mathbb{R}^n$ we denote the $\mathcal{L}_2$ norm by $\LtNorm{\boldsymbol{f}} \vcentcolon= (\int_0^T \|\boldsymbol{f}(t)\|^2_2\,\mathrm{d}t)^{1/2}$ with Euclidean vector norm $\|\cdot\|_2$. The switched system \eqref{eq:switchedSystem} is written as $\system = (\A_i,\B_i,\C_i,\D_i\mid i=1,\ldots,\switchNr)$ and each subsystem
\begin{displaymath}
	\dot{\state}(t) = \A_j\state(t) + \B_j\inVar(t),\qquad \outVar(t) = \C_j\state(t) + \D_j\inVar(t)
\end{displaymath}
is denoted by $\system_j = (\A_j,\B_j,\C_j,\D_j)$ with transfer function $H_j(s) = \C_j(sI_\stateDim - \A_j)^{-1}\B_j + \D_j$. We say that $\system_j$ is stable, if all eigenvalues of $\A_j$ have non-positive real part and say the $\system_j$ is asymptotically stable, if all eigenvalues of $\A_j$ have negative real part. In the latter case, there exist matrices $\mathcal{P}_j = \mathcal{P}_j^T\geq 0$ and $\mathcal{Q}_j = \mathcal{Q}_j^T\geq 0$ that satisfy the Lyapunov equations
\begin{align*}
	\A_j\mathcal{P}_j + \mathcal{P}_j\A_j^T + \B_j\B_j^T = 0\qquad\text{and}\qquad
	\A_j^T\mathcal{Q}_j + \mathcal{Q}_j\A_j + \C_j^T\C_j = 0.
\end{align*}
The matrix $\mathcal{P}_j$ is called controllability Gramian of $\system_j$, while $\mathcal{Q}_j$ is called observability Gramian of $\system_j$. Moreover, the square roots of the eigenvalues of $\mathcal{P}_j\mathcal{Q}_j$ are the Hankel singular values of $\system_j$. We say that $\system_j$ is controllable or observable, if the conditions $\rank(\begin{bmatrix}
	\lambda I_n -\A_j & \B_j
\end{bmatrix}) = n$ or $\rank(\begin{bmatrix}
	\lambda I_n - \A_j^T & \C_j^T
\end{bmatrix}) = n$ hold for all $\lambda\in\mathbb{C}$, respectively. The system $\system_j$ is called minimal, if it is controllable and observable. For corresponding definitions for the switched system \eqref{eq:switchedSystem} we refer to \cite{SunGL02}. The $\mathcal{H}_\infty$ norm of $\system_j$ is the $\mathcal{L}_2$--$\mathcal{L}_2$ induced norm and denoted by $\HinfNorm{\system_j}$.

\section{Problem setting and motivation}

Starting from the LSS \eqref{eq:switchedSystem} we want to construct a reduced order model
\begin{equation}
	\label{eq:switchedROM}
	\systemRed = \left\{
	\begin{aligned}
		\dot{\stateRed}(t) &= \Ared_\switch\stateRed(t) + \Bred_\switch\inVar(t),\\
		\outVarRed(t) &= \Cred_\switch\stateRed(t) + \Dred_\switch\inVar(t),
	\end{aligned}
	\right.\qquad\qquad t\geq0
\end{equation}
of the same form as \eqref{eq:switchedSystem} with $\Ared_i\in\mathbb{R}^{\stateDimRed\times\stateDimRed}$, $\Bred_i\in\mathbb{R}^{\stateDimRed\times\inDim}$, $\Cred_i\in\mathbb{R}^{\outDim\times\stateDimRed}$, and $\Dred_i\in\mathbb{R}^{\outDim\times\inDim}$ for $i=1,\ldots,\switchNr$ and $\stateDimRed\ll\stateDim$, such that the approximation error $\outVar-\outVarRed$ is small (in some appropriate norm).
If we want to emphasize the dependency of the state or the output on the input variable, we write $\state(\inVar)$ or $\outVar(\inVar)$.

As is common for many model reduction techniques \cite{Ant05}, the reduced matrices $\Ared_i$, $\Bred_i$, $\Cred_i$, and $\Dred_i$ are constructed via Petrov-Galerkin projection. 
More precisely, we determine matrices $\leftProj_i,\rightProj_i\in\mathbb{R}^{\stateDim\times\stateDimRed}$ with $\leftProj_i^T\rightProj_i$ nonsingular and compute the reduced model $\systemRed$ via
\begin{align}
\label{eq:projection}
	\Ared_i &\vcentcolon= \left(\leftProj_i^T\rightProj_i\right)^{-1}\leftProj_i^T\A_i\rightProj_i, & \Bred_i &\vcentcolon= \left(\leftProj_i^T\rightProj_i\right)^{-1}\leftProj_i^T\B_i, &
	\Cred_i &\vcentcolon= \C_i\rightProj_i, &\Dred_i &\vcentcolon= \D_i
\end{align}
for $i=1,\ldots,\switchNr$. Note that if we use different $\leftProj_i$ and $\rightProj_i$ for each subsystem then we assume $\rightProj_i\stateRed \approx \state$ for each subsystem and thus the reduced state $\stateRed \approx (\leftProj_i^T\rightProj_i)^{-1}\leftProj_i^T\state$ depends on the subsystem. Accordingly we need a state transition transformation. If we switch from subsystem $i$ to subsystem $j$ we need to transform the state via $\stateRed \mapsto \stateTransRed_{ij}\stateRed$ with
\begin{equation}
	\label{eq:stateTransformation}
	\stateTransRed_{ij} = \left(\leftProj_j^T\rightProj_j\right)^{-1}\leftProj_j^T\rightProj_i.
\end{equation}
A standard approach in the literature \cite{MazVBB08,PapP14,PapP16} is to reduce the subsystems independently. The following example shows that this may result in large approximation errors.
\begin{example}
	\label{ex:independentReduction}
	We consider the RLC circuit given in \Cref{fig:RLC}, where the input is given by the entering current $i$, while the output of interest is $i_L$, the current of the branch containing the inductance.
	The corresponding control system is given in \cite{SelBA13_ppt}, but here we modify the system by allowing a switch in the inductance.
	Precisely, we set $R=1$, $C=1$, $L_1=1/2$, and $L_2=1$.
	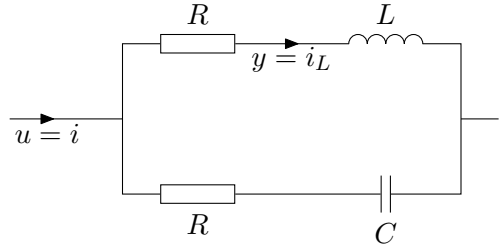
\begin{figure}[ht]
		\centering
		\begin{tikzpicture}[circuit ee IEC]
  			\draw (0,0) to ++(1.5,0)
  			  	to ++(0,1)
  			  	to ++(0.5,0)
  			  	to [resistor={info ={$R$}}]             ++(1, 0) 
  			  	to ++(1.5,0)
  			  	to [inductor={info=$L$}]                  ++(1,0)
  			  	to ++(0.5,0)
  			  	to ++(0,-2)
  			  	to ++(-0.5,0)
  			  	to [capacitor={info={$C$}}]                  ++(-1,0)
  			  	to ++(-1.5,0)
  			  	to [resistor={info ={$R$}}]             ++(-1, 0) 
  			  	to ++(-0.5,0)
  			  	to ++(0,1);
  			  	\draw (6,0) to ++(0.5,0);
  			  	\node[] at (0.5,-0.2) {$u=i$};
  			  	\node[current direction={black}] at (0.5,0){};
  			  	\node[current direction={black}] at (3.75,1){};
  			  	\node[] at (3.75,0.8) {$y=i_L$};
		\end{tikzpicture}
		\caption{A simple RLC circuit \cite{SelBA13_ppt}}
		\label{fig:RLC}
	\end{figure}	
	The resulting switched system has two modes, i.\,e. $\switchNr = 2$, given by
	\begin{displaymath}
		\A_1 = \begin{bmatrix}
			0 & -1\\
			2 & -4
		\end{bmatrix},\qquad \A_2 = \begin{bmatrix}
			0 & -1\\
			1 & -2
		\end{bmatrix},\qquad \B_1 = \begin{bmatrix}
			1\\ 2
		\end{bmatrix},\qquad \B_2 = \begin{bmatrix}
			1\\ 1
		\end{bmatrix},\qquad \C_1^T = \C_2^T = \begin{bmatrix}
			0\\ 1
		\end{bmatrix},
	\end{displaymath}
	i.\,e., the switching occurs only in the matrices $A_i$ and $B_i$ but not in the output matrix. The transfer functions of the two subsystems are given by
	\begin{displaymath}
		\H_1(s) = \frac{2s+2}{s^2+4s+2}\qquad\text{and}\qquad
		\H_2(s) = \frac{1}{s+1}.	
	\end{displaymath}	 
	Notice that the first subsystem is minimal but the second one is not controllable. 
	We only reduce the dimension of the second subsystem by truncating the states that are not controllable (e.\,g. via Kalman decomposition), i.\,e., we make no error with respect to the transfer function. 
	This can, for instance, be achieved using the projection matrices
	\begin{displaymath}
		\rightProj_2^T = \begin{bmatrix}
			1 & 1
		\end{bmatrix}, \quad
		\leftProj_2^T = \begin{bmatrix}
			2 & -1
		\end{bmatrix}
	\end{displaymath}
	and via \eqref{eq:projection} the reduced model and its transfer functions are given by
	\begin{align*}
		\Ared_2 &= -1, & \Bred_2 &= 1, & \Cred_2 &= 1, & \Hred_2(s) &= \frac{1}{s+1}.
	\end{align*}
	Now let us suppose that we have the switching function
	\begin{displaymath}
		\switch:[0,2]\to\{1,2\},\qquad \switch(t) = \begin{cases}
			1, & \text{if}\ t\in[0,1],\\
			2, & \text{if}\ t\in(1,2]
		\end{cases}	
	\end{displaymath}		
	and we are interested in the output $\outVar$ of the switched system at the final time $t=2$. 
	We fix $\xi\in\mathbb{R}$ and choose $\smash{\inVar\big|_{[0,1]}}$ such that $\state(1) = \smash{\begin{bmatrix}
		\xi & 2\xi
	\end{bmatrix}^T}$. 
	This is possible since the first system is controllable. 
	Afterwards we do not control the system anymore, i.\,e. $\smash{\inVar\big|_{(1,2]}\equiv0}$. 
	This yields $\state(2) = \smash{\begin{bmatrix}
		0 & \xi\exp(-1)
	\end{bmatrix}^T}$ and hence $\outVar(2) = \xi\exp(-1)$. 
	However, if we use the reduced model for the computation, we obtain $\outVarRed(2) = 0$, thus the error can be arbitrarily large, although no approximation error in the input-output mapping of both subsystems was made.
\end{example}

The reason for the behavior in the previous example is the fact that in a minimal realization of a switched system the subsystems might not be minimal \cite{Pet11}. This can be seen from the solution formula for switched systems. More precisely, assume that the switching signal $\switch$ depends only on time and the switching time points are $0 = t_0 < t_1 < \ldots < t_s$ with switching index sequence $i_k \vcentcolon= \sigma(t_k)$. Then for $t\geq t_s$ the solution \cite{SunGL02} of the state equation in \eqref{eq:switchedSystem} is given by (recall that we assume $\state_0 = 0$) 
\begin{equation}
	\label{eq:solutionFormula}
	\begin{aligned}
		\state(t;\inVar) &= \mathrm{e}^{\A_{i_s}(t-t_s)}\cdots\mathrm{e}^{\A_{i_1}(t_2-t_1)}\int_{t_0}^{t_1}\mathrm{e}^{A_{i_0}(t_1-\tau)}\B_{i_0}\inVar(\tau)\mathrm{d}\tau + \cdots \\
		&\quad + \mathrm{e}^{\A_{i_s}(t-t_s)}\int_{t_{s-1}}^{t_s}\mathrm{e}^{\A_{i_{s-1}}(t_s-\tau)}\B_{i_{s-1}}\inVar(\tau)\mathrm{d}\tau + \int_{t_s}^t\mathrm{e}^{\A_{i_s}(t-\tau)}\B_{i_s}\inVar(\tau)\mathrm{d}\tau.
	\end{aligned}
\end{equation}
If we reduce each subsystem independently, we expect a good approximation of 
\begin{displaymath}
	\int_{t_{k}}^{t} \C_{i_k}\mathrm{e}^{\A_{i_k}(t-\tau)}\B_{i_k}\inVar(\tau)\mathrm{d}\tau + \D_{i_k}\inVar(t)\qquad\text{for each}\ 0 \leq k \leq s.
\end{displaymath}
However this does not necessarily imply a good approximation of the interaction of the subsystems, which is described by the terms
\begin{multline*}
	\mathrm{e}^{\A_{i_k}(t-t_k)}\cdots\mathrm{e}^{\A_{i_1}(t_2-t_1)}\int_{t_0}^{t_1}\mathrm{e}^{A_{i_0}(t_1-\tau)}\B_{i_0}\inVar(\tau)\mathrm{d}\tau + \cdots\\
	 + \mathrm{e}^{\A_{i_k}(t-t_k)}\int_{t_{k-1}}^{t_k}\mathrm{e}^{\A_{i_{k-1}}(t_k-\tau)}\B_{i_{k-1}}\inVar(\tau)\mathrm{d}\tau.
\end{multline*}
The interaction terms can be interpreted as (non-zero) initial condition at each switching time point.
Note that also the techniques offered in \cite{HeiRA11,BeaGM17} cannot be used, since it is a priori unclear which initial conditions are important. 

A standard question in approximation theory is about the approximation quality in form of error bounds or error estimators. 
The following toy example illustrates what we can expect to happen if the switching signal depends on the state or the output of the system.
\begin{example}
	\label{ex:outputDependentSwitching}
	Consider the switched system $\system = (\A_i,\B_i,\C_i,\D_i\mid i=1,\ldots,\switchNr)$ where the first subsystem $\system_1 = (\A_1,\B_1,\C_1,\D_1)$  is given by
	\begin{equation}
		\label{eq:MORfailureSwitching}
		\A_1 = \begin{bmatrix}
			-1 & 0\\
			0 & -1
		\end{bmatrix}, \qquad \B_1 = \begin{bmatrix}
			1\\1
		\end{bmatrix}, \qquad \C_1^T = \begin{bmatrix}
			1\\p
		\end{bmatrix}, \qquad \D_1 = 0
	\end{equation}
	with parameter $p>0$. The approximation $\systemRed_1 = (\Ared_1,\Bred_1,\Cred_1,\Dred_1)$ with
	\begin{equation}
		\label{eq:MORfailureSwitchingRed}
		\Ared_1 = -1, \qquad \Bred_1 = 1, \qquad \Cred_1 = 1, \qquad \Dred_1 = 0
	\end{equation}
	satisfies (without switching) $\outVarRed = (1+p)^{-1}\outVar$, and thus the approximation error becomes arbitrarily small for $p\to 0$. 
	Suppose the switching condition and the input are given by
	\begin{displaymath}
		\switch(t,\outVar) = \begin{cases}
			1, & \outVar \geq \left(1+\frac{p}{2}\right)(1-\exp(-1)),\\
			2, & \text{otherwise},
		\end{cases}, \qquad
		\inVar(t) = \begin{cases}
			1, & \text{if}\ t \leq 1,\\
			0, & \text{otherwise},
		\end{cases}	
	\end{displaymath}	
	Then for $p>0$, we have $\outVarRed(t)<\left(1+\frac{p}{2}\right)(1-\exp(-1))$ for all $t>0$ whereas there exists a $t_1\in(0,1)$ with $\outVar(t_1) = \left(1+\frac{p}{2}\right)(1-\exp(-1))$. The situation is depicted in \Cref{fig:outputDependendSwitching}.
	\begin{figure}[ht]
		\centering
		\input{outputDependendSwitching}
		\caption{Output-dependent switching: full model and approximation}
		\label{fig:outputDependendSwitching}
	\end{figure}
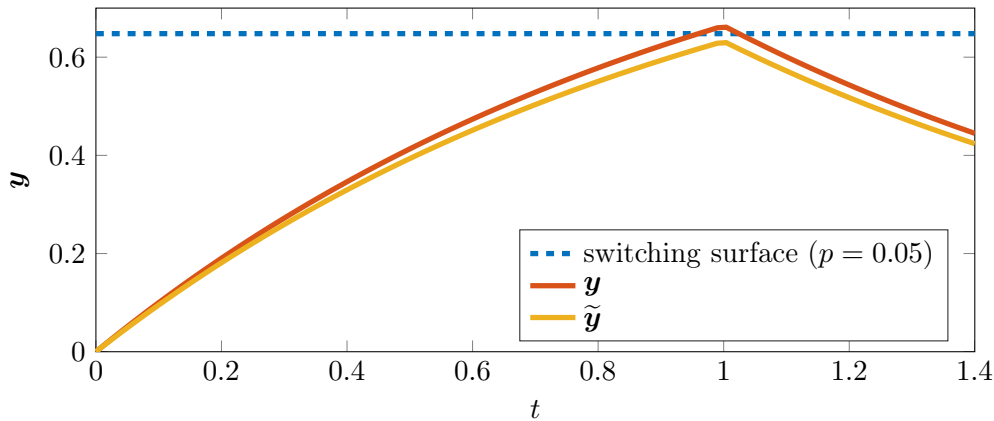		
	Thus, the approximation \eqref{eq:MORfailureSwitchingRed} driven with above input and switching condition never switches the system, while the original model \eqref{eq:MORfailureSwitching} switches to the second (not further specified) subsystem at $t_1$.
\end{example}

The failure of the approximation in the previous example is caused by the fact that for small $p$ the output approaches the switching surface approximately tangentially, which makes the system ill-posed for this input and one may need a regularization \cite{KunM17ppt}. 
Since we do not know the input a priori, we cannot exclude such cases. 
Anyhow, the previous example shows that independent from the approximation quality of the subsystem, the error may become arbitrarily large, if the switching depends on the state or the output. 
Thus, we cannot expect to have error estimators for output-dependent switching laws, implying that we need the following assumption to derive an error bound.

\begin{assumption}
	\label{ass:switching}
	The switching depends only on time, i.\,e. $\switch:[0,\infty)\to\{1,\ldots,\switchNr\}$.
\end{assumption}

\section{Envelope system and model reduction}
\label{sec:envelopeSystem}
Consider the switched system \eqref{eq:switchedSystem} with switching time points $0 = t_0 < t_1 < \ldots < t_s$ and switching sequence $i_k \vcentcolon= \sigma(t_k)$. 
In order to circumvent the problems that come with switching between different subsystems, we propose the following view:
For a given input $\inVar$, is it possible to modify $\inVar$ on the interval $(t_1,t_2]$ such that the unswitched system driven with the modified input produces the same output as the switched system with the original input in the interval $[0,t_2]$? 
If this is the case, then we could run the first subsystem without the need to switch and still observe the same behavior.

However, it is in general not possible to excite the first subsystem such that it behaves as the second subsystem. 
For an example we refer to \cite{SunGL02}. 
Instead we construct an \emph{envelope system} of the first subsystem with extended in- and outputs that is able to mimic the behavior of all other modes of the switched system. 
Note that a similar viewpoint was considered for parametric MOR in \cite{BauBB14} and for MOR of systems with time delay in \cite{WouMB15}.

Let us describe for $i=2,\ldots,\switchNr$ the difference of the $i$th subsystem to the first subsystem via the matrices
\begin{equation}
	\label{eq:differenceSwitching}
	\begin{aligned}
		\Delta_{\A,i} &\vcentcolon= \A_1 - \A_i \in\mathbb{R}^{\stateDim\times\stateDim}, & 
		\Delta_{\B,i} &\vcentcolon= \B_1 - \B_i\in\mathbb{R}^{\stateDim\times\inDim},\\
		\Delta_{\C,i} &\vcentcolon= \C_1 - \C_i\in\mathbb{R}^{\outDim\times\stateDim}, & 
		\Delta_{\D,i} &\vcentcolon= \D_1 - \D_i\in\mathbb{R}^{\outDim\times\inDim}
	\end{aligned}
\end{equation}
and we additionally assume a decomposition
\begin{equation}
	\label{eq:lowRankPerturbation}
	\Delta_{\A,i} = \lowRankLeft_{i}\midMatr_{i}\lowRankRight_{i}^T\qquad\text{with}\qquad \lowRankLeft_{i},\lowRankRight_{i}\in\mathbb{R}^{\stateDim\times\perturbationDimA_i},\quad \midMatr_{i}\in\mathbb{R}^{\perturbationDimA_i\times\perturbationDimA_i}
\end{equation}
with $\perturbationDimA_i = \rank(\Delta_{\A,i})$. Notice that we can always obtain such a decomposition via singular value decomposition (SVD) or Schur decomposition. Using the decomposition \eqref{eq:lowRankPerturbation}, we introduce for $i=2,\ldots,\switchNr$ the new variables
\begin{align*}
	\switchVar_{i}(t) &\vcentcolon= \lowRankRight_{i}^T\state(t) \in\mathbb{R}^{\perturbationDimA_i}.
\end{align*}
Thus, using the Kronecker delta $\delta_{i,j}$, \eqref{eq:switchedSystem} can be rewritten as the descriptor system
\begin{equation}
	\left\{\begin{aligned}
		\dot{\state}(t) &= \A_1\state(t) - \sum_{i=2}^\switchNr \lowRankLeft_i\delta_{i,\switch}\midMatr_i\switchVar_i(t) + \left(\B_1 - \sum_{i=2}^\switchNr \delta_{i,\switch}\Delta_{\B,i}\right)\inVar(t),\\
		\switchVar_i(t) &= \lowRankRight^T_{i}\state(t),\\
		\outVar(t) &= \left(\C_1  - \sum_{i=2}^\switchNr \delta_{i,\switch}\Delta_{\C,i}\right)\state(t)  + \left(\D_1 - \sum_{i=2}^\switchNr \delta_{i,\switch}\Delta_{\D,i}\right)\inVar(t).
	\end{aligned}\right.
\end{equation}

Viewing $\delta_{i,\switch}\midMatr_i\switchVar_{i}$ as feedback control law, we can define the following abstraction of the switched system \eqref{eq:switchedSystem}.

\begin{defn}[Envelope system]
\label{def:envelopeSystem}
	Consider the switched system \eqref{eq:switchedSystem} and matrices $\Delta_{\B,i}$, $\Delta_{\C,i}$, $\Delta_{\D,i}$, $\lowRankLeft_{i}$, and $\lowRankRight_{i}$ as in \eqref{eq:differenceSwitching} and \eqref{eq:lowRankPerturbation}. Define 
	\begin{align*}
		\eA &\vcentcolon= \A_1\in\mathbb{R}^{\stateDimE\times\stateDimE},\\
		\eB &\vcentcolon= \begin{bmatrix}
			\B_1 & \Delta_{\B,2} & \cdots & \Delta_{\B,\switchNr} & \lowRankLeft_2 & \cdots & \lowRankLeft_{\switchNr}
		\end{bmatrix}\in\mathbb{R}^{\stateDimE\times\inDimE},\\
		\eC^T &\vcentcolon= \begin{bmatrix}
			\C_1^T & \Delta_{\C,2}^T & \cdots & \Delta_{\C,\switchNr}^T & \lowRankRight_2 & \cdots & \lowRankRight_\switchNr
		\end{bmatrix}\in\mathbb{R}^{\stateDimE\times\outDimE},\\
		\eD &\vcentcolon= \blkdiag(\D_1,\Delta_{\D,2},\ldots,\Delta_{\D,\switchNr},0_{\mathfrak{L},\mathfrak{L}})\in\mathbb{R}^{\outDimE\times\inDimE},
	\end{align*}
	with dimensions $\stateDimE\vcentcolon= \stateDim$, $\inDimE \vcentcolon= \switchNr\inDim + \mathfrak{L}$, $\outDimE \vcentcolon= \switchNr\outDim + \mathfrak{L}$, and $\mathfrak{L} \vcentcolon=\sum_{i=2}^\switchNr \perturbationDimA_i$. Then we call 
	\begin{equation*}
		\label{eq:envelopeSystem}
		\eSystem = \left\{\begin{aligned}
			\deState(t) &= \eA\eState(t) + \eB\eInVar(t),\\
			\eOutVar(t) &= \eC\eState(t) + \eD\eInVar(t),\\
		\end{aligned}\right.\qquad t\geq 0
	\end{equation*}
	an \emph{envelope system} for the switched system \eqref{eq:switchedSystem}.
\end{defn}

Notice that we have included the $\lowRankRight_i$ matrices in the output matrix $\eC$, since in terms of model reduction it is not sufficient to have a good approximation of the output variable $\outVar$, but we need in addition a good approximation of $\switchVar_i$ to compute the feedback control law.

\begin{remark}
	\label{rem:computationSVD}
	In a large-scale setting the computation of \eqref{eq:lowRankPerturbation} via a full SVD is computationally expensive. However, the decomposition \eqref{eq:lowRankPerturbation} requires only a skinny SVD, which can be computed efficiently, see \cite{Ber92}.
\end{remark}

\begin{remark}
	\label{rem:inputOutputCompression}
	Notice that the envelope system is not unique. Besides different low-rank factorizations \eqref{eq:lowRankPerturbation}, also the ordering of the subsystems in \eqref{eq:switchedSystem} is an arbitrary choice that effects the construction of the envelope system. Moreover, it may be computationally advantageous to compute the difference of the subsystems to a hypothesized system $\system=(\A,\B,\C,\D)$. This may be particularly useful, if neither of the subsystems is asymptotically stable but all subsystems are low-rank perturbations of an asymptotically stable system. Moreover, it is a priori not clear that $\eB$ and $\eC$ have full rank and thus a rank-revealing decomposition of $\eB$ and $\eC$ may be used to compress the number of inputs and outputs.
\end{remark}

The connection of the envelope system \eqref{eq:envelopeSystem} to the original switched system \eqref{eq:switchedSystem} is as follows. For a given input $\inVar$, consider the feedback control $\eInVar = \feedback(\switch)\eOutVar + \feedback_0(\switch)\inVar$ with
\begin{equation}
	\label{eq:envelopeFeedback}
	\begin{aligned}
	\feedback(\switch) &\vcentcolon= \blkdiag(0_{\switchNr\inDim,\switchNr\outDim},-\delta_{2,\switch}\midMatr_2,\ldots,-\delta_{\switchNr,\switch}\midMatr_\switchNr)\in\mathbb{R}^{\inDimE\times\outDimE},\\
	\feedback_0(\switch) &\vcentcolon= \begin{bmatrix}
		I_\inDim & -\delta_{2,\switch}I_\inDim & \cdots & -\delta_{\switchNr,\switch}I_\inDim & 0_{\inDim,\mathfrak{L}}
	\end{bmatrix}^T \in \mathbb{R}^{\inDimE\times\inDim},
	\end{aligned}
\end{equation}
and the matrix function
\begin{equation}
	\label{eq:envelopeOutput}
	\C_0(\switch) \vcentcolon = \begin{bmatrix}
		I_\outDim & -\delta_{2,\switch}I_\outDim & \cdots & -\delta_{\switchNr,\switch}I_\outDim & 0_{\outDim,\mathfrak{L}}
	\end{bmatrix}\in \mathbb{R}^{\outDim\times\outDimE}.
\end{equation}
Then by construction, the envelope system \eqref{eq:envelopeSystem} with feedback law \eqref{eq:envelopeFeedback} reproduces the output of the LSS \eqref{eq:switchedSystem}. More precisely, we have the following result.

\begin{prop}
	\label{lem:envelopeFeedback}
	Consider the LSS \eqref{eq:switchedSystem} and its envelope system \eqref{eq:envelopeSystem}. For a given input $\inVar$ let $\outVar(t;\inVar)$ denote the output of \eqref{eq:switchedSystem} under this input, and accordingly, let $\eOutVar(t;\eInVar)$ denote the output of the envelope system with input $\eInVar$. Then
	\begin{displaymath}
		\outVar(t;\inVar) = \C_0(\switch)\eOutVar\left(t;\feedback(\switch)\eOutVar + \feedback_0(\switch)\inVar\right),
	\end{displaymath}
	where $\feedback$, $\feedback_0$, and $\C_0$ are defined as in \eqref{eq:envelopeFeedback} and \eqref{eq:envelopeOutput}, respectively.
\end{prop}

\begin{remark}
	If the envelope system is not minimal, then neither is the switched system and we can remove the uncontrollable and unobservable parts of the envelope system. Unfortunately, the converse is not true. Consider for example the switched system $\system = (\A_i,\B_i,\C_i,0\mid i=1,2)$ with
	\begin{displaymath}
		\A_1 = \begin{bmatrix}
			-1 & 0\\
			0 & -2
		\end{bmatrix},\qquad \Delta_{A,2} = \begin{bmatrix}
			1 & 0\\
			1 & 0
		\end{bmatrix} = \begin{bmatrix}
			1\\1
		\end{bmatrix}\begin{bmatrix}
			1 & 0
		\end{bmatrix},\qquad \B_1 = \B_2 = \begin{bmatrix}
			0\\0
		\end{bmatrix}.
	\end{displaymath}
	The reachable set of this system from the zero initial condition under arbitrary switching is $\{0\}$ and following \cite{SunGL02} thus not controllable. However, the envelope system
	\begin{displaymath}
		\eA = \begin{bmatrix}
			-1 & 0\\
			0 & -2
		\end{bmatrix},\qquad \eB = \begin{bmatrix}
			1\\1
		\end{bmatrix}
	\end{displaymath}
	is controllable.
\end{remark}

The advantage of the envelope system \eqref{eq:envelopeSystem} over the switched system \eqref{eq:switchedSystem} is that we can apply all standard MOR methods for MIMO LTI systems, for example balanced truncation \cite{Moo81} or IRKA \cite{GugAB08}. Since the construction of the envelope system is computationally tractable (provided that the matrices $\Delta_{\A,i}$ have low rank), MOR of switched systems within our framework is possible, whenever MOR for one of the subsystems is computationally feasible. 

Assume that the MOR method of choice has determined projection matrices $\leftProj,\rightProj\in\mathbb{R}^{\stateDim,\stateDimRed}$ for the envelope system \eqref{eq:envelopeSystem}. Then the reduced envelope system is given via
\begin{equation}
	\label{eq:envelopeSystemRed}
	\eAred = (\leftProj^T\rightProj)^{-1}\leftProj^T\eA\rightProj, \quad
	\eBred = (\leftProj^T\rightProj)^{-1}\leftProj^T\eB, \quad
	\eCred = \eC\rightProj, \quad 
	\eDred = \eD.
\end{equation}
On the other hand, we can reduce the switched system \eqref{eq:switchedSystem} with $\leftProj$ and $\rightProj$ according to \eqref{eq:projection} via
\begin{equation}
	\label{eq:switchedROM2}
	\Ared_i = (\leftProj^T\rightProj)^{-1}\leftProj^T\A_i\rightProj, \quad
	\Bred_i = (\leftProj^T\rightProj)^{-1}\leftProj^T\B_i, \quad
	\Cred_i = \C_i\rightProj, \quad 
	\Dred_i = \D_i
\end{equation}
for $i=1,\ldots,\switchNr$. We infer from
\begin{align*}
	\Delta_{\Ared,i} \vcentcolon= \Ared_1-\Ared_i = (\leftProj^T\rightProj)^{-1}\leftProj^T\Delta_{\A,i}\rightProj = (\leftProj^T\rightProj)^{-1}\leftProj^T \lowRankLeft_{i}\midMatr_{i}\lowRankRight_{i}^T\rightProj
\end{align*}
that the reduced envelope system \eqref{eq:envelopeSystemRed} is an envelope system for the reduced system \eqref{eq:switchedROM2} and thus \Cref{lem:envelopeFeedback} applies. 

Note that we can drive the envelope system with other input signals than the feedback law from \Cref{lem:envelopeFeedback} and thus can generate dynamics that the switched system cannot produce.
Similarly, standard LTI MOR methods like balanced truncation and IRKA aim for a good approximation for all possible input signals and are not tailored to a restricted input space, which in our case is parameterized due to the feedback law. 

\begin{remark}
	The larger the part of the system, that is affected by the switching, i.\,e. the larger $\sum_{i=2}^\switchNr \perturbationDimA_i$, the more we have to extend the input and the output dimensions of the envelope system thus making the system harder to reduce (provided $\rank(\eB) = \inDimE$ and $\rank(\eC) = \outDimE$, see \Cref{rem:inputOutputCompression}). 
	Notably, the controllability Gramian $\mathcal{P}_\mathrm{E}$ of the envelope system $\eSystem$ satisfies
	\begin{displaymath}
		0 = \eA\mathcal{P}_\mathrm{E} + \mathcal{P}_\mathrm{E}\eA^T + \eB\eB^T = \A_1\mathcal{P}_\mathrm{E} + \mathcal{P}_\mathrm{E}\A_1^T + \B_1\B_1^T + \mathcal{Q}
	\end{displaymath}
	for some $\mathcal{Q} = \mathcal{Q}^T\geq 0$ and is thus a generalized Gramian for the first subsystem, cf. \cite{ShaW09}. 
	In particular, the Hankel singular values of the envelope system are greater than or equal to the Hankel singular values of the first subsystem.
\end{remark}

\begin{remark}
	As outlined in \Cref{rem:inputOutputCompression}, the construction of the envelope system is not unique. 
	In particular, the decomposition \eqref{eq:lowRankPerturbation} allows to weight the subsystems differently (for example via scaling of $\lowRankLeft_{i},\midMatr_{i}$, and $\lowRankRight_{i}$) for the model reduction procedure. 
	This may be exploited, if the switching sequence is known beforehand. 
	A theoretical discussion and optimal scaling for the modes is however beyond the scope of this paper. 
	It is worth to mention that scaling issues are not unique to our methodology, but appear in other MOR methods, e.\,g. balanced truncation, as well. 
	Heuristics for the scaling are, for instance, proposed in \cite{HeiRA11}.
\end{remark}

As outlined in \Cref{ex:outputDependentSwitching} it is in general not possible to derive an error bound or error estimator for LSS with state- or output-dependent switching. 
Thus, we need to invoke \Cref{ass:switching}, which guarantees that the full-order model and the reduced model switch at the same time to the same subsystem. 
This in turn allows us to use error bounds that we obtain based on the model reduction of the envelope system. 
For instance, if we assume that the envelope system is minimal and asymptotically stable, then the error bound for balanced truncation \cite{Ant05} for square-integrable inputs $\eInVar$ is given by
\begin{equation*}
	\label{eq:errorBoundBT}
	\LtNorm{\eOutVar(\eInVar) - \eOutVarRed(\eInVar)} \leq \left(2\sum_{i=k}^q \sigma_i\right)\LtNorm{\eInVar},
\end{equation*}
where $\sigma_1> \sigma_2 > \ldots > \sigma_q>0$ denote the distinct Hankel singular values of $\eSystem$ with multiplicities $\nu_1,\ldots,\nu_q\in\mathbb{N}$ and $\sigma_k$ denotes the largest neglected Hankel singular value. 
Recall that $\eOutVar(\eInVar)$ denotes the solution of the envelope system \eqref{eq:envelopeSystem} with input $\eInVar$ and $\eOutVarRed(\eInVar)$ denotes the solution of the reduced envelope system \eqref{eq:envelopeSystemRed} with the same input $\eInVar$. 
Unfortunately, this error bound is not carried over to the error of the switched system, since the feedback in \Cref{lem:envelopeFeedback} is different for the envelope system and for the reduced envelope system. 
Thus we need to compare $\eOutVar(\eInVar)$ and $\eOutVarRed(\eInVarRed)$. 
Moreover, any error bound that is obtained in this way is of a posteriori type, since the input for the envelope system depends on the actual solution of the system.

\begin{theorem}
	\label{thm:errorBound}
	Let the switching signal satisfy \Cref{ass:switching} and for any square-integrable input $\inVar$ let $\eInVarRed \vcentcolon= \feedback(\switch)\eOutVarRed + \feedback_0(\switch)\inVar$ denote the feedback law from \Cref{lem:envelopeFeedback} for the reduced envelope system. Moreover assume that
	\begin{equation}
		\label{eq:HinfCondition}
		\max_{i=2,\ldots,\switchNr} \|\midMatr_i\|_2 \HinfNorm{\eSystem}<1.
	\end{equation}
	Then there exists a constant $\eta = \eta(\eSystem)>0$ independent of $\eSystemRed$ and $\inVar$ such that
	\begin{align}
		\label{eq:errorBoundHinf}
		\LtNorm{\outVar(\inVar) - \outVarRed(\inVar)} &\leq \eta \HinfNorm{\eSystem-\eSystemRed} \LtNorm{\eInVarRed}.
	\end{align}
	holds.
\end{theorem}

\begin{proof}
	First note that due to \Cref{ass:switching}, the matrix functions $C_0$, $\feedback$, and $\feedback_0$ are identical for the envelope system and the reduced envelope system. Let $\eInVar = \feedback(\switch)\eOutVar + \feedback_0(\switch)\inVar$ denote the feedback law from \Cref{lem:envelopeFeedback} for the envelope system. From \Cref{lem:envelopeFeedback} we infer $\outVar(\inVar) = C_0(\switch)\eOutVar(\eInVar)$ and $\outVarRed(\inVar) = C_0(\switch)\eOutVarRed(\eInVarRed)$. Since for any $\switch\in\{1,\ldots,\switchNr\}$ the largest singular value of $C_0$ is $\sqrt{2}$, we obtain
	\begin{align*}
		\LtNorm{\outVar(\inVar) - \outVarRed(\inVar)} &= \LtNorm{C_0(\switch)\left(\eOutVar(\eInVar) - \eOutVarRed(\eInVarRed)\right)} \leq \sqrt{2}\LtNorm{\eOutVar(\eInVar) - \eOutVarRed(\eInVarRed)}\\
		&\leq \sqrt{2}\LtNorm{\eOutVar(\eInVar) - \eOutVar(\eInVarRed)} + \sqrt{2}\LtNorm{\eOutVar(\eInVarRed)-\eOutVarRed(\eInVarRed)}\\
		&\leq \sqrt{2}\LtNorm{\eOutVar(\eInVar) - \eOutVar(\eInVarRed)} + \sqrt{2}\HinfNorm{\eSystem-\eSystemRed} \LtNorm{\eInVarRed}.
	\end{align*}
	Due to the definitions of $\eInVar$ and $\eInVarRed$ and \Cref{ass:switching} we get
	\begin{align*}
		\LtNorm{\eOutVar(\eInVar)-\eOutVar(\eInVarRed)} &\leq \HinfNorm{\eSystem}\LtNorm{\eInVar-\eInVarRed} = \HinfNorm{\eSystem}\LtNorm{\feedback(\sigma)\left(\eOutVar(\eInVar) - \eOutVarRed(\eInVarRed)\right)}\\
		&\leq \HinfNorm{\eSystem} \max_{i=2,\ldots,\switchNr} \|\midMatr_i\|_2 \LtNorm{\eOutVar(\eInVar) - \eOutVarRed(\eInVarRed)}.
	\end{align*}	
	From the triangle inequality and \eqref{eq:HinfCondition} we obtain
	\begin{align*}
		\LtNorm{\eOutVar(\eInVar)-\eOutVar(\eInVarRed)} \leq\frac{\displaystyle\max_{i=2,\ldots,\switchNr} \|\midMatr_i\|_2\HinfNorm{\eSystem}}{1 -\displaystyle\max_{i=2,\ldots,\switchNr} \|\midMatr_i\|_2\HinfNorm{\eSystem}} \HinfNorm{\eSystem-\eSystemRed}\LtNorm{\eInVarRed}
	\end{align*}
	and thus
	\begin{align*}
		\LtNorm{\outVar(\inVar) - \outVarRed(\inVar)} &\leq \frac{\sqrt{2}}{1 -\displaystyle\max_{i=2,\ldots,\switchNr} \|\midMatr_i\|_2\HinfNorm{\eSystem}} \HinfNorm{\eSystem-\eSystemRed} \LtNorm{\eInVarRed}.\qedhere
	\end{align*}
\end{proof}

Note that the switching information is encoded in $\LtNorm{\eInVarRed}$ and thus explicitly taken into account in the error bound. 
Nevertheless, the bound \eqref{eq:errorBoundHinf} seems rather restrictive, since in the proof we bound the output error $\outVar-\outVarRed$ by adding up all output errors of the envelope system $\eOutVar-\eOutVarRed$. 

\begin{remark}
	\label{rem:lowRankE}
	Although the theory presented above is tailored to systems in standard state-space form, it is easily generalized to generalized state-space form, i.\,e., a switched system of the form
	\begin{displaymath}
		\E_\switch\dot{\state}(t) = \A_\switch\state(t) + \B_\switch\inVar(t), \qquad \outVar(t) = \C_\switch\state(t) + \D_\switch\inVar(t)
	\end{displaymath}
	with nonsingular matrices $\E_i\in\mathbb{R}^{\stateDim\times\stateDim}$. Setting $\Delta_{\E,i} \vcentcolon= \E_1-\E_i$, we have 
	\begin{displaymath}
		\E_i^{-1} = (\E_1 - \Delta_{\E,i})^{-1} = \E_1^{-1} + \E_1^{-1}\Delta_{\E,i}\E_i^{-1}	
	\end{displaymath}
	for $i=2,\ldots,\switchNr$. Transformation to standard state-space form thus transforms the  perturbations in \eqref{eq:lowRankPerturbation} as
	\begin{align*}
		\Delta_{\A,i} &\mapsto \E_1^{-1}\Delta_{\A,i} - \E_1^{-1}\Delta_{\E,i}\E_i^{-1}\A_i, &
		\Delta_{\B,i} &\mapsto \E_1^{-1}\Delta_{\B,i} - \E_1^{-1}\Delta_{\E,i}\E_i^{-1}\B_i.
	\end{align*}
	In particular, if the perturbations have a low-rank structure, then the transformed perturbations are also low-rank and thus the framework presented above can be applied.
\end{remark}

\section{Structure preservation}
In the last decades, a lot of research in model reduction is devoted to structure preservation \cite{BeaG09,SchUBG18,GugPBS12,Fre08,JarDM13,LalKM03} and preservation of system properties like stability or passivity. The latter properties are directly encoded in so called \emph{port-Hamiltonian} (pH) systems \cite{SchJ14}, which can be written in the form 
\begin{equation}
	\label{eq:pH}
	\left\{
	\begin{aligned}
		\dot{\state}(t) &= \left(\J-\R\right)\Q\state(t) + \B\inVar(t),\\
		\outVar(t) &= \B^TQ\state(t),\\
	\end{aligned}
	\right.
\end{equation}
with $\J=-\J^T$, $\R=\R^T\geq0$, and $\Q=\Q^T>0$. 
In this context any square matrix $\A$ of the form $\A = (\J-\R)\Q$ is called a \emph{dissipative Hamiltonian} matrix \cite{GilS17,MehMW18}. The special structure in \eqref{eq:pH} allows us to use the Hamiltonian $\Ham(\state)=1/2\state^T\Q\state$, which often represents the total energy of the system, as a Lyapunov function, thus showing that pH systems are stable. 
Let us mention the recent work \cite{BeaMXZ17_ppt,MehMW18} for a more general form of \eqref{eq:pH}.

In terms of switched systems, stability is usually not defined for each subsystem separately, but in terms of a joint Lyapunov function. More precisely, an LSS \eqref{eq:switchedSystem} is called \emph{quadratic Lyapunov stable} \cite{Lib03} if there exists a symmetric positive definite matrix $\lyap\in\mathbb{R}^{\stateDim\times\stateDim}$ that satisfies the linear matrix inequalities
\begin{equation}
	\label{eq:quadraticStability}
	\A_i^T\lyap + \lyap A_i < 0\qquad\text{for all}\ i\in\{1,\ldots,\switchNr\}.
\end{equation}
Following \cite[Lemma~2]{GilS17}, we immediately have the following connection between quadratic Lyapunov stable LSS and dissipative Hamiltonian systems.
\begin{prop}
	\label{prop:stableDH}
	The LSS \eqref{eq:switchedSystem} is quadratic Lyapunov stable, if and only if there exist some $\lyap\in\mathbb{R}^{\stateDim\times\stateDim}$ with $\lyap = \lyap^T>0$ and matrices $\J_i, \R_i\in\mathbb{R}^{\stateDim\times\stateDim}$ with $\J_i^T = -\J_i$ and $\R_i = \R_i^T> 0$ such that $\A_i = (\J_i-\R_i)\lyap$ for $i=1,\ldots,\switchNr$.
\end{prop}

\begin{proof}
	Suppose that the switched system \eqref{eq:switchedSystem} is quadratic Lyapunov stable, i.\,e. there exists $\lyap\in\mathbb{R}^{\stateDim\times\stateDim}$ with $\lyap = \lyap^T>0$ such that \eqref{eq:quadraticStability} holds. For $i=1,\ldots,\switchNr$ define
	\begin{displaymath}
		\J_i \vcentcolon= \frac{1}{2}\left(\A_i\lyap^{-1} - \lyap^{-1}\A_i^T\right)\qquad\text{and}\qquad
		\R_i \vcentcolon= -\frac{1}{2}\left(\A_i\lyap^{-1} + \lyap^{-1}\A_i^T\right).
	\end{displaymath}
	Then $\J_i^T = -\J_i$, $\R_i = \R_i^T$ and $\A_i = (\J_i-\R_i)\lyap$. The $\R_i$s are positive definite if and only if \eqref{eq:quadraticStability} holds, since
	\begin{align*}
		\state^T \R_i \state &= -\frac{1}{2}\state^T\left(\A_i\lyap^{-1} + \lyap^{-1}\A_i^T\right)\state = -\frac{1}{2}(\lyap^{-1}\state)^T\left(\A_i^T\lyap + \lyap\A_i\right)\lyap^{-1} \state
	\end{align*}
	for each $\state\in\mathbb{R}^{\stateDim}$, which also shows the converse direction.
\end{proof}
Using \Cref{prop:stableDH}, we see that for a quadratic Lyapunov stable LSS the envelope system \eqref{eq:envelopeSystem} is also a dissipative Hamiltonian system \cite{GilS17} and hence we can use any of the techniques derived for structure-preserving model reduction for port-Hamiltonian systems \cite{ChaBG16,GugPBS12,WolLEK10} to ensure quadratic stability in the reduced switched system as well. 
Following \cite{GugPBS12}, we obtain for instance the following result.
\begin{theorem}
	\label{thm:quadraticStability}
	Suppose that the LSS \eqref{eq:switchedSystem} is quadratic Lyapunov stable, i.\,e. \eqref{eq:switchedSystem} satisfies \eqref{eq:quadraticStability} for some $\lyap\in\mathbb{R}^{\stateDim\times\stateDim}$ with $\lyap = \lyap^T>0$. 
	For any $\rightProj\in\mathbb{R}^{\stateDim\times\stateDimRed}$ with full column rank, set $\leftProj = \lyap\rightProj(\rightProj^T\lyap\rightProj)^{-1}$. 
	Then the reduced switched system $\systemRed = (\Ared_i,\Bred_i,\Cred_i,\Dred_i\mid i=1,\ldots,\switchNr)$ with $\Ared_i,\Bred_i,\Cred_i,\Dred_i$ obtained as in \eqref{eq:switchedROM2} is quadratic Lyapunov stable.
\end{theorem}
\begin{proof}
	For $i=1,\ldots,\switchNr$ we have
	\begin{align*}
		\Ared_i^T\rightProj^T\lyap\rightProj + \rightProj^T\lyap\rightProj\Ared_i &= 
		\rightProj^T\A_i^T\leftProj\rightProj^T\lyap\rightProj +  \rightProj^T\lyap\rightProj\leftProj^T\A_i\rightProj\\
		&= \rightProj^T(\A_i^T\lyap + \lyap\A_i)\rightProj < 0.\qedhere
	\end{align*}
\end{proof}
From a modeling perspective, the matrix $\lyap$ is usually not known beforehand and hence we may only assume that each mode of the switched system exhibits a port-Hamiltonian structure, i.\,e.
\begin{equation}
	\label{eq:pHLSS}
	\left\{
	\begin{aligned}
	\dot{\state}(t) &= \left(\J_\switch-\R_\switch\right)\Q_\switch\state(t) + \B_\switch\inVar(t),\\
		\outVar(t) &= \B^T_\switch\Q_\switch\state(t),\\
	\end{aligned}
	\right.
\end{equation}
with $\J_i = -\J_i^T$, $\R_i = \R_i^T\geq 0$, and $\Q_i = \Q_i^T>0$. Instead of asking to preserve potential quadratic stability of the switched system, we can also aim for preserving the port-Hamiltonian structure of each subsystem, which in turn implies that each subsystem is stable and passive. 
Note that stability of the subsystems does not imply quadratic stability of the switched system, for a counter example we refer to \cite{Lib03}.

\begin{remark}
	If for $i=1,\ldots,\switchNr$ the matrices $\R_i$ in \eqref{eq:pHLSS} are positive definite and $\Q_i = \alpha_i\Q_0$ for some $\alpha_i>0$ and symmetric positive definite matrix $\Q_0$, then the previous discussion immediately implies that the LSS \eqref{eq:pHLSS} is quadratic Lyapunov stable. 
\end{remark}

A structure-preserving model reduction scheme for the LSS \eqref{eq:pHLSS} can be obtained by a straightforward extension of \Cref{thm:quadraticStability} and \cite{GugPBS12}. This is summarized in the following result.
\begin{theorem}
	\label{thm:LSSpH}
	Consider the  LSS \eqref{eq:pHLSS}. For any $\rightProj_i\in\mathbb{R}^{\stateDim\times\stateDimRed}$ with full column rank, set $\leftProj_i = \Q_i\rightProj_i(\rightProj_i^T\Q_i\rightProj_i)^{-1}$ for $i=1,\ldots,\switchNr$. Then each mode of the reduced switched system in \eqref{eq:switchedROM} exhibits a port-Hamiltonian structure.
\end{theorem}

\Cref{thm:quadraticStability} and \Cref{thm:LSSpH} both do not further specify the projection matrices $\rightProj$ and $\rightProj_i$. Following our framework from \cref{sec:envelopeSystem}, we can compute the envelope system and apply any MOR method of choice to determine projection matrices $\leftProj,\rightProj\in\mathbb{R}^{\stateDim\times\stateDimRed}$. In general, the left projection matrix $\leftProj$ does not satisfy the condition in \Cref{thm:quadraticStability} and \Cref{thm:LSSpH} and thus needs to be modified accordingly. Let us mention that one can modify the construction of $\rightProj$ within the IRKA framework, such that some necessary $\mathcal{H}_2$ optimality conditions are satisfied while at the same time the port-Hamiltonian structure is preserved \cite{GugPBS12}.

So far, we have considered the cases when either a common quadratic Lyapunov function or a port-Hamiltonian structure of the subsystems of \eqref{eq:switchedSystem} is known. 
However, if neither of them is available we can still preserve potential stability of the subsystems, for example, by formulating different envelope systems for the different modes with $\eA=A_i$ etc.\ for $i=2,\ldots,\switchNr$ and reducing each of them separately with a stability-preserving model reduction scheme. 
In this case the transformation matrices in \eqref{eq:stateTransformation} are in general not identity matrices. However, we postpone a detailed analysis of this situation to future work.
An alternative way of preserving stability is to write each stable $A_i$ as $A_i=(J_i-R_i)Q_i$, cf.\ \cite[Lemma~2]{GilS17}, and afterwards apply a projection as in \Cref{thm:LSSpH}.

\section{Numerical examples}

We illustrate the proposed method by means of numerical examples and compare the reduced-order models with the respective full-order model. 
In the figure legends, we use the notation FOM for the full-order model and $\eSystemRed$ for the reduced-order models obtained by the method presented in \Cref{sec:envelopeSystem}. 
The supplement IRKA refers to the usage of IRKA \cite{AntBG10} for reducing the envelope system and BT  to balanced truncation \cite{Moo81}. 
In \Cref{ex:random} we also compare our approach with the method introduced in \cite{BasPWL16}, which we denote with \emph{nice selection} in the figure legends.

For constructing the envelope system, we use an SVD $\Delta_{\A,i} = \svdLeft_{i}\svdMid_{i}\svdRight_i^T$ for the decomposition in \eqref{eq:lowRankPerturbation} and we assign $\lowRankLeft_{i}=\svdLeft_{i}$, $\midMatr_{i}=I_{\perturbationDimA_i}$, and $\lowRankRight_{i} = \svdRight_i\svdMid_{i}$. 
Furthermore, all time simulations are performed with the MATLAB solver \texttt{ode45} with default settings except for \Cref{ex:twoRoomsStateDependentSwitching} where we used a backward Euler scheme. 
The source code for the numerical examples is available from \cite{SchU18_code}.

\begin{example}
	\label{ex:random}
	Consider the first numerical example from \cite{BasPWL16}, which is a single-input single-output (SISO) LSS with two subsystems ($\switchNr = 2$) and $\stateDim = 11$. 
	The system matrices are random matrices and thus from a model reduction perspective we cannot expect a good approximation quality with a smaller dimension. 
	Anyhow, using randomly generated systems appears to be a benchmark example in the MOR literature for switched systems \cite{ShaW09,ShaW12,BasPWL14b,BasPWL16}. 
	Notice that for a randomly generated system the difference matrices $\Delta_{\A,i}$ have full rank and thus the input and output matrices of the envelope system both have rank $\stateDim$. 
	\begin{figure}[ht]
		\centering
		\input{IEEETacExample1-HankelSingularValues}
		\caption{Normalized Hankel singular values of the two subsystems ($\system_1$ and $\system_2$) and of the envelope system $\eSystem$ for \Cref{ex:random}}
		\label{fig:randomHankelSingularValues}
	\end{figure}
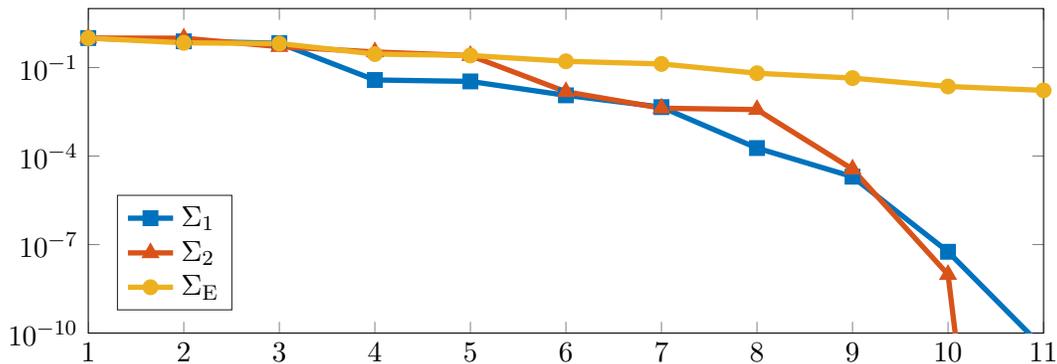		
	As expected the decay of the Hankel singular values of the two subsystems and of the envelope system (cf. \Cref{fig:randomHankelSingularValues}) is slow and hence we do not expect a good approximation. 
	We consider two switching signals
	\begin{displaymath}
		\switch_1(t) = \begin{cases}
			2, & t\in[0,0.2]\cup (0.6,0.8],\\
			1, & t\in(0.2,0.6]\cup (0.8,1.2],
		\end{cases}\qquad
		\switch_2(t) = \begin{cases}
			1, & t\in[0,0.4]\cup (0.7,0.9],\\
			2, & t\in(0.4,0.7]\cup (0.9,1.2]
		\end{cases}
	\end{displaymath}
	and input functions $\inVar_1(t) = \sin(2\pi t)$ and $\inVar_2(t) = \exp(-t)$. For the model reduction, we pick $\stateDimRed = 8$ and perform the reduction based on the envelope system $\eSystem$ with balanced truncation and IRKA. The time simulations for above switching and input signals are presented in \Cref{fig:randomTimeSimulation}. 
	\begin{figure}[ht]
		\centering
		\begin{subfigure}[b]{1\linewidth}
			\centering
			\input{IEEETacExample1-TimeDomainSimulation-r8-v1}
			\caption{Time domain simulation with switching signal $\switch_1$ and input $\inVar_1$}
			\label{fig:randomTimeSimulation1}
		\end{subfigure}\\[.5em]
		\begin{subfigure}[b]{1\linewidth}
			\centering
			\input{IEEETacExample1-TimeDomainSimulation-r8-v2}
			\caption{Time domain simulation with switching signal $\switch_2$ and input $\inVar_2$}
			\label{fig:randomTimeSimulation2}
		\end{subfigure}
		\caption{Comparison of the full order model (FOM) and the reduced models of dimension $\stateDimRed=8$ for \Cref{ex:random}}
		\label{fig:randomTimeSimulation}
	\end{figure}
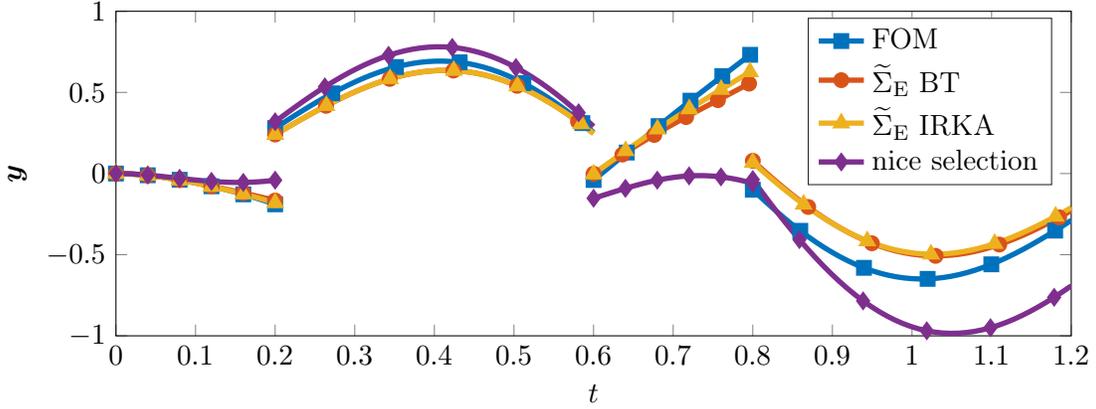
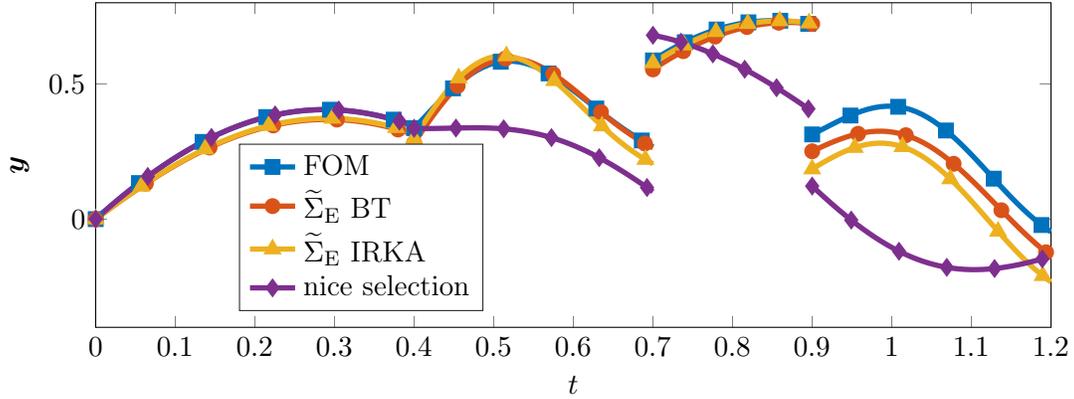
	As expected from the singular value decay, there is a significant error between the full-order model (FOM) and the envelope reduced systems. 
	\begin{table}[ht]
		\centering
		\caption{$\mathcal{L}_2$ and $\mathcal{L}_\infty$ errors for the output of the reduced models in \Cref{ex:random}}
		\label{tab:exrandom}
		\begin{tabular}{lrrrr}
			\toprule
			& \multicolumn{2}{c}{$\inVar_1$} & \multicolumn{2}{c}{$\inVar_2$}\\\cmidrule(l){2-3} \cmidrule(l){4-5}
			& $\LtNorm{\cdot}$ & $\LinfNorm{\cdot}$ & $\LtNorm{\cdot}$ & $\LinfNorm{\cdot}$\\\midrule
			$\eSystemRed$ BT $r=8$ & 1.15e-02 & 3.29e-01 & 3.06e-03 & 2.97e-01\\
			$\eSystemRed$ IRKA $r=8$ & 1.09e-02 & 3.82e-01 & 7.94e-03 & 3.09e-01\\
			nice selection $r=8$ & 7.42e-02 & 8.05e-01 & 6.67e-02 & 5.44e-01\\\bottomrule
		\end{tabular}
	\end{table}
	Still the envelope systems capture the qualitative behavior of the switched system quite accurately compared to the methodology from \cite{BasPWL16}. For a quantitative assessment see \Cref{tab:exrandom}.
\end{example}

Since our approach clearly outperforms the method from \cite{BasPWL16} for both input signals in \Cref{ex:random} and in addition needs considerably less user choices for the reduction, we use in the following examples only reduced models based on the envelope system.

\begin{example}
	\label{ex:twoRooms2}
	Continuing with \Cref{ex:twoRooms}, we ignore effects of convection and model the time evolution of the temperature distribution by the heat equation
	\begin{align}
	\label{eq:heatEquation}
		\rhoc(\xi,\switch)\partial_t z(t,\xi) - \partial_{\xi} \left(k(\xi,\switch)\partial_{\xi}z(t,\xi)\right) &= 0 && \text{for } (t,\xi)\in [0,T]\times \Omega,
	\end{align}
	where $k$ is the \emph{thermal conductivity}, $\rhoc$ the \emph{volumetric heat capacity}, and $z$ is the relative temperature difference with respect to the ambient temperature $\theta_{\infty}$, i.\,e., $z(t,\xi)=\theta(t,\xi)-\theta_{\infty}$ with absolute temperature $\theta$. The thermal conductivity and the volumetric heat capacity are assumed to be homogeneous over the total domain apart from the fact that the properties in the second subdomain $\Omega_2$ are affected by a switch, i.\,e.,
	\begin{alignat*}{3}
	k(\xi,\switch) &= \begin{cases} k_{\Omega_2}\left(\switch\right), &\text{if } \xi\in\Omega_2,\\
								  k_2,&\text{otherwise},\end{cases}\qquad
	&&\rhoc(\xi,\switch) = \begin{cases} \rhoc_{\Omega_2}\left(\switch\right), &\text{if } \xi\in\Omega_2,\\
										 \rhoc_2,&\text{otherwise},\end{cases}\\
	k_{\Omega_2}\left(\switch\right) &=\begin{cases} k_1, & \text{if } \switch=1,\\
													 k_2, & \text{if } \switch=2,\end{cases}\qquad	
	&&\rhoc_{\Omega_2}\left(\switch\right) =\begin{cases} \rhoc_1, & \text{if } \switch=1,\\
																 \rhoc_2, & \text{if } \switch=2.\end{cases}									
	\end{alignat*}	
The heater at the left boundary acts as a boundary control and we assume that at the right side of room~2 there is a heat transfer to the environment, i.\,e., we assume homogeneous Robin boundary conditions at this point, cf. \cite{Oez68}. More precisely, we have the boundary conditions
	\begin{alignat}{3}
	\label{eq:BCleft}
		-k_2\partial_\xi z(t,0) &= \inVar(t),\quad && \text{for}\ t\in[0,T],\\
		\label{eq:BCright}
		k_2\partial_{\xi} z(t,L)+hz(t,L) &= 0,\quad && \text{for}\ t\in[0,T],
	\end{alignat}
	where $L$ is the total length of the computational domain $\Omega$ and $h$ denotes the heat transfer coefficient at the right boundary. We furthermore assume that the initial temperature distribution in both rooms and the door equals the ambient temperature yielding the initial condition $z(0,\xi) = 0$ for $\xi\in\Omega$. As output we use the average temperature in the second room, which is
	\begin{equation*}
		\outVar(t) = \frac{1}{|\Omega_3|}\int_{\Omega_3} z(t,\xi)\,\mathrm{d}\xi,\quad \text{for } t\in[0,T].
	\end{equation*}
	Discretization with finite volumes leads to a switched system of the form \eqref{eq:switchedSystem} with coefficient matrices specified in \Cref{sec:appendix}, where the first mode corresponds to the closed door and the second one to the open door. 
	The used physical, geometry, and discretization parameters are listed in \Cref{tab:twoRoomsPhysics} and \Cref{tab:twoRoomsGeometry}. 
	Here we assume that the (closed) door acts like an isolator, i.\,e., with low thermal conductivity $k_1$ and high heat capacity $\rhoc_1$.
{\renewcommand{\arraystretch}{1.2}
\begin{table}[ht]
	\centering
	\caption{\Cref{ex:twoRooms2} -- Physical parameters}
	\label{tab:twoRoomsPhysics}
	\begin{tabular}{ccccc}
		$\rhoc_1\,\left[\frac{\si{J}}{\si{m^3K}}\right]$ & $\rhoc_2\,\left[\frac{\si{J}}{\si{m^3K}}\right]$ & $k_1\,\left[\frac{\si{W}}{\si{mK}}\right]$ & $k_2\,\left[\frac{\si{W}}{\si{mK}}\right]$ & $h\,\left[\frac{\si{W}}{\si{m^2K}}\right]$\\\midrule
		$2\times 10^6$ & $700$ & $0.015$ & $3$ & $100$
	\end{tabular}
\end{table}
}
\begin{table}[ht]
	\centering
	\caption{\Cref{ex:twoRooms2} -- Geometry and discretization parameters}
	\label{tab:twoRoomsGeometry}
	\begin{tabular}{lccc}
		& $\left| \Omega_i\right|\,\left[m\right]$ & $\Delta \xi_i\,\left[m\right]$ & $n_i$\\\toprule
		$i\in\left\lbrace 1,3\right\rbrace$ & $5$ & $0.1$ & $50$\\\hline
		$i=2$ & $0.3$ & $0.1$ & $3$
	\end{tabular}
\end{table}

The Hankel singular values of the two subsystems and of the envelope system are depicted in \Cref{fig:twoRoomsHankelSingularValues}. 
Since the Hankel singular values decay is fast for all three systems we expect (in contrast to \Cref{ex:random}) a good approximation quality with a small dimension.
\begin{figure}[ht]
		\centering
		\input{TwoRoomsStatic-HankelSingularValues}
		\caption{\Cref{ex:twoRooms2} -- Normalized Hankel singular values of the two subsystems ($\system_1$ and $\system_2$) and of the envelope system $\eSystem$}
		\label{fig:twoRoomsHankelSingularValues}
	\end{figure}
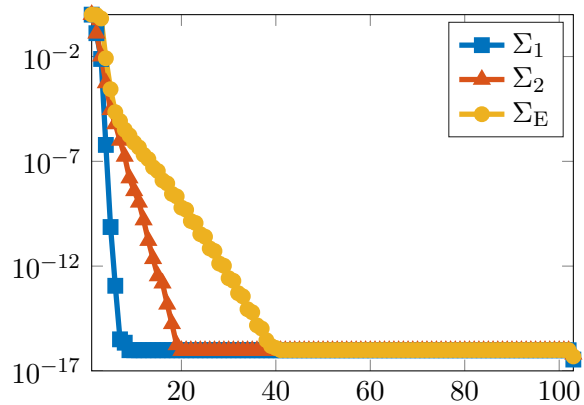
We construct reduced-order models based on IRKA and based on balanced truncation of the envelope system with reduced dimensions $r=6$ and $r=10$. As input we use a constant heat flux density of $\inVar(t)\equiv 1\si{W}/(\si{m^2})$ and the switching sequence is given by
\begin{equation*}
		\switch(t) = \begin{cases}
			2, & t\in[0,1.1]\cup (1.6,1.7],\\
			1, & t\in(1.1,1.6]\cup (1.7,6].
		\end{cases}
	\end{equation*}
\Cref{fig:twoRoomsStaticSimulation} shows the time evolution of the output of the full-order model and of the reduced-order models. 
At the beginning the door is open and the temperature in the second room increases until the system is switched the first time. 
Afterwards, the output temperature decreases since the isolating door separates the second room from the heated first one. 
After each switching the output dynamics changes abruptly depending on whether the door is open or not. 
This behavior of the full-order model ($n=103$) is captured by the reduced-order models of dimension $10$, whose graphs lie on top of the one of the full-order model (the $\mathcal{L}_\infty$ errors are less than $10^{-2}$). 
On the other hand, the reduced-order models of dimension $6$ show clear quantitative deviations from the full-order model while the qualitative behavior is still captured reasonably well. 
The significant difference between the reduced models of dimension $6$ and those of dimension $10$ can be explained by the strong decay between the sixth and the tenth Hankel singular value of the envelope system, cf. \Cref{fig:twoRoomsHankelSingularValues}.
\begin{figure}[ht]
		\centering
		\input{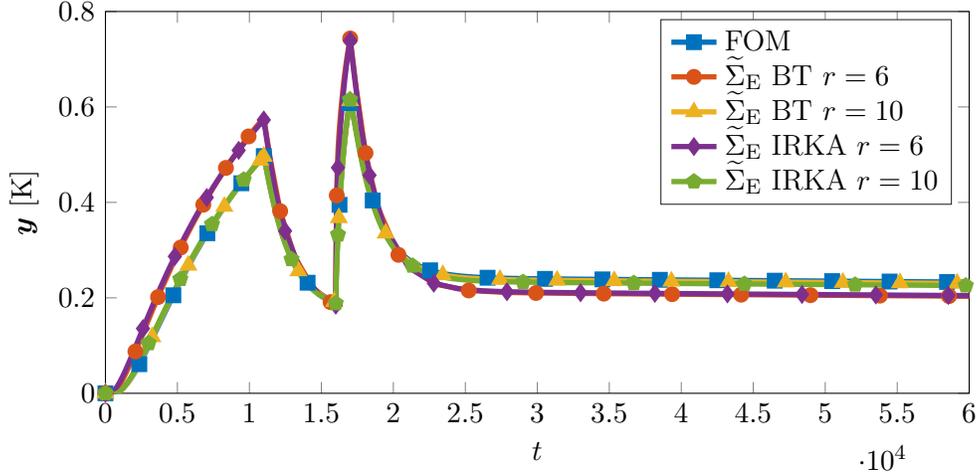}
		\caption{\Cref{ex:twoRooms2} -- Time domain simulation of the full-order model (FOM) and the reduced-order models}
		\label{fig:twoRoomsStaticSimulation}
	\end{figure}
\end{example}

In \Cref{ex:outputDependentSwitching} we demonstrated with a toy example that output dependent switching may pose a severe challenge for MOR. Indeed, the next example illustrates that using an output dependent switching law in the previous example may result in inaccurate approximations.

\begin{example}
	\label{ex:twoRoomsStateDependentSwitching}
	We revisit the previous \Cref{ex:twoRooms2} and consider an output-dependent switching signal, that closes the door whenever the average temperature in the second room becomes too high and opens the door, if the average temperature falls below a certain threshold. The corresponding switching law reads
\begin{equation*}
		\begin{cases}
			\mathrm{if}\left(\sigma=1\land \outVar\left(t\right)<\vartheta_{1}\right), & \mbox{switch to }\sigma=2,\\
			\mathrm{if}\left(\sigma=2\land \outVar\left(t\right)>\vartheta_{2}\right), & \mbox{switch to }\sigma=1.
		\end{cases}
\end{equation*}
To avoid a permanent switching between the modes, the threshold temperatures $\vartheta_1$ and $\vartheta_2$ should be chosen such that $\vartheta_1<\vartheta_2$. 
Here, we choose $\vartheta_1=0.2\si{K}$ and $\vartheta_2=0.5\si{K}$. 
The other parameters and the input signal are chosen as before, see \Cref{ex:twoRooms2}. Moreover, the initial value for $\sigma$ is set to $2$ again, which refers to an open door.

In \Cref{fig:twoRoomsDynamicSimulation}, the simulated time evolution of the output of the full-order model and of the reduced-order models from \Cref{ex:twoRoomsStateDependentSwitching} is depicted. 
\begin{figure}[ht]
		\centering
		\input{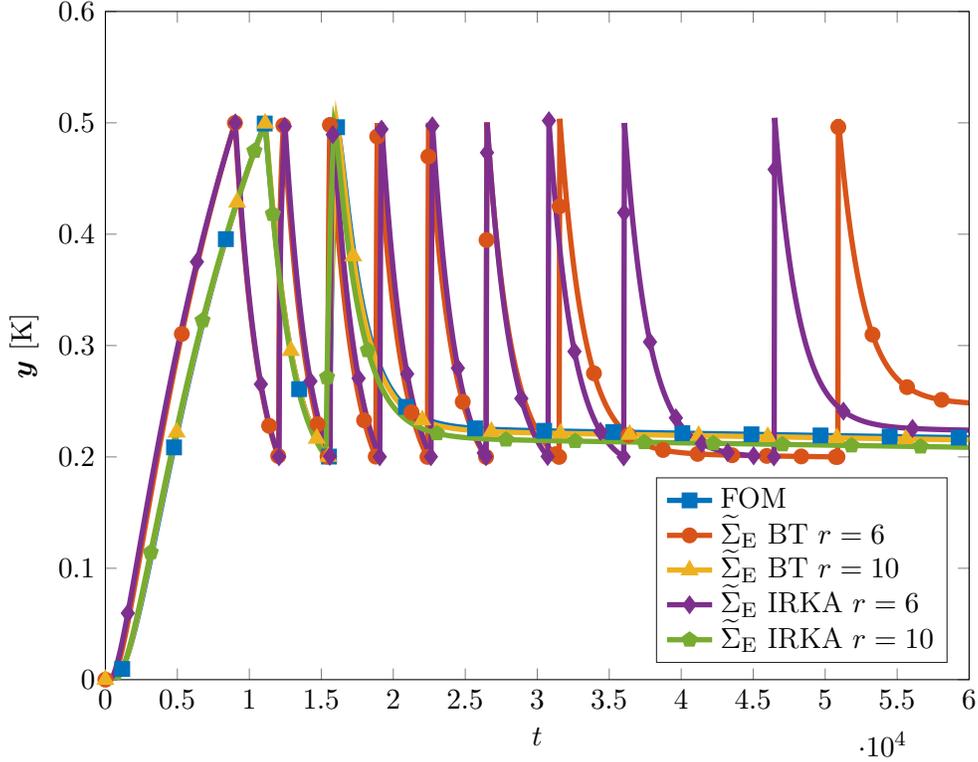}
		\caption{\Cref{ex:twoRoomsStateDependentSwitching} -- Time domain simulation of the full-order model (FOM) and the reduced-order models}
		\label{fig:twoRoomsDynamicSimulation}
	\end{figure}
The dynamics of the full-order model look similar as in the case of the time-dependent switching. 
The ROM of dimension $10$ is again sufficient to approximate the dynamics very well, although the approximation error is larger compared to \Cref{ex:twoRooms2}. 
While in \Cref{ex:twoRooms2} the ROM of dimension $6$ was at least able to capture the qualitative behavior, in the output-dependent switching setting, the output of the ROM of order $6$ reveals more switches than the full-order model and consequently the error is large, see \Cref{fig:twoRoomsDynamicSimulation}. 
\end{example}

\begin{example}
	\label{ex:CDplayer}
	Similar as in \cite{GosPAF17} we consider the CD player example \cite{ChaV02} and use the different columns and rows of the input and output matrix to create a SISO switched system of dimension $\stateDim = 120$. More precisely, the first mode consists of the second column of the input matrix and the first row of the output matrix (multiplied by $2/1000$) and the second mode is defined via the first columns of the input matrix and the second row of the output matrix (multiplied by $5$). The scaling of the subsystems is performed to adjust the magnitude of the output of the different subsystems. Notice that the switching affects only the $\B$ and $\C$ matrices and the envelope system is given by the original CD player example. We drive the system with $\inVar(t) = \exp(-5t)$ and switching signal
\begin{equation*}
	\switch(t) = \begin{cases}
		2, & t\in[0,0.5]\cup (1,1.5],\\
		1, & t\in(0.5,1]\cup (1.5,2].
	\end{cases}
\end{equation*}
\begin{figure}[ht]
	\centering
	\input{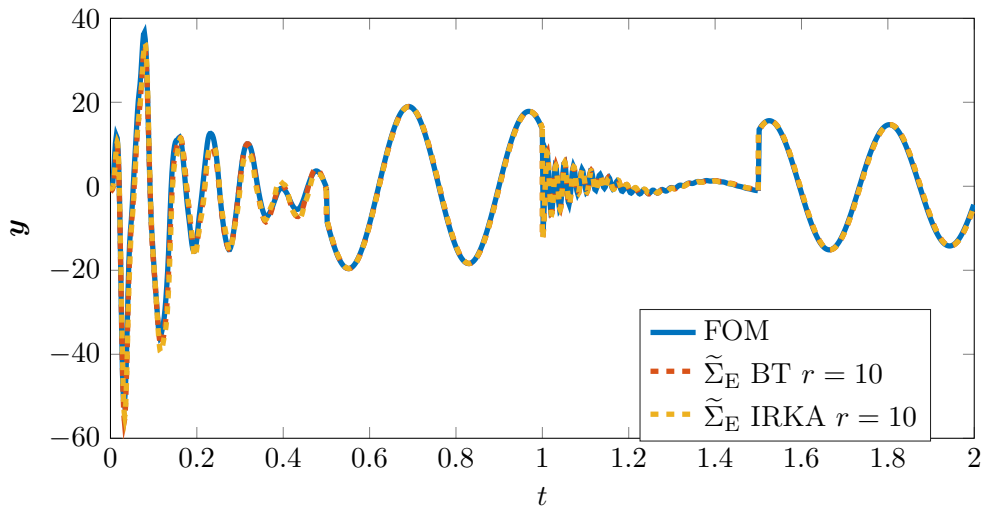}
	\caption{\Cref{ex:CDplayer} -- Time domain simulation of the full-order model (FOM) and the reduced-order models}
	\label{fig:CDplayerTimeSimulation}
\end{figure}%
The time simulation of the FOM together with the envelope reduced systems of order $\stateDimRed = 10$ are presented in \Cref{fig:CDplayerTimeSimulation}. Again, the envelope reduced systems capture the dynamics very accurately.
\end{example}

\section{Summary} 
For switched LTI systems, we introduced in \Cref{def:envelopeSystem} a so-called envelope system, which is a standard LTI system without switching and which allows to reproduce the output of the original switched system by using a suitable feedback law (\Cref{lem:envelopeFeedback}). 
The envelope system can be reduced by classical model order reduction techniques and the reduced envelope system can be transformed back to a linear switched system by the same feedback law as for the full-order envelope system.
Our theoretical findings are confirmed by several numerical examples.
The strengths of this new model order reduction framework for switched systems are summarized in the following:

\begin{itemize}
	\item The model reduction is applied to the non-switched envelope system, which allows the usage of standard model reduction techniques and, hence, existing implementations. 
	\item The approach allows for efficient treatment of large-scale systems since the state dimension of the envelope system equals the state dimension of the modes of the switched system. The main cost for constructing the envelope system is due to the matrix decompositions in \eqref{eq:lowRankPerturbation} which can be computed efficiently, cf.\ \Cref{rem:computationSVD}.
	\item The linear time-invariant envelope system is easier accessible for analysis than the original switched system. Especially, an error bound for the output error of the reduced-order switched system has been presented.
	\item For the case that the original switched system is quadratic Lyapunov stable or consists of port-Hamiltonian subsystems, we can adapt the model reduction scheme such that these properties are preserved in the reduced switched system.
\end{itemize}

In contrast to the strengths are the following weaknesses:
\begin{itemize}
	\item While the envelope system behaves like the original switched system when applying a certain feedback law, also other inputs are possible allowing for different dynamics. Thus the envelope reduction framework may be improved, if one modifies existing MOR methods to handle restricted input spaces.
	\item The derived error bound seems to be quite conservative, since it is based on a sum of the output errors of the envelope system.
\end{itemize}

An interesting direction for future work is to analyze the approximation quality of the reduced order model with respect to the freedoms in the construction of the envelope system. 
In particular, it is an open task to determine optimal scaling strategies for the low-rank decomposition in \eqref{eq:lowRankPerturbation} and the effect of the different orderings of the subsystems.

\bibliographystyle{plain}
\bibliography{Refs}

\appendix
\section{Spatial discretization of \texorpdfstring{\Cref{ex:twoRooms2}}{Example 5.2}}
\label{sec:appendix}

The partial differential equation \eqref{eq:heatEquation} associated with the boundary conditions \eqref{eq:BCleft} and \eqref{eq:BCright} is discretized in space using a cell-centered finite volume scheme. 
Each of the three subdomains is discretized with a uniform mesh with mesh widths $\Delta x_i$ and numbers of cells $n_i$ for $i=1,\ldots,3$. 
The integrals over the finite volumes are approximated using the midpoint rule and the temperature gradients at the cell interfaces are approximated by central finite differences. 
Moreover, the effective thermal conductivity at the cell interfaces between the different subdomains is approximated based on the values at the cell centers as described in \cite{Pat80}, i.\,e., 
\begin{align*}
k_{1+1/2}\left(\switch\right) = \left(\frac{\Delta x_1}{k_2}+\frac{\Delta x_{2}}{k_{\Omega_2}\left(\switch\right)}\right)^{-1}\left(\Delta x_1+\Delta x_{2}\right),\\
k_{2+1/2}\left(\switch\right) = \left(\frac{\Delta x_3}{k_2}+\frac{\Delta x_{2}}{k_{\Omega_2}\left(\switch\right)}\right)^{-1}\left(\Delta x_3+\Delta x_{2}\right).
\end{align*}
The resulting semi-discretized system of dimension $n=n_1+n_2+n_3$ can be written in the form

\begin{equation}
\label{eq:switchedDescriptorSystem}
	\left\{
	\begin{aligned}
		\hat\E_\switch\dot{\state}(t) &= \hat\A_\switch\state(t) + \hat\B\inVar(t),\\
		\outVar(t) &= \hat\C\state(t),
	\end{aligned}
	\right.\qquad\qquad t>0
\end{equation}
with coefficient matrices
\allowdisplaybreaks
\begin{align*}
\hat\E_\switch &=\blkdiag\left(\rhoc_2\Delta x_1I_{n_1},\;\rhoc_{\Omega_2}\left(\switch\right)\Delta x_2I_{n_2},\;\rhoc_2\Delta x_3I_{n_3}\right),\quad
\hat\A_\switch = \begin{bmatrix}
\hat\A_{11,\switch} & \hat\A_{12,\switch} & 0\\
\hat\A_{12,\switch}^T & \hat\A_{22,\switch} & \hat\A_{23,\switch}\\
0 & \hat\A_{23,\switch}^T & \hat\A_{33,\switch}
\end{bmatrix},\\
\hat\B_\switch &= \begin{bmatrix}
1\\
0\\
\vdots\\
0
\end{bmatrix}\in\mathbb{R}^{n\times1},\quad
\hat\C_\switch = \frac{1}{n_3}\begin{bmatrix}
0_{1,\left(n_1+n_2\right)} & 1 & \cdots & 1
\end{bmatrix}\in\mathbb{R}^{1\times n},
\end{align*}
where the block matrices of $\hat\A_\switch$ are given by
\begin{align*}
\hat\A_{11,\switch} &= \frac{k_2}{\Delta x_1}\mathrm{tridiag}\left(1,-2,1\right)+\frac{k_2}{\Delta x_1}\begin{bmatrix}
1 & 0 & \cdots & 0\\
0 & 0 & \cdots & 0\\
\vdots & \vdots & \ddots & \vdots\\
0 & 0 & \cdots & 0
\end{bmatrix}\\
&+\left(\frac{k_2}{\Delta x_1}-\frac{2k_{1+1/2}\left(\switch\right)}{\Delta x_1+\Delta x_2}\right)\begin{bmatrix}
0 & \cdots & 0 & 0\\
\vdots & \ddots & \vdots & \vdots\\
0 & \cdots & 0 & 0\\
0 & \cdots & 0 & 1
\end{bmatrix}\in\mathbb{R}^{n_1\times n_1},\\
\hat\A_{12,\switch} &= \frac{2k_{1+1/2}\left(\switch\right)}{\Delta x_1+\Delta x_2}\begin{bmatrix}
0 & 0 & \cdots & 0\\
\vdots & \vdots & \ddots & \vdots\\
0 & 0 & \cdots & 0 \\
1 & 0 & \cdots & 0 
\end{bmatrix}\in\mathbb{R}^{n_1\times n_2},\\
\hat\A_{22,\switch} &= \frac{k_{\Omega_2}\left(\switch\right)}{\Delta x_2}\mathrm{tridiag}\left(1,-2,1\right)+\left(\frac{k_{\Omega_2}\left(\switch\right)}{\Delta x_2}-\frac{2k_{1+1/2}\left(\switch\right)}{\Delta x_1+\Delta x_2}\right)\begin{bmatrix}
1 & 0 & \cdots & 0\\
0 & 0 & \cdots & 0\\
\vdots & \vdots & \ddots & \vdots\\
0 & 0 & \cdots & 0
\end{bmatrix}\\
&+\left(\frac{k_{\Omega_2}\left(\switch\right)}{\Delta x_2}-\frac{2k_{2+1/2}\left(\switch\right)}{\Delta x_3+\Delta x_2}\right)\begin{bmatrix}
0 & \cdots & 0 & 0\\
\vdots & \ddots & \vdots & \vdots\\
0 & \cdots & 0 & 0\\
0 & \cdots & 0 & 1
\end{bmatrix}\in\mathbb{R}^{n_2\times n_2},\\
\hat\A_{23,\switch} &= \frac{2k_{2+1/2}\left(\switch\right)}{\Delta x_3+\Delta x_2}\begin{bmatrix}
0 & 0 & \cdots & 0\\
\vdots & \vdots & \ddots & \vdots\\
0 & 0 & \cdots & 0 \\
1 & 0 & \cdots & 0 
\end{bmatrix}\in\mathbb{R}^{n_2\times n_3},\\
\hat\A_{33,\switch} &= \frac{k_{2}}{\Delta x_3}\mathrm{tridiag}\left(1,-2,1\right)+\left(\frac{k_{2}}{\Delta x_3}-\frac{2k_{2+1/2}\left(\switch\right)}{\Delta x_3+\Delta x_2}\right)\begin{bmatrix}
1 & 0 & \cdots & 0\\
0 & 0 & \cdots & 0\\
\vdots & \vdots & \ddots & \vdots\\
0 & 0 & \cdots & 0
\end{bmatrix}\\
&+\left(\frac{k_{2}}{\Delta x_3}-\frac{h}{2}\right)\begin{bmatrix}
0 & \cdots & 0 & 0\\
\vdots & \ddots & \vdots & \vdots\\
0 & \cdots & 0 & 0\\
0 & \cdots & 0 & 1
\end{bmatrix}\in\mathbb{R}^{n_3\times n_3}.
\end{align*}

The notation $\mathrm{tridiag}(a,b,c)$ denotes a tridiagonal matrix with constant values $a$, $b$, and $c$ on the first lower, the main, and the first upper diagonal, respectively. 
Since $\hat\E_\switch$ is a diagonal matrix with positive entries only, it is invertible and we obtain a system of the form \eqref{eq:switchedSystem} which is equivalent to \eqref{eq:switchedDescriptorSystem} and whose coefficient matrices are given by $\A_\switch=\hat\E_\switch^{-1}\hat\A_\switch$, $\B_\switch = \hat\E_\switch^{-1}\hat\B_\switch$, $\C_\switch = \hat\C_\switch$, and $\D_\switch=0$. 
It is noteworthy that only $\A_\switch$ is affected by switching whereas $\B_\switch$, $\C_\switch$, and $\D_\switch$ are constant matrices. 
\end{document}

%% file: outputDependendSwitching.tex
%
\definecolor{mycolor1}{rgb}{0.00000,0.44700,0.74100}%
\definecolor{mycolor2}{rgb}{0.85000,0.32500,0.09800}%
\definecolor{mycolor3}{rgb}{0.92900,0.69400,0.12500}%
\begin{tikzpicture}

\begin{axis}[%
width=4.6in,
height=1.8in,
at={(0.758in,0.481in)},
scale only axis,
xmin=0,
xmax=1.4,
xlabel={$t$},
ymin=0,
ymax=0.7,
ylabel={$\outVar$},
axis background/.style={fill=white},
legend style={at={(0.97,0.03)},anchor=south east,legend cell align=left,align=left,draw=white!15!black}
]
\addplot [color=mycolor1,dashed, line width=2pt]
  table[row sep=crcr]{%
0	0.647923572799272\\
0.0141414141414141	0.647923572799272\\
0.0282828282828283	0.647923572799272\\
0.0424242424242424	0.647923572799272\\
0.0565656565656566	0.647923572799272\\
0.0707070707070707	0.647923572799272\\
0.0848484848484848	0.647923572799272\\
0.098989898989899	0.647923572799272\\
0.113131313131313	0.647923572799272\\
0.127272727272727	0.647923572799272\\
0.141414141414141	0.647923572799272\\
0.155555555555556	0.647923572799272\\
0.16969696969697	0.647923572799272\\
0.183838383838384	0.647923572799272\\
0.197979797979798	0.647923572799272\\
0.212121212121212	0.647923572799272\\
0.226262626262626	0.647923572799272\\
0.24040404040404	0.647923572799272\\
0.254545454545455	0.647923572799272\\
0.268686868686869	0.647923572799272\\
0.282828282828283	0.647923572799272\\
0.296969696969697	0.647923572799272\\
0.311111111111111	0.647923572799272\\
0.325252525252525	0.647923572799272\\
0.339393939393939	0.647923572799272\\
0.353535353535354	0.647923572799272\\
0.367676767676768	0.647923572799272\\
0.381818181818182	0.647923572799272\\
0.395959595959596	0.647923572799272\\
0.41010101010101	0.647923572799272\\
0.424242424242424	0.647923572799272\\
0.438383838383838	0.647923572799272\\
0.452525252525252	0.647923572799272\\
0.466666666666667	0.647923572799272\\
0.480808080808081	0.647923572799272\\
0.494949494949495	0.647923572799272\\
0.509090909090909	0.647923572799272\\
0.523232323232323	0.647923572799272\\
0.537373737373737	0.647923572799272\\
0.551515151515151	0.647923572799272\\
0.565656565656566	0.647923572799272\\
0.57979797979798	0.647923572799272\\
0.593939393939394	0.647923572799272\\
0.608080808080808	0.647923572799272\\
0.622222222222222	0.647923572799272\\
0.636363636363636	0.647923572799272\\
0.65050505050505	0.647923572799272\\
0.664646464646465	0.647923572799272\\
0.678787878787879	0.647923572799272\\
0.692929292929293	0.647923572799272\\
0.707070707070707	0.647923572799272\\
0.721212121212121	0.647923572799272\\
0.735353535353535	0.647923572799272\\
0.749494949494949	0.647923572799272\\
0.763636363636364	0.647923572799272\\
0.777777777777778	0.647923572799272\\
0.791919191919192	0.647923572799272\\
0.806060606060606	0.647923572799272\\
0.82020202020202	0.647923572799272\\
0.834343434343434	0.647923572799272\\
0.848484848484849	0.647923572799272\\
0.862626262626263	0.647923572799272\\
0.876767676767677	0.647923572799272\\
0.890909090909091	0.647923572799272\\
0.905050505050505	0.647923572799272\\
0.919191919191919	0.647923572799272\\
0.933333333333333	0.647923572799272\\
0.947474747474747	0.647923572799272\\
0.961616161616161	0.647923572799272\\
0.975757575757576	0.647923572799272\\
0.98989898989899	0.647923572799272\\
1.0040404040404	0.647923572799272\\
1.01818181818182	0.647923572799272\\
1.03232323232323	0.647923572799272\\
1.04646464646465	0.647923572799272\\
1.06060606060606	0.647923572799272\\
1.07474747474747	0.647923572799272\\
1.08888888888889	0.647923572799272\\
1.1030303030303	0.647923572799272\\
1.11717171717172	0.647923572799272\\
1.13131313131313	0.647923572799272\\
1.14545454545455	0.647923572799272\\
1.15959595959596	0.647923572799272\\
1.17373737373737	0.647923572799272\\
1.18787878787879	0.647923572799272\\
1.2020202020202	0.647923572799272\\
1.21616161616162	0.647923572799272\\
1.23030303030303	0.647923572799272\\
1.24444444444444	0.647923572799272\\
1.25858585858586	0.647923572799272\\
1.27272727272727	0.647923572799272\\
1.28686868686869	0.647923572799272\\
1.3010101010101	0.647923572799272\\
1.31515151515151	0.647923572799272\\
1.32929292929293	0.647923572799272\\
1.34343434343434	0.647923572799272\\
1.35757575757576	0.647923572799272\\
1.37171717171717	0.647923572799272\\
1.38585858585859	0.647923572799272\\
1.4	0.647923572799272\\
};
\addlegendentry{switching surface ($p = 0.05$)};

\addplot [color=mycolor2,solid, line width=2pt]
  table[row sep=crcr]{%
0	0\\
0.0141414141414141	0.0147439887159664\\
0.0282828282828283	0.0292809439050217\\
0.0424242424242424	0.0436137727100087\\
0.0565656565656566	0.0577453414519778\\
0.0707070707070707	0.0716784762034035\\
0.0848484848484848	0.0854159633533502\\
0.098989898989899	0.0989605501647022\\
0.113131313131313	0.112314945323569\\
0.127272727272727	0.125481819480978\\
0.141414141414141	0.138463805786958\\
0.155555555555556	0.151263500417122\\
0.16969696969697	0.163883463091862\\
0.183838383838384	0.176326217588247\\
0.197979797979798	0.188594252244737\\
0.212121212121212	0.200690020458806\\
0.226262626262626	0.21261594117758\\
0.24040404040404	0.224374399381588\\
0.254545454545455	0.235967746561712\\
0.268686868686869	0.247398301189452\\
0.282828282828283	0.258668349180574\\
0.296969696969697	0.269780144352258\\
0.311111111111111	0.280735908873823\\
0.325252525252525	0.291537833711121\\
0.339393939393939	0.302188079064693\\
0.353535353535354	0.312688774801774\\
0.367676767676768	0.32304202088223\\
0.381818181818182	0.33324988777851\\
0.395959595959596	0.343314416889711\\
0.41010101010101	0.35323762094982\\
0.424242424242424	0.363021484430225\\
0.438383838383838	0.372667963936579\\
0.452525252525252	0.382178988600086\\
0.466666666666667	0.391556460463291\\
0.480808080808081	0.400802254860463\\
0.494949494949495	0.409918220792623\\
0.509090909090909	0.418906181297317\\
0.523232323232323	0.427767933813192\\
0.537373737373737	0.43650525053945\\
0.551515151515151	0.44511987879026\\
0.565656565656566	0.453613541344193\\
0.57979797979798	0.461987936788741\\
0.593939393939394	0.470244739860017\\
0.608080808080808	0.47838560177766\\
0.622222222222222	0.486412150575064\\
0.636363636363636	0.494325991424947\\
0.65050505050505	0.502128706960363\\
0.664646464646465	0.509821857591199\\
0.678787878787879	0.517406981816234\\
0.692929292929293	0.52488559653081\\
0.707070707070707	0.532259197330182\\
0.721212121212121	0.539529258808619\\
0.735353535353535	0.546697234854293\\
0.749494949494949	0.553764558940029\\
0.763636363636364	0.560732644409982\\
0.777777777777778	0.567602884762277\\
0.791919191919192	0.574376653927686\\
0.806060606060606	0.581055306544391\\
0.82020202020202	0.587640178228888\\
0.834343434343434	0.594132585843088\\
0.848484848484849	0.600533827757665\\
0.862626262626263	0.606845184111713\\
0.876767676767677	0.613067917068744\\
0.890909090909091	0.619203271069108\\
0.905050505050505	0.625252473078853\\
0.919191919191919	0.631216732835101\\
0.933333333333333	0.637097243087972\\
0.947474747474747	0.642895179839117\\
0.961616161616161	0.648611702576896\\
0.975757575757576	0.654247954508255\\
0.98989898989899	0.659805062787351\\
1.0040404040404	0.661050273521378\\
1.01818181818182	0.651767875641868\\
1.03232323232323	0.642615820209589\\
1.04646464646465	0.633592276969711\\
1.06060606060606	0.624695441367649\\
1.07474747474747	0.615923534188183\\
1.08888888888889	0.607274801199641\\
1.1030303030303	0.59874751280309\\
1.11717171717172	0.590339963686441\\
1.13131313131313	0.582050472483415\\
1.14545454545455	0.573877381437303\\
1.15959595959596	0.565819056069436\\
1.17373737373737	0.557873884852325\\
1.18787878787879	0.550040278887377\\
1.2020202020202	0.542316671587147\\
1.21616161616162	0.534701518362041\\
1.23030303030303	0.527193296311431\\
1.24444444444444	0.519790503919099\\
1.25858585858586	0.512491660752956\\
1.27272727272727	0.505295307168986\\
1.28686868686869	0.498200004019339\\
1.3010101010101	0.491204332364524\\
1.31515151515151	0.484306893189651\\
1.32929292929293	0.477506307124648\\
1.34343434343434	0.470801214168411\\
1.35757575757576	0.464190273416828\\
1.37171717171717	0.45767216279462\\
1.38585858585859	0.451245578790947\\
1.4	0.444909236198735\\
};
\addlegendentry{$\outVar$};

\addplot [color=mycolor3,solid,line width=2pt]
  table[row sep=crcr]{%
0	0\\
0.0141414141414141	0.0140418940152061\\
0.0282828282828283	0.0278866132428779\\
0.0424242424242424	0.0415369263904845\\
0.0565656565656566	0.0549955632875979\\
0.0707070707070707	0.0682652154318128\\
0.0848484848484848	0.0813485365270002\\
0.098989898989899	0.0942481430140021\\
0.113131313131313	0.106966614593876\\
0.127272727272727	0.119506494743789\\
0.141414141414141	0.131870291225674\\
0.155555555555556	0.144060476587735\\
0.16969696969697	0.156079488658916\\
0.183838383838384	0.167929731036426\\
0.197979797979798	0.179613573566416\\
0.212121212121212	0.19113335281791\\
0.226262626262626	0.202491372550076\\
0.24040404040404	0.213689904172941\\
0.254545454545455	0.224731187201631\\
0.268686868686869	0.23561742970424\\
0.282828282828283	0.246350808743404\\
0.296969696969697	0.256933470811675\\
0.311111111111111	0.267367532260784\\
0.325252525252525	0.277655079724877\\
0.339393939393939	0.287798170537803\\
0.353535353535354	0.297798833144547\\
0.367676767676768	0.307659067506885\\
0.381818181818182	0.317380845503342\\
0.395959595959596	0.326966111323534\\
0.41010101010101	0.336416781856971\\
0.424242424242424	0.345734747076405\\
0.438383838383838	0.35492187041579\\
0.452525252525252	0.363979989142939\\
0.466666666666667	0.372910914726944\\
0.480808080808081	0.381716433200441\\
0.494949494949495	0.390398305516784\\
0.509090909090909	0.398958267902207\\
0.523232323232323	0.40739803220304\\
0.537373737373737	0.415719286228048\\
0.551515151515151	0.423923694085962\\
0.565656565656566	0.432012896518279\\
0.57979797979798	0.439988511227373\\
0.593939393939394	0.447852133200016\\
0.608080808080808	0.455605335026343\\
0.622222222222222	0.463249667214347\\
0.636363636363636	0.47078665849995\\
0.65050505050505	0.478217816152726\\
0.664646464646465	0.485544626277333\\
0.678787878787879	0.492768554110699\\
0.692929292929293	0.499891044315057\\
0.707070707070707	0.50691352126684\\
0.721212121212121	0.513837389341542\\
0.735353535353535	0.520664033194564\\
0.749494949494949	0.527394818038122\\
0.763636363636364	0.534031089914268\\
0.777777777777778	0.540574175964073\\
0.791919191919192	0.547025384693035\\
0.806060606060606	0.553386006232754\\
0.82020202020202	0.559657312598941\\
0.834343434343434	0.565840557945798\\
0.848484848484849	0.571936978816824\\
0.862626262626263	0.577947794392107\\
0.876767676767677	0.583874206732137\\
0.890909090909091	0.589717401018198\\
0.905050505050505	0.595478545789384\\
0.919191919191919	0.601158793176286\\
0.933333333333333	0.606759279131402\\
0.947474747474747	0.612281123656302\\
0.961616161616161	0.617725431025615\\
0.975757575757576	0.623093290007862\\
0.98989898989899	0.628385774083191\\
1.0040404040404	0.629571689067979\\
1.01818181818182	0.620731310135112\\
1.03232323232323	0.612015066866275\\
1.04646464646465	0.60342121616163\\
1.06060606060606	0.594948039397761\\
1.07474747474747	0.586593842083983\\
1.08888888888889	0.578356953523467\\
1.1030303030303	0.570235726479134\\
1.11717171717172	0.56222853684423\\
1.13131313131313	0.554333783317538\\
1.14545454545455	0.546549887083145\\
1.15959595959596	0.538875291494701\\
1.17373737373737	0.531308461764119\\
1.18787878787879	0.523847884654645\\
1.2020202020202	0.516492068178235\\
1.21616161616162	0.509239541297182\\
1.23030303030303	0.502088853629934\\
1.24444444444444	0.495038575161046\\
1.25858585858586	0.488087295955196\\
1.27272727272727	0.481233625875225\\
1.28686868686869	0.474476194304132\\
1.3010101010101	0.467813649870975\\
1.31515151515151	0.46124466018062\\
1.32929292929293	0.454767911547284\\
1.34343434343434	0.44838210873182\\
1.35757575757576	0.442085974682694\\
1.37171717171717	0.43587825028059\\
1.38585858585859	0.429757694086617\\
1.4	0.423723082094033\\
};
\addlegendentry{$\outVarRed$};

\end{axis}
\end{tikzpicture}%

%% file: IEEETacExample1-HankelSingularValues.tex
%
\definecolor{mycolor1}{rgb}{0.00000,0.44700,0.74100}%
\definecolor{mycolor2}{rgb}{0.85000,0.32500,0.09800}%
\definecolor{mycolor3}{rgb}{0.92900,0.69400,0.12500}%
\begin{tikzpicture}

\begin{axis}[%
width=5in,
height=1.7in,
at={(0.758in,0.481in)},
scale only axis,
xmin=1,
xmax=11,
ymode=log,
ymin=1e-10,
ymax=10,
yminorticks=true,
axis background/.style={fill=white},
legend style={at={(0.03,0.07)},anchor=south west, legend cell align=left,align=left,draw=white!15!black}
]
\addplot [color=mycolor1,line width=2,mark=square*]
  table[row sep=crcr]{%
1	1\\
2	0.774259454869587\\
3	0.670837565687198\\
4	0.0379395890753344\\
5	0.0339386298313121\\
6	0.0114531021778059\\
7	0.00456188709080935\\
8	0.000186947392053302\\
9	1.9999232106993e-05\\
10	5.7120006794784e-08\\
11	3.10954975355928e-11\\
};
\addlegendentry{$\system_1$};

\addplot [color=mycolor2,line width=2,mark=triangle*]
  table[row sep=crcr]{%
1	1\\
2	0.989522390707091\\
3	0.511519470122391\\
4	0.337999108234459\\
5	0.254536642192015\\
6	0.0151443150666238\\
7	0.00419594260493886\\
8	0.00376444012124156\\
9	3.59411563328752e-05\\
10	9.64217601095739e-09\\
11	3.76930854503305e-35\\
};
\addlegendentry{$\system_2$};

\addplot [color=mycolor3,line width=2,mark=*]
  table[row sep=crcr]{%
1	1\\
2	0.688327067269352\\
3	0.652278208667937\\
4	0.287355323012747\\
5	0.256854668449004\\
6	0.163066563391507\\
7	0.132816926977504\\
8	0.0642230710253048\\
9	0.0439204302427028\\
10	0.0228395117501187\\
11	0.0168521703713514\\
};
\addlegendentry{$\eSystem$};

\end{axis}
\end{tikzpicture}%

%% file: IEEETacExample1-TimeDomainSimulation-r8-v1.tex
%
\definecolor{mycolor1}{rgb}{0.00000,0.44700,0.74100}%
\definecolor{mycolor2}{rgb}{0.85000,0.32500,0.09800}%
\definecolor{mycolor3}{rgb}{0.92900,0.69400,0.12500}%
\definecolor{mycolor4}{rgb}{0.49400,0.18400,0.55600}%
\begin{tikzpicture}

\begin{axis}[%
width=5in,
height=1.7in,
at={(0.758in,0.481in)},
scale only axis,
xmin=0,
xmax=1.2,
ymin=-1,
ymax=1,
xlabel={$t$},
ylabel={$\outVar$},
axis background/.style={fill=white},
legend style={legend cell align=left,align=left,draw=white!15!black}
]
\addplot [color=mycolor1,solid,forget plot,line width=2,mark=square*,mark repeat=8]
  table[row sep=crcr]{%
0	0\\
0.005	-0.00017573861293956\\
0.01	-0.000697282250307673\\
0.015	-0.00155561945200414\\
0.02	-0.00274106788069632\\
0.025	-0.0042433355471935\\
0.03	-0.00605154736440023\\
0.035	-0.00815447433554931\\
0.04	-0.0105406623243361\\
0.045	-0.0131984798625116\\
0.05	-0.0161161240129054\\
0.055	-0.0192818964891713\\
0.06	-0.0226843415489534\\
0.065	-0.0263122745875246\\
0.07	-0.0301547689324855\\
0.075	-0.0342014392926288\\
0.08	-0.0384425692362979\\
0.085	-0.0428691173890881\\
0.09	-0.0474726898445384\\
0.095	-0.0522457896643849\\
0.1	-0.0571819154394493\\
0.105	-0.0622755448893814\\
0.11	-0.0675221000174755\\
0.115	-0.0729181247638594\\
0.12	-0.0784613394380291\\
0.125	-0.0841506045088988\\
0.13	-0.0899858871622002\\
0.135	-0.0959683372388387\\
0.14	-0.102100287485262\\
0.145	-0.108385202967608\\
0.15	-0.114827658120544\\
0.155	-0.121433292835914\\
0.16	-0.128208754890153\\
0.165	-0.13516164240381\\
0.17	-0.142300499745227\\
0.175	-0.149634649687539\\
0.18	-0.157174081285963\\
0.185	-0.164929393898284\\
0.19	-0.172911818506217\\
0.195	-0.181132936579626\\
0.2	-0.189604523277966\\
};
\addplot [color=mycolor1,solid,forget plot,line width=2,mark=square*,mark repeat=8]
  table[row sep=crcr]{%
0.2	0.279917749180419\\
0.204404641536935	0.293981678129148\\
0.209298687689084	0.309537088976058\\
0.221745710765159	0.348649909810348\\
0.231745710765159	0.379460014389376\\
0.241745710765159	0.409570066082774\\
0.251745710765159	0.438839612752088\\
0.261745710765159	0.467130039513542\\
0.271745710765159	0.494305180948389\\
0.281745710765159	0.520232324634952\\
0.291745710765159	0.544782707419204\\
0.301745710765159	0.567831841314939\\
0.311745710765159	0.589260076279037\\
0.321745710765159	0.608953495428496\\
0.331745710765159	0.626804351754516\\
0.341745710765159	0.6427113332098\\
0.351745710765159	0.656580040666696\\
0.361745710765159	0.668323716489595\\
0.371745710765159	0.677863593160315\\
0.381745710765159	0.685129080592598\\
0.391745710765159	0.690058129376392\\
0.401745710765159	0.69259774449614\\
0.411745710765159	0.692704222803279\\
0.421745710765159	0.690343250580651\\
0.431745710765159	0.685490129500956\\
0.441745710765159	0.678130040633947\\
0.451745710765159	0.668258154693761\\
0.461745710765159	0.655879633604592\\
0.471745710765159	0.641009705309781\\
0.481745710765159	0.623673658924978\\
0.491745710765159	0.603906819751852\\
0.501745710765159	0.581754454710327\\
0.511745710765159	0.557271691683822\\
0.521745710765159	0.530523243734031\\
0.531745710765159	0.50158324948967\\
0.541745710765159	0.470535088425052\\
0.551745710765159	0.437471150220524\\
0.561745710765159	0.402492303257936\\
0.571745710765159	0.365707609610888\\
0.578809283073869	0.338699516011996\\
0.585872855382579	0.310892156366092\\
0.59293642769129	0.282330783117962\\
0.6	0.253062322716777\\
};
\addplot [color=mycolor1,solid,forget plot,line width=2,mark=square*,mark repeat=8]
  table[row sep=crcr]{%
0.6	-0.0396554421835247\\
0.605041396206302	-0.0197706469864986\\
0.611721861608403	0.00691179748687599\\
0.616721861608403	0.0270844277299098\\
0.621721861608403	0.0473928170647342\\
0.626721861608403	0.0678055399319481\\
0.631721861608403	0.0882930877494307\\
0.636721861608403	0.108827776201385\\
0.641721861608403	0.129384116680663\\
0.646721861608403	0.149938931693649\\
0.651721861608403	0.170471282921845\\
0.656721861608403	0.190962381619651\\
0.661721861608403	0.211395779371646\\
0.666721861608403	0.231757389607338\\
0.671721861608403	0.252035393419228\\
0.676721861608403	0.272220168452379\\
0.681721861608403	0.292304275122005\\
0.686721861608403	0.312282381540886\\
0.691721861608403	0.332151160220854\\
0.696721861608403	0.351909248963969\\
0.701721861608403	0.371557032573569\\
0.706721861608403	0.391096479829046\\
0.711721861608403	0.410531045371807\\
0.716721861608403	0.429865672580435\\
0.721721861608403	0.449106394675285\\
0.726721861608403	0.468260102680917\\
0.731721861608403	0.487334466268286\\
0.736721861608403	0.506337983828432\\
0.741721861608403	0.525279447935754\\
0.746721861608403	0.544167671320056\\
0.751721861608403	0.563011438270578\\
0.756721861608403	0.581819601672484\\
0.761721861608403	0.600600473612722\\
0.766721861608403	0.619361541307179\\
0.771721861608403	0.638109457126597\\
0.776721861608403	0.656850176863657\\
0.781721861608403	0.675588345013073\\
0.786721861608403	0.69432703380184\\
0.790041396206303	0.706768904043082\\
0.793360930804202	0.719211864288196\\
0.796680465402101	0.731655661713264\\
0.8	0.744099659713777\\
};
\addplot [color=mycolor1,solid,line width=2,mark=square*,mark repeat=8]
  table[row sep=crcr]{%
0.8	-0.0995331138730465\\
0.804953229162545	-0.122614565749623\\
0.806604305550061	-0.130239530430978\\
0.814859687487636	-0.167819597322416\\
0.823115069425212	-0.204434940840437\\
0.831370451362788	-0.240010209755288\\
0.839625833300363	-0.274472741536951\\
0.849625833300363	-0.314628051534492\\
0.859625833300363	-0.352929921437078\\
0.869625833300363	-0.389267201806886\\
0.879625833300363	-0.423536419639362\\
0.889625833300363	-0.455641956892397\\
0.899625833300363	-0.485496357725297\\
0.909625833300363	-0.513021030779387\\
0.919625833300363	-0.538146559387711\\
0.929625833300363	-0.560812787167789\\
0.939625833300363	-0.580968984778079\\
0.949625833300363	-0.598574285524605\\
0.959625833300363	-0.613597860570434\\
0.969625833300363	-0.626018906210283\\
0.979625833300363	-0.635826658127681\\
0.989625833300363	-0.643020542672112\\
0.999625833300363	-0.647610209806964\\
1.00962583330036	-0.6496154229168\\
1.01962583330036	-0.649065918319837\\
1.02962583330036	-0.646001271507687\\
1.03962583330036	-0.640470789195044\\
1.04962583330036	-0.632533308662973\\
1.05962583330036	-0.622256910145073\\
1.06962583330036	-0.609718514465181\\
1.07962583330036	-0.59500364413116\\
1.08962583330036	-0.578206144907879\\
1.09962583330036	-0.559427768044663\\
1.10962583330036	-0.538777531953019\\
1.11962583330036	-0.516371370263043\\
1.12962583330036	-0.492331793194219\\
1.13962583330036	-0.466787364849742\\
1.14962583330036	-0.439871875785533\\
1.15962583330036	-0.411723902805934\\
1.16971937497527	-0.382208254245644\\
1.17981291665018	-0.351733608003191\\
1.18990645832509	-0.320453893383677\\
1.2	-0.288525472321972\\
};
\addlegendentry{FOM};

\addplot [color=mycolor2,solid,forget plot,line width=2,mark=*,mark repeat=8]
  table[row sep=crcr]{%
0	0\\
0.005	-0.000193680239075705\\
0.01	-0.000766044227066033\\
0.015	-0.00170347834060893\\
0.02	-0.00299157091699983\\
0.025	-0.00461517289680971\\
0.03	-0.00655842072950159\\
0.035	-0.00880500951692524\\
0.04	-0.0113383370685613\\
0.045	-0.0141415516765919\\
0.05	-0.0171975575318309\\
0.055	-0.0204893253033281\\
0.06	-0.0240000416346644\\
0.065	-0.027713139114905\\
0.07	-0.031612286138663\\
0.075	-0.0356816971972579\\
0.08	-0.0399062696600342\\
0.085	-0.0442715932386367\\
0.09	-0.0487639287271765\\
0.095	-0.0533704794007823\\
0.1	-0.0580794980326714\\
0.105	-0.0628802756843195\\
0.11	-0.0677631157433913\\
0.115	-0.0727195312385855\\
0.12	-0.0777423080554431\\
0.125	-0.0828254754430395\\
0.13	-0.0879642829978112\\
0.135	-0.0931552960791022\\
0.14	-0.0983964057930957\\
0.145	-0.103686785905069\\
0.15	-0.10902688074662\\
0.155	-0.11441838148944\\
0.16	-0.119864179179741\\
0.165	-0.12536831452763\\
0.17	-0.130935984091589\\
0.175	-0.136573392818871\\
0.18	-0.142287652620171\\
0.185	-0.148086732572436\\
0.19	-0.153979489210281\\
0.195	-0.159975404113978\\
0.2	-0.166084436544465\\
};
\addplot [color=mycolor2,solid,forget plot,line width=2,mark=*,mark repeat=8]
  table[row sep=crcr]{%
0.2	0.240550821620865\\
0.20551155689755	0.256349972674768\\
0.211135594548111	0.272439985232698\\
0.213947613373392	0.280465365628228\\
0.223947613373392	0.308844185728346\\
0.233947613373392	0.336865979367348\\
0.243947613373392	0.364391971841526\\
0.253947613373392	0.391284490043426\\
0.263947613373392	0.41740732527343\\
0.273947613373392	0.442626376447086\\
0.283947613373392	0.466810549556261\\
0.293947613373392	0.489832205898662\\
0.303947613373392	0.511567499075289\\
0.313947613373392	0.531896998044168\\
0.323947613373392	0.5507064753241\\
0.333947613373392	0.5678872990421\\
0.343947613373392	0.583336722725775\\
0.353947613373392	0.596958449622906\\
0.363947613373392	0.608663255464295\\
0.373947613373392	0.618369297737229\\
0.383947613373392	0.626002339917875\\
0.393947613373392	0.631496222205534\\
0.403947613373392	0.634793275249898\\
0.413947613373392	0.635844525381106\\
0.423947613373392	0.634609839115958\\
0.433947613373392	0.631058271357272\\
0.443947613373392	0.62516823991898\\
0.453947613373392	0.616927612140913\\
0.463947613373392	0.606333760579093\\
0.473947613373392	0.59339376742417\\
0.483947613373392	0.578124345068033\\
0.493947613373392	0.560551797176829\\
0.503947613373392	0.540711982119372\\
0.513947613373392	0.518650361381008\\
0.523947613373392	0.494421667824632\\
0.533947613373392	0.468089742401489\\
0.543947613373392	0.439727407726841\\
0.553947613373392	0.409416358041221\\
0.563947613373392	0.377246593180182\\
0.573947613373392	0.343316140152789\\
0.580460710030044	0.320320106924185\\
0.586973806686696	0.296652924444274\\
0.593486903343348	0.272346822270843\\
0.6	0.247435246710376\\
};
\addplot [color=mycolor2,solid,forget plot,line width=2,mark=*,mark repeat=8]
  table[row sep=crcr]{%
0.6	-0.00113571488991581\\
0.603532699186387	0.0106259200512068\\
0.607948573169371	0.025283971240737\\
0.612364447152355	0.0398788319861229\\
0.616780321135339	0.0543963924579689\\
0.621196195118323	0.068823885847382\\
0.626196195118323	0.0850367303936423\\
0.631196195118323	0.101104844716021\\
0.636196195118323	0.117015829260631\\
0.641196195118323	0.132759532737771\\
0.646196195118323	0.148327976608746\\
0.651196195118323	0.163715311582156\\
0.656196195118323	0.178917782805606\\
0.661196195118323	0.193933657250997\\
0.666196195118323	0.208763140337996\\
0.671196195118323	0.223408360395183\\
0.676196195118323	0.237873156363127\\
0.681196195118323	0.252162928831821\\
0.686196195118323	0.266284559932936\\
0.691196195118323	0.280246433598837\\
0.696196195118323	0.294058063790188\\
0.701196195118323	0.307729883711063\\
0.706196195118323	0.321273180003241\\
0.711196195118323	0.334700153121811\\
0.716196195118323	0.348023421324643\\
0.721196195118323	0.361255769404633\\
0.726196195118323	0.374410106887331\\
0.731196195118323	0.387499568729555\\
0.736196195118323	0.400536943934636\\
0.741196195118323	0.413534409722259\\
0.746196195118323	0.426503520998279\\
0.751196195118323	0.439455347310653\\
0.756196195118323	0.452399883477765\\
0.761196195118323	0.465345796949046\\
0.766196195118323	0.478300452536496\\
0.771196195118323	0.491270077590916\\
0.776196195118323	0.50425921448536\\
0.781196195118323	0.517270507782409\\
0.785897146338742	0.529524562074963\\
0.790598097559162	0.541798280467473\\
0.795299048779581	0.554089354471624\\
0.8	0.566393729148291\\
};
\addplot [color=mycolor2,solid,line width=2,mark=*,mark repeat=8]
  table[row sep=crcr]{%
0.8	0.0785956283191812\\
0.804787570469528	0.0572484337904771\\
0.809575140939056	0.0360984559940948\\
0.819575140939056	-0.00737043448547884\\
0.829575140939056	-0.0497614936984634\\
0.839575140939056	-0.0909338906794314\\
0.849575140939056	-0.130752580578333\\
0.859575140939056	-0.169088585348518\\
0.869575140939056	-0.20581951107475\\
0.879575140939056	-0.240830266818163\\
0.889575140939056	-0.274013412476595\\
0.899575140939056	-0.305269365807434\\
0.909575140939056	-0.334506811127062\\
0.919575140939056	-0.361643200367443\\
0.929575140939056	-0.386604990884705\\
0.939575140939056	-0.409327766294774\\
0.949575140939056	-0.429756510886916\\
0.959575140939056	-0.447845864248628\\
0.969575140939056	-0.463560235505065\\
0.979575140939056	-0.476873830842833\\
0.989575140939056	-0.487770776308733\\
0.999575140939056	-0.496245113368408\\
1.00957514093906	-0.502300784016758\\
1.01957514093906	-0.505951563697215\\
1.02957514093906	-0.507221024577986\\
1.03957514093906	-0.50614227611003\\
1.04957514093906	-0.502757823706909\\
1.05957514093906	-0.497119411668122\\
1.06957514093906	-0.489287828950105\\
1.07957514093906	-0.479332415120703\\
1.08957514093906	-0.467330803343673\\
1.09957514093906	-0.453368683515877\\
1.10957514093906	-0.437539462358189\\
1.11957514093906	-0.419943570093407\\
1.12957514093906	-0.400688105779112\\
1.13957514093906	-0.379886535819466\\
1.14957514093906	-0.357658229278182\\
1.15957514093906	-0.334127613193227\\
1.16957514093906	-0.309423744422528\\
1.17718135570429	-0.289929915226358\\
1.18478757046953	-0.269894520667039\\
1.19239378523476	-0.249378680640158\\
1.2	-0.228444391020615\\
};
\addlegendentry{$\eSystemRed$ BT};

\addplot [color=mycolor3,solid,forget plot,line width=2,mark=triangle*,mark repeat=8]
  table[row sep=crcr]{%
0	0\\
0.005	-0.000193043837808789\\
0.01	-0.000763918515462472\\
0.015	-0.0016997096945871\\
0.02	-0.00298679759052218\\
0.025	-0.00461091731805518\\
0.03	-0.00655717703195856\\
0.035	-0.00881034864683133\\
0.04	-0.0113550190426017\\
0.045	-0.01417563515585\\
0.05	-0.0172565029169777\\
0.055	-0.0205821102095348\\
0.06	-0.0241372799689297\\
0.065	-0.0279071951490502\\
0.07	-0.0318773811287925\\
0.075	-0.0360340200432382\\
0.08	-0.0403640863918244\\
0.085	-0.04485534955654\\
0.09	-0.049496345020226\\
0.095	-0.0542766391832472\\
0.1	-0.0591869297478136\\
0.105	-0.0642190262831631\\
0.11	-0.0693658176373588\\
0.115	-0.0746214510102083\\
0.12	-0.079981383072154\\
0.125	-0.0854423418474426\\
0.13	-0.0910022988664162\\
0.135	-0.0966605351291693\\
0.14	-0.102417634259171\\
0.145	-0.108275431393632\\
0.15	-0.114236998766869\\
0.155	-0.120306583326951\\
0.16	-0.126489539637164\\
0.165	-0.132792273246964\\
0.17	-0.139222247484652\\
0.175	-0.145787791565397\\
0.18	-0.152497977733995\\
0.185	-0.159362567531613\\
0.19	-0.166392045691739\\
0.195	-0.173597312056468\\
0.2	-0.180989513850015\\
};
\addplot [color=mycolor3,solid,forget plot,line width=2,mark=triangle*,mark repeat=8]
  table[row sep=crcr]{%
0.2	0.240210773509795\\
0.20571928874942	0.256654635199476\\
0.212868399686195	0.277155210837676\\
0.220017510622969	0.297548099472285\\
0.227166621559744	0.317782495218931\\
0.234315732496519	0.337807654378689\\
0.244315732496519	0.365371225101889\\
0.254315732496519	0.392288780661413\\
0.264315732496519	0.418424495156906\\
0.274315732496519	0.443644862049345\\
0.284315732496519	0.46781904708025\\
0.294315732496519	0.490819526399484\\
0.304315732496519	0.512522935936579\\
0.314315732496519	0.532810494734467\\
0.324315732496519	0.551568320521271\\
0.334315732496519	0.568688026738629\\
0.344315732496519	0.584067432870003\\
0.354315732496519	0.597610917779156\\
0.364315732496519	0.609229677921638\\
0.374315732496519	0.6188422463544\\
0.384315732496519	0.626375014770468\\
0.394315732496519	0.631762492718803\\
0.404315732496519	0.634947492117629\\
0.414315732496519	0.635881536259313\\
0.424315732496519	0.63452515637754\\
0.434315732496519	0.630848038681404\\
0.444315732496519	0.624829123532802\\
0.454315732496519	0.616456879385902\\
0.464315732496519	0.605729351220728\\
0.474315732496519	0.592654184551923\\
0.484315732496519	0.577248633060622\\
0.494315732496519	0.559539680890884\\
0.504315732496519	0.539563836296398\\
0.514315732496519	0.517367029722469\\
0.524315732496519	0.493004529496541\\
0.534315732496519	0.466540905412866\\
0.544315732496519	0.438049577435164\\
0.554315732496519	0.407612593093173\\
0.564315732496519	0.375320456668792\\
0.574315732496519	0.341271936939004\\
0.584315732496519	0.30557339670398\\
0.594315732496519	0.268338462568293\\
0.595736799372389	0.262929363631089\\
0.597157866248259	0.257492022746744\\
0.59857893312413	0.252026799795027\\
0.6	0.246534057244766\\
};
\addplot [color=mycolor3,solid,forget plot,line width=2,mark=triangle*,mark repeat=8]
  table[row sep=crcr]{%
0.6	-0.00381053368691148\\
0.605714910451362	0.0169551739198242\\
0.608889860702118	0.0284667185392661\\
0.615239761203632	0.0514085827007457\\
0.620239761203632	0.0693747603214495\\
0.625239761203632	0.0872342134037326\\
0.630239761203632	0.104971108029148\\
0.635239761203632	0.122571947309686\\
0.640239761203632	0.140025493731278\\
0.645239761203632	0.157322709339426\\
0.650239761203632	0.17445678348746\\
0.655239761203632	0.191423084367593\\
0.660239761203632	0.208219069046087\\
0.665239761203632	0.224844249836999\\
0.670239761203632	0.241300030777288\\
0.675239761203632	0.257589574362427\\
0.680239761203632	0.273717711282661\\
0.685239761203632	0.289690942739747\\
0.690239761203632	0.305517098776113\\
0.695239761203632	0.321205133348595\\
0.700239761203632	0.336765046100508\\
0.705239761203632	0.352207926807293\\
0.710239761203632	0.367545468394905\\
0.715239761203632	0.382789711569469\\
0.720239761203632	0.397952989885206\\
0.725239761203632	0.41304801811755\\
0.730239761203632	0.428087309138319\\
0.735239761203632	0.443082895150999\\
0.740239761203632	0.458046304944088\\
0.745239761203632	0.472988693259941\\
0.750239761203632	0.487920221810096\\
0.755239761203632	0.502849787077441\\
0.760239761203632	0.517785035227419\\
0.765239761203632	0.532732524987108\\
0.770239761203632	0.547697137702479\\
0.775239761203632	0.562681841308989\\
0.780239761203632	0.577687744277483\\
0.785239761203632	0.592714280646789\\
0.790239761203632	0.607758711571998\\
0.795239761203632	0.622815951442262\\
0.796429820902724	0.626400879857765\\
0.797619880601816	0.629986018025987\\
0.798809940300908	0.633571241948832\\
0.8	0.637156420084051\\
};
\addplot [color=mycolor3,solid,line width=2,mark=triangle*,mark repeat=8]
  table[row sep=crcr]{%
0.8	0.0679952790523712\\
0.80518072705095	0.0453447383554826\\
0.810937090440894	0.0204580014371707\\
0.813815272135866	0.00813377054321401\\
0.823815272135866	-0.0340108122510958\\
0.833815272135866	-0.0749999161905726\\
0.843815272135866	-0.114697580638675\\
0.853815272135866	-0.1529738215849\\
0.863815272135866	-0.189704904824359\\
0.873815272135866	-0.224773852653095\\
0.883815272135866	-0.258071134697272\\
0.893815272135866	-0.289495002470717\\
0.903815272135866	-0.318951687119806\\
0.913815272135866	-0.346355793629254\\
0.923815272135866	-0.371630771504694\\
0.933815272135866	-0.394709138009826\\
0.943815272135866	-0.415532588527192\\
0.953815272135866	-0.434052253564134\\
0.963815272135866	-0.450228922278951\\
0.973815272135866	-0.464033141681737\\
0.983815272135866	-0.475445233139063\\
0.993815272135866	-0.484455395983482\\
1.00381527213587	-0.491063672944442\\
1.01381527213587	-0.49527992059344\\
1.02381527213587	-0.497123731391581\\
1.03381527213587	-0.496624377240072\\
1.04381527213587	-0.493820522539402\\
1.05381527213587	-0.488760069406468\\
1.06381527213587	-0.481499990560534\\
1.07381527213587	-0.472106116110647\\
1.08381527213587	-0.460652616161799\\
1.09381527213587	-0.447221732307148\\
1.10381527213587	-0.431903532200949\\
1.11381527213587	-0.414795552638841\\
1.12381527213587	-0.396002088332079\\
1.13381527213587	-0.375633828378463\\
1.14381527213587	-0.353807548233128\\
1.15381527213587	-0.330645631088432\\
1.16381527213587	-0.306275211718542\\
1.17381527213587	-0.280827742605129\\
1.1803614541019	-0.263650856075834\\
1.18690763606793	-0.24610919091067\\
1.19345381803397	-0.228242029770596\\
1.2	-0.21008910006526\\
};
\addlegendentry{$\eSystemRed$ IRKA};

\addplot [color=mycolor4,forget plot,solid,line width=2,mark=diamond*,mark repeat=8]
  table[row sep=crcr]{%
0	0\\
0.005	-0.00017352327534162\\
0.01	-0.000679646509113549\\
0.015	-0.00149630275904892\\
0.02	-0.0026008956776976\\
0.025	-0.00397032665613991\\
0.03	-0.00558101441985145\\
0.035	-0.00740905045084271\\
0.04	-0.00943027140679793\\
0.045	-0.011620284786567\\
0.05	-0.0139544885777157\\
0.055	-0.0164082301264463\\
0.06	-0.0189568782105477\\
0.065	-0.0215758465460834\\
0.07	-0.0242406137334548\\
0.075	-0.0269268782223575\\
0.08	-0.0296106268247674\\
0.085	-0.0322681555235448\\
0.09	-0.0348760900421213\\
0.095	-0.0374115297809895\\
0.1	-0.0398521097485752\\
0.105	-0.0421760182860479\\
0.11	-0.0443620186328099\\
0.115	-0.0463895752247782\\
0.12	-0.048238906330125\\
0.125	-0.0498909984637305\\
0.13	-0.0513276293342536\\
0.135	-0.052531470667686\\
0.14	-0.0534861292472988\\
0.145	-0.0541761579492333\\
0.15	-0.0545870804364043\\
0.155	-0.0547054656689175\\
0.16	-0.0545189555451646\\
0.165	-0.0540162726406525\\
0.17	-0.0531872469602875\\
0.175	-0.052022858477652\\
0.18	-0.0505152501361649\\
0.185	-0.0486577325105615\\
0.19	-0.0464448128447229\\
0.195	-0.0438722032756952\\
0.2	-0.0409368185572335\\
};
\addplot [color=mycolor4,solid,forget plot,line width=2,mark=diamond*,mark repeat=8]
  table[row sep=crcr]{%
0.2	0.319160320144601\\
0.20479396010162	0.336729291628647\\
0.208218217317063	0.349208779740878\\
0.216778860355671	0.380109314637172\\
0.225339503394278	0.410507270070112\\
0.233900146432886	0.440307525619731\\
0.242460789471493	0.469415838016009\\
0.252460789471493	0.50241756800487\\
0.262460789471493	0.534201771482212\\
0.272460789471493	0.564625681914751\\
0.282460789471493	0.593550545117986\\
0.292460789471493	0.620841950035495\\
0.302460789471493	0.646370395638831\\
0.312460789471493	0.670012240764565\\
0.322460789471493	0.6916501625393\\
0.332460789471493	0.711173430121254\\
0.342460789471493	0.728478396229345\\
0.352460789471493	0.743469278698725\\
0.362460789471493	0.75605853089785\\
0.372460789471493	0.766167040649764\\
0.382460789471493	0.773724515178875\\
0.392460789471493	0.778670043905729\\
0.402460789471493	0.780952356650954\\
0.412460789471493	0.780529934729735\\
0.422460789471493	0.777371265001083\\
0.432460789471493	0.771455146404766\\
0.442460789471493	0.762770818551656\\
0.452460789471493	0.75131797758056\\
0.462460789471493	0.737106883387318\\
0.472460789471493	0.72015838795491\\
0.482460789471493	0.700503924910639\\
0.492460789471493	0.678185428797767\\
0.502460789471493	0.65325528892533\\
0.512460789471493	0.625776094661758\\
0.522460789471493	0.595820485187798\\
0.532460789471493	0.56347097698444\\
0.542460789471493	0.528819767492485\\
0.552460789471493	0.491968210137479\\
0.562460789471493	0.453026532272764\\
0.57184559210362	0.414685124072966\\
0.581230394735746	0.374712338961462\\
0.590615197367873	0.333218916993815\\
0.6	0.290321174452241\\
};
\addplot [color=mycolor4,solid,forget plot,line width=2,mark=diamond*,mark repeat=8]
  table[row sep=crcr]{%
0.6	-0.152340179081358\\
0.605	-0.144698699470346\\
0.61	-0.136994157925086\\
0.615	-0.129261153092876\\
0.62	-0.121533902579016\\
0.625	-0.113846216553402\\
0.63	-0.106231480179345\\
0.635	-0.0987224481153074\\
0.64	-0.0913511554791364\\
0.645	-0.0841488966465955\\
0.65	-0.077146209157213\\
0.655	-0.0703726889197741\\
0.66	-0.0638569127773936\\
0.665	-0.0576264227556291\\
0.67	-0.0517077107301243\\
0.675	-0.0461260623471171\\
0.68	-0.0409054943558403\\
0.685	-0.0360687443371622\\
0.69	-0.0316372554056811\\
0.695	-0.0276310555595292\\
0.7	-0.0240687122915834\\
0.705	-0.0209673276139728\\
0.71	-0.0183425221031468\\
0.715	-0.0162083548506949\\
0.72	-0.0145772971231057\\
0.725	-0.0134602323009705\\
0.73	-0.0128664386565578\\
0.735	-0.0128035533539936\\
0.74	-0.0132775661293335\\
0.745	-0.0142928235928411\\
0.75	-0.0158520102499959\\
0.755	-0.0179561581655523\\
0.76	-0.0206046608135015\\
0.765	-0.023795281052119\\
0.77	-0.027524130054629\\
0.775	-0.0317857223805711\\
0.78	-0.0365730093310816\\
0.785	-0.0418773898037453\\
0.79	-0.0476886869804538\\
0.795	-0.0539952467142996\\
0.8	-0.060783988950773\\
};
\addplot [color=mycolor4,solid,line width=2,mark=diamond*,mark repeat=8]
  table[row sep=crcr]{%
0.8	-0.0373937685013989\\
0.804760339881337	-0.069510065376573\\
0.810710764733009	-0.109228269808708\\
0.81666118958468	-0.148436146341259\\
0.822611614436351	-0.187096142565991\\
0.828562039288023	-0.225171480125298\\
0.838562039288023	-0.287743251326623\\
0.848562039288023	-0.348397113985638\\
0.858562039288023	-0.406975147684475\\
0.868562039288023	-0.463327601933484\\
0.878562039288023	-0.517313096118993\\
0.888562039288023	-0.56879893058531\\
0.898562039288023	-0.617662027419362\\
0.908562039288023	-0.663789340299333\\
0.918562039288023	-0.707077960641289\\
0.928562039288023	-0.747435294080704\\
0.938562039288023	-0.784779716033681\\
0.948562039288023	-0.819040837536801\\
0.958562039288023	-0.850159510770559\\
0.968562039288023	-0.878087857078398\\
0.978562039288023	-0.9027896122417\\
0.988562039288023	-0.924240237755195\\
0.998562039288023	-0.942426825180525\\
1.00856203928802	-0.957347971428718\\
1.01856203928802	-0.969013806218612\\
1.02856203928802	-0.977445947155732\\
1.03856203928802	-0.982677308661156\\
1.04856203928802	-0.984751830070419\\
1.05856203928802	-0.983724195843984\\
1.06856203928802	-0.979659641866637\\
1.07856203928802	-0.972633680616774\\
1.08856203928802	-0.962731697094831\\
1.09856203928802	-0.950048389851764\\
1.10856203928802	-0.93468744452503\\
1.11856203928802	-0.916761192095395\\
1.12856203928802	-0.896390096165218\\
1.13856203928802	-0.873701958650293\\
1.14856203928802	-0.848831484174521\\
1.15856203928802	-0.821919892385174\\
1.16856203928802	-0.793114327126954\\
1.17856203928802	-0.762566883706576\\
1.18856203928802	-0.730434093904875\\
1.19142152946602	-0.720976702589312\\
1.19428101964401	-0.711406603091098\\
1.19714050982201	-0.701727621259375\\
1.2	-0.691943601640834\\
};
\addlegendentry{nice selection};

\end{axis}
\end{tikzpicture}%

%% file: IEEETacExample1-TimeDomainSimulation-r8-v2.tex
%
\definecolor{mycolor1}{rgb}{0.00000,0.44700,0.74100}%
\definecolor{mycolor2}{rgb}{0.85000,0.32500,0.09800}%
\definecolor{mycolor3}{rgb}{0.92900,0.69400,0.12500}%
\definecolor{mycolor4}{rgb}{0.49400,0.18400,0.55600}%
\begin{tikzpicture}

\begin{axis}[%
width=5in,
height=1.7in,
at={(0.758in,0.481in)},
scale only axis,
xmin=0,
xmax=1.2,
ymin=-0.4,
ymax=0.8,
xlabel={$t$},
ylabel={$\outVar$},
axis background/.style={fill=white},
legend style={at={(0.15,0.05)},anchor=south west,legend cell align=left,align=left,draw=white!15!black}
]
\addplot [color=mycolor1,solid,forget plot,line width=2,mark=square*,mark repeat=8]
  table[row sep=crcr]{%
0	0\\
0.00482772026777773	0.0127505429947475\\
0.00969437376352142	0.0254351699578776\\
0.0145610272592651	0.0379480947491389\\
0.0194276807550088	0.0502870251226944\\
0.0242943342507525	0.0624496694285892\\
0.0342943342507525	0.0868785266170972\\
0.0442943342507525	0.110533543613826\\
0.0542943342507525	0.133395039838707\\
0.0642943342507525	0.15544373556669\\
0.0742943342507525	0.176660733978353\\
0.0842943342507525	0.197027448803063\\
0.0942943342507525	0.216525929526413\\
0.104294334250752	0.235138975962258\\
0.114294334250752	0.252850119242495\\
0.124294334250752	0.269643558686708\\
0.134294334250752	0.285504423325653\\
0.144294334250752	0.300418861108975\\
0.154294334250753	0.314374018812792\\
0.164294334250753	0.327357987449607\\
0.174294334250752	0.339360004638072\\
0.184294334250753	0.35037052016387\\
0.194294334250753	0.360381174913522\\
0.204294334250752	0.369384754305741\\
0.214294334250753	0.377375335721324\\
0.224294334250753	0.384348332103141\\
0.234294334250753	0.3903004701268\\
0.244294334250753	0.395229751240424\\
0.254294334250753	0.39913554827037\\
0.264294334250753	0.40201862878054\\
0.274294334250753	0.403881132767536\\
0.284294334250753	0.404726540952919\\
0.294294334250753	0.404559724734179\\
0.304294334250753	0.403386951099442\\
0.314294334250753	0.401215860188702\\
0.324294334250753	0.398055440502116\\
0.334294334250753	0.393916036515197\\
0.344294334250753	0.388809337042319\\
0.354294334250753	0.382748353038507\\
0.364294334250753	0.375747399378809\\
0.374294334250753	0.367822064655402\\
0.384294334250753	0.358989184985283\\
0.388220750688064	0.355276802357186\\
0.392147167125376	0.351428423156536\\
0.396073583562688	0.347445219956752\\
0.4	0.343328392933633\\
};
\addplot [color=mycolor1,solid,forget plot,line width=2,mark=square*,mark repeat=8]
  table[row sep=crcr]{%
0.4	0.33645884975644\\
0.404480092765995	0.350258178997988\\
0.409051615996602	0.364490474165097\\
0.411337377611905	0.371639577206934\\
0.418837377611905	0.395110575788187\\
0.426337377611905	0.418341779246229\\
0.433837377611905	0.44100875705782\\
0.441337377611905	0.462807848358053\\
0.448837377611905	0.483456233175439\\
0.456337377611905	0.50268946973878\\
0.463837377611905	0.520271917239609\\
0.471337377611905	0.536001102219111\\
0.478837377611905	0.549706120335298\\
0.486337377611905	0.561244160827213\\
0.493837377611905	0.570507864328918\\
0.501337377611905	0.577427335753268\\
0.508837377611905	0.581967300646863\\
0.516337377611905	0.58412349360376\\
0.523837377611905	0.583925230361712\\
0.531337377611905	0.581434625384638\\
0.538837377611905	0.576743240137735\\
0.546337377611905	0.569969240072081\\
0.553837377611905	0.561254683133467\\
0.561337377611905	0.550762211878921\\
0.568837377611905	0.538672052175625\\
0.576337377611905	0.525180638985319\\
0.583837377611905	0.510493460463288\\
0.591337377611905	0.4948201294288\\
0.598837377611905	0.478372490382743\\
0.606337377611905	0.461365032072583\\
0.613837377611905	0.444005233289958\\
0.621337377611905	0.426488314299184\\
0.628837377611905	0.408996915434689\\
0.636337377611905	0.391703154144015\\
0.643837377611905	0.374759013177574\\
0.651337377611905	0.358292129568337\\
0.658837377611905	0.342407097004736\\
0.666337377611905	0.327188605913545\\
0.673837377611905	0.312694349094514\\
0.681337377611905	0.29895291261051\\
0.686003033208929	0.290786298212748\\
0.690668688805953	0.28290860733583\\
0.695334344402976	0.275311757739985\\
0.7	0.267983841041342\\
};
\addplot [color=mycolor1,solid,forget plot,line width=2,mark=square*,mark repeat=8]
  table[row sep=crcr]{%
0.7	0.586730781759912\\
0.705793723305565	0.59771058018613\\
0.709932097095255	0.605302094272783\\
0.714932097095255	0.614193247447126\\
0.719932097095255	0.622775290812145\\
0.724932097095255	0.631046537843957\\
0.729932097095255	0.639005399695356\\
0.734932097095255	0.646650384941529\\
0.739932097095255	0.653980098515348\\
0.744932097095255	0.660993246470645\\
0.749932097095255	0.667688637670598\\
0.754932097095255	0.674065183496621\\
0.759932097095255	0.680121896900118\\
0.764932097095255	0.685857896223795\\
0.769932097095255	0.691272406500407\\
0.774932097095255	0.696364759127265\\
0.779932097095255	0.701134391031624\\
0.784932097095255	0.705580847583359\\
0.789932097095255	0.709703783518573\\
0.794932097095255	0.71350296258234\\
0.799932097095255	0.716978256804894\\
0.804932097095255	0.72012964854018\\
0.809932097095255	0.7229572310294\\
0.814932097095255	0.725461208014801\\
0.819932097095255	0.727641893123729\\
0.824932097095255	0.729499711068936\\
0.829932097095255	0.731035197867791\\
0.834932097095255	0.73224900043003\\
0.839932097095255	0.733141876046653\\
0.844932097095255	0.733714692789329\\
0.849932097095255	0.733968429401653\\
0.854932097095255	0.733904174863902\\
0.859932097095255	0.733523127983607\\
0.864932097095255	0.732826597032924\\
0.869932097095255	0.731815999328986\\
0.874932097095255	0.730492860778813\\
0.879932097095255	0.728858815568216\\
0.884932097095255	0.726915605077543\\
0.889932097095255	0.724665077168857\\
0.892449072821441	0.72341649604737\\
0.894966048547627	0.722090791550591\\
0.897483024273814	0.720688229785103\\
0.9	0.71920908354183\\
};
\addplot [color=mycolor1,solid,line width=2,mark=square*,mark repeat=8]
  table[row sep=crcr]{%
0.9	0.313219052315188\\
0.906764325510035	0.323073507124773\\
0.910522284126722	0.328677318948369\\
0.914280242743408	0.334338246658901\\
0.918038201360094	0.340028147625808\\
0.925538201360094	0.351357990529101\\
0.933038201360094	0.362460846863716\\
0.940538201360094	0.37310842505442\\
0.948038201360094	0.383078489622531\\
0.955538201360094	0.392156098814629\\
0.963038201360094	0.400132768484873\\
0.970538201360094	0.406815126498832\\
0.978038201360094	0.412029313869385\\
0.985538201360094	0.415620817535183\\
0.993038201360094	0.417452329076982\\
1.00053820136009	0.417411635978413\\
1.00803820136009	0.41541479858593\\
1.01553820136009	0.411404624360386\\
1.02303820136009	0.405347775152951\\
1.03053820136009	0.397239872869722\\
1.03803820136009	0.387106639852858\\
1.04553820136009	0.375001410603858\\
1.05303820136009	0.361002233134513\\
1.06053820136009	0.345212885883641\\
1.06803820136009	0.32776169462896\\
1.07553820136009	0.308798722991484\\
1.08303820136009	0.288493611583479\\
1.09053820136009	0.267032254368173\\
1.09803820136009	0.244613601568068\\
1.10553820136009	0.221447248630644\\
1.11303820136009	0.197752563278365\\
1.12053820136009	0.173751879663664\\
1.12803820136009	0.149666109702969\\
1.13553820136009	0.125713342575516\\
1.14303820136009	0.102109475607615\\
1.15053820136009	0.079059667043142\\
1.15803820136009	0.0567538761710895\\
1.16553820136009	0.0353668209385108\\
1.17303820136009	0.0150599410567039\\
1.18053820136009	-0.00402672077465401\\
1.18803820136009	-0.0217779218819159\\
1.19102865102007	-0.0284621252643061\\
1.19401910068005	-0.0349140121870299\\
1.19700955034002	-0.04112935655379\\
1.2	-0.0471045843040652\\
};
\addlegendentry{FOM};

\addplot [color=mycolor2,solid,forget plot,line width=2,mark=*,mark repeat=8]
  table[row sep=crcr]{%
0	0\\
0.00519138547662594	0.0114586636468717\\
0.010403619890909	0.0228702569691202\\
0.0130097370980505	0.0285391514502144\\
0.0230097370980505	0.0500470842606419\\
0.0330097370980506	0.0711356758660123\\
0.0430097370980506	0.0917639361511788\\
0.0530097370980506	0.111892246626126\\
0.0630097370980506	0.131482351762305\\
0.0730097370980506	0.15049731875057\\
0.0830097370980506	0.168901700802619\\
0.0930097370980506	0.18666159633516\\
0.103009737098051	0.203744637185808\\
0.113009737098051	0.22011995052259\\
0.123009737098051	0.235758291312582\\
0.133009737098051	0.250632087390841\\
0.143009737098051	0.264715424918334\\
0.153009737098051	0.277984013146078\\
0.163009737098051	0.29041528409447\\
0.173009737098051	0.301988422982947\\
0.183009737098051	0.312684351349242\\
0.193009737098051	0.322485695275613\\
0.203009737098051	0.331376851231191\\
0.213009737098051	0.33934400172993\\
0.223009737098051	0.346375096574238\\
0.233009737098051	0.352459825043755\\
0.243009737098051	0.357589647749664\\
0.253009737098051	0.361757797766807\\
0.263009737098051	0.364959260494949\\
0.273009737098051	0.367190750195341\\
0.283009737098051	0.368450708568417\\
0.293009737098051	0.368739291962095\\
0.303009737098051	0.368058350360433\\
0.313009737098051	0.366411408532996\\
0.323009737098051	0.363803632855624\\
0.333009737098051	0.360241805522893\\
0.343009737098051	0.35573430317694\\
0.353009737098051	0.350291082812691\\
0.363009737098051	0.343923619097472\\
0.373009737098051	0.336644866804628\\
0.379757302823538	0.331225785365004\\
0.386504868549025	0.325403135679276\\
0.393252434274513	0.319182004516977\\
0.4	0.312567757650885\\
};
\addplot [color=mycolor2,solid,forget plot,line width=2,mark=*,mark repeat=8]
  table[row sep=crcr]{%
0.4	0.281703178272372\\
0.404844927553228	0.302966979742288\\
0.409689855106456	0.323951797064455\\
0.417189855106456	0.355713258965064\\
0.424689855106456	0.386355992647652\\
0.432189855106455	0.415621226912739\\
0.439689855106455	0.44326878548236\\
0.447189855106455	0.469077198021014\\
0.454689855106455	0.492841763530063\\
0.462189855106455	0.514383108061017\\
0.469689855106455	0.533550793842796\\
0.477189855106455	0.550222108932359\\
0.484689855106455	0.564299297067542\\
0.492189855106455	0.575715703377225\\
0.499689855106455	0.584437535984426\\
0.507189855106455	0.590461609260149\\
0.514689855106455	0.593812387059065\\
0.522189855106455	0.594544268303852\\
0.529689855106455	0.592741077688602\\
0.537189855106455	0.588513291714612\\
0.544689855106455	0.581995595299433\\
0.552189855106455	0.573344761033306\\
0.559689855106455	0.562736982764076\\
0.567189855106455	0.55036524303547\\
0.574689855106455	0.536437968027085\\
0.582189855106455	0.521172997186393\\
0.589689855106455	0.504793388897715\\
0.597189855106455	0.487525559969771\\
0.604689855106455	0.46959936656881\\
0.612189855106456	0.451239594980442\\
0.619689855106455	0.43266123770675\\
0.627189855106455	0.414068855269848\\
0.634689855106455	0.395658071741541\\
0.642189855106456	0.377606616220175\\
0.649689855106455	0.360070199949915\\
0.657189855106455	0.343183259807222\\
0.664689855106456	0.327061517143641\\
0.672189855106456	0.311794715235966\\
0.679689855106456	0.297444088975187\\
0.684767391329842	0.288267430548127\\
0.689844927553228	0.279532130773382\\
0.694922463776614	0.271236941143605\\
0.7	0.263374885167896\\
};
\addplot [color=mycolor2,solid,forget plot,line width=2,mark=*,mark repeat=8]
  table[row sep=crcr]{%
0.7	0.553575580155589\\
0.704519323956955	0.562328866104162\\
0.70903864791391	0.570883164143449\\
0.713557971870865	0.57923541985886\\
0.71807729582782	0.587382667529908\\
0.72307729582782	0.596154114810123\\
0.72807729582782	0.604667335395417\\
0.73307729582782	0.612918693395817\\
0.73807729582782	0.620904688672299\\
0.74307729582782	0.628621956581174\\
0.74807729582782	0.636067267283699\\
0.75307729582782	0.643237527979846\\
0.75807729582782	0.650129783627416\\
0.76307729582782	0.656741216649904\\
0.76807729582782	0.663069146315735\\
0.77307729582782	0.669111030326248\\
0.77807729582782	0.674864465259832\\
0.78307729582782	0.680327186245561\\
0.78807729582782	0.685497066413889\\
0.79307729582782	0.690372117841945\\
0.79807729582782	0.694950491725375\\
0.80307729582782	0.699230478020186\\
0.80807729582782	0.703210504966294\\
0.81307729582782	0.706889139397124\\
0.81807729582782	0.710265086643486\\
0.82307729582782	0.713337190146228\\
0.82807729582782	0.716104431053541\\
0.83307729582782	0.718565927905684\\
0.83807729582782	0.720720936276773\\
0.84307729582782	0.722568848360981\\
0.84807729582782	0.724109192644056\\
0.85307729582782	0.725341632977615\\
0.85807729582782	0.726265967966262\\
0.86307729582782	0.72688213053017\\
0.86807729582782	0.727190187650807\\
0.87307729582782	0.727190338852745\\
0.87807729582782	0.726882915344946\\
0.883557971870865	0.726193140367181\\
0.88903864791391	0.725135158634539\\
0.894519323956955	0.723709946055308\\
0.9	0.721918662797868\\
};
\addplot [color=mycolor2,solid,line width=2,mark=*,mark repeat=8]
  table[row sep=crcr]{%
0.9	0.250844215568183\\
0.904393761268584	0.256981653350585\\
0.913181283805753	0.268924478793251\\
0.920504219253393	0.278414816430898\\
0.927827154701034	0.287354961094383\\
0.935150090148674	0.295623486430203\\
0.942650090148674	0.303270480887443\\
0.950150090148674	0.309959828762481\\
0.957650090148674	0.315570293675393\\
0.965150090148674	0.319989347134373\\
0.972650090148674	0.323113480281634\\
0.980150090148674	0.324847360772221\\
0.987650090148674	0.325108662936085\\
0.995150090148674	0.323830270618387\\
1.00265009014867	0.320959828247848\\
1.01015009014867	0.316458322071759\\
1.01765009014867	0.310303911703108\\
1.02515009014867	0.302493231292959\\
1.03265009014867	0.29304027397705\\
1.04015009014867	0.281974731357552\\
1.04765009014867	0.269343870988784\\
1.05515009014867	0.255212600158461\\
1.06265009014867	0.239661939980718\\
1.07015009014867	0.222787520774506\\
1.07765009014867	0.204699002689901\\
1.08515009014867	0.185518858217085\\
1.09265009014867	0.165380794527913\\
1.10015009014867	0.144428770784853\\
1.10765009014867	0.122814039679616\\
1.11515009014867	0.100692911184615\\
1.12265009014867	0.0782254987727041\\
1.13015009014867	0.0555755042028617\\
1.13765009014867	0.0329055275918499\\
1.14515009014867	0.0103743220626167\\
1.15265009014867	-0.0118638339132123\\
1.16015009014867	-0.0336585200066048\\
1.16765009014867	-0.0548685263612882\\
1.17515009014867	-0.0753644708642237\\
1.18136256761151	-0.091716569514133\\
1.18757504507434	-0.107435935801529\\
1.19378752253717	-0.122468136340552\\
1.2	-0.136766425195519\\
};
\addlegendentry{$\eSystemRed$ BT};

\addplot [color=mycolor3,solid,forget plot,line width=2,mark=triangle*,mark repeat=8]
  table[row sep=crcr]{%
0	0\\
0.00543159277359727	0.0121028504616607\\
0.0109069887147235	0.0241968322065264\\
0.0163823846558498	0.0361769560857759\\
0.0218577805969761	0.0480363502113046\\
0.0273331765381024	0.0597682520811956\\
0.0373331765381024	0.0808439758206533\\
0.0473331765381024	0.101432735283786\\
0.0573331765381024	0.121496239933022\\
0.0673331765381024	0.140997601016222\\
0.0773331765381024	0.159901321993228\\
0.0873331765381024	0.178173259032761\\
0.0973331765381024	0.195780775255239\\
0.107333176538102	0.212692795741299\\
0.117333176538102	0.228879794959772\\
0.127333176538102	0.244313759649824\\
0.137333176538102	0.258968311561786\\
0.147333176538102	0.272818748171335\\
0.157333176538102	0.285842027488222\\
0.167333176538102	0.29801673399219\\
0.177333176538102	0.309323168223611\\
0.187333176538102	0.319743372810464\\
0.197333176538102	0.329261115089867\\
0.207333176538102	0.337861856667724\\
0.217333176538102	0.345532809111818\\
0.227333176538102	0.352262945284488\\
0.237333176538102	0.358042980248503\\
0.247333176538102	0.362865344911451\\
0.257333176538102	0.366724207956017\\
0.267333176538102	0.36961547284117\\
0.277333176538102	0.371536757489839\\
0.287333176538102	0.372487372362792\\
0.297333176538102	0.372468309602943\\
0.307333176538102	0.371482226413186\\
0.317333176538102	0.369533424035641\\
0.327333176538102	0.366627830479521\\
0.337333176538102	0.362772958638882\\
0.347333176538102	0.357977877068495\\
0.357333176538102	0.352253188763334\\
0.367333176538102	0.345611018643908\\
0.377333176538102	0.338064943112842\\
0.387333176538103	0.329629949524789\\
0.390499882403577	0.32677611849283\\
0.393666588269051	0.323835329198234\\
0.396833294134526	0.320808143987845\\
0.4	0.31769513839542\\
};
\addplot [color=mycolor3,solid,forget plot,line width=2,mark=triangle*,mark repeat=8]
  table[row sep=crcr]{%
0.4	0.297987612600981\\
0.40546449223761	0.323792983610422\\
0.41092898447522	0.349048210923835\\
0.41842898447522	0.382600514386598\\
0.42592898447522	0.414598772678185\\
0.43342898447522	0.444765245478283\\
0.44092898447522	0.472846901068319\\
0.44842898447522	0.498615196028904\\
0.45592898447522	0.521863521125831\\
0.46342898447522	0.542416779899162\\
0.47092898447522	0.560135212595048\\
0.47842898447522	0.574912637896554\\
0.48592898447522	0.586673054446176\\
0.49342898447522	0.595376915248295\\
0.50092898447522	0.601022593693327\\
0.50842898447522	0.60364349837546\\
0.51592898447522	0.603304641969498\\
0.52342898447522	0.600104099038302\\
0.53092898447522	0.594171750494046\\
0.53842898447522	0.585665963701924\\
0.54592898447522	0.574770959583226\\
0.55342898447522	0.561693093063337\\
0.56092898447522	0.546657170957075\\
0.56842898447522	0.529903503715377\\
0.57592898447522	0.511686716506391\\
0.58342898447522	0.492267745275238\\
0.59092898447522	0.471908616816906\\
0.59842898447522	0.450870595420252\\
0.60592898447522	0.429414737294028\\
0.61342898447522	0.407791543588209\\
0.62092898447522	0.386235473236436\\
0.62842898447522	0.364964645422624\\
0.63592898447522	0.34418301243796\\
0.64342898447522	0.324070193349792\\
0.65092898447522	0.304777063405532\\
0.65842898447522	0.286427094200356\\
0.66592898447522	0.269119621538626\\
0.67342898447522	0.25292235075495\\
0.68092898447522	0.237869109368432\\
0.685696738356415	0.228896971516099\\
0.69046449223761	0.220384020487742\\
0.695232246118805	0.212319930421185\\
0.7	0.204689439694851\\
};
\addplot [color=mycolor3,solid,forget plot,line width=2,mark=triangle*,mark repeat=8]
  table[row sep=crcr]{%
0.7	0.577472972406354\\
0.70412808180009	0.585198022715255\\
0.70912808180009	0.594315663241275\\
0.71412808180009	0.603167980173417\\
0.71912808180009	0.611751468517894\\
0.72412808180009	0.620062757888548\\
0.72912808180009	0.628098612246755\\
0.73412808180009	0.635855929234629\\
0.73912808180009	0.643331742242533\\
0.74412808180009	0.650523221060425\\
0.74912808180009	0.657427671582902\\
0.75412808180009	0.664042535211958\\
0.75912808180009	0.670365390289332\\
0.76412808180009	0.676393952478087\\
0.76912808180009	0.682126074435074\\
0.77412808180009	0.687559745284785\\
0.77912808180009	0.69269309142144\\
0.78412808180009	0.697524376624191\\
0.78912808180009	0.702052001699419\\
0.79412808180009	0.706274504026797\\
0.79912808180009	0.710190557739718\\
0.80412808180009	0.713798973579046\\
0.80912808180009	0.717098698507826\\
0.81412808180009	0.720088815330192\\
0.81912808180009	0.722768542262326\\
0.82412808180009	0.725137232531177\\
0.82912808180009	0.727194373964253\\
0.83412808180009	0.728939588681584\\
0.83912808180009	0.73037263207283\\
0.84412808180009	0.731493392148788\\
0.84912808180009	0.732301889109045\\
0.85412808180009	0.732798275106763\\
0.85912808180009	0.732982832650824\\
0.86412808180009	0.732855973719268\\
0.86912808180009	0.732418239307674\\
0.87412808180009	0.731670299266105\\
0.87912808180009	0.730612950148253\\
0.88412808180009	0.729247114097213\\
0.888096061350067	0.727944333439458\\
0.892064040900045	0.726448506160352\\
0.896032020450023	0.724760260376537\\
0.9	0.722880272444879\\
};
\addplot [color=mycolor3,solid,line width=2,mark=triangle*,mark repeat=8]
  table[row sep=crcr]{%
0.9	0.185139836196137\\
0.904004904549246	0.192314729771306\\
0.912014713647739	0.20633672848787\\
0.916019618196986	0.21313168784524\\
0.923519618196986	0.225355187623901\\
0.931019618196986	0.236777418540762\\
0.938519618196986	0.247227565079426\\
0.946019618196986	0.256541535544922\\
0.953519618196986	0.264562819258322\\
0.961019618196986	0.271141781680975\\
0.968519618196986	0.276142287163745\\
0.976019618196986	0.279444998683297\\
0.983519618196986	0.280947208551112\\
0.991019618196986	0.28056119481702\\
0.998519618196986	0.278220163063437\\
1.00601961819699	0.273880605577943\\
1.01351961819699	0.267521127004938\\
1.02101961819699	0.25914025112316\\
1.02851961819699	0.24876019233493\\
1.03601961819699	0.236427669555053\\
1.04351961819699	0.222211996550732\\
1.05101961819699	0.206202870818605\\
1.05851961819699	0.188510990359627\\
1.06601961819699	0.169267086651611\\
1.07351961819699	0.148619734633281\\
1.08101961819699	0.126733668644924\\
1.08851961819699	0.103787015999076\\
1.09601961819699	0.0799687461745876\\
1.10351961819699	0.0554767309348172\\
1.11101961819699	0.0305170111241116\\
1.11851961819699	0.00529824236380713\\
1.12601961819699	-0.0199718518630668\\
1.13351961819699	-0.0450876937060505\\
1.14101961819699	-0.0698454369713479\\
1.14851961819699	-0.0940500245390822\\
1.15601961819699	-0.117518892321946\\
1.16351961819699	-0.140082152440665\\
1.17101961819699	-0.161581114777629\\
1.17851961819699	-0.181875179496433\\
1.18601961819699	-0.200844810539319\\
1.18951471364774	-0.209203750034975\\
1.19300980909849	-0.217244194735576\\
1.19650490454925	-0.224958509298614\\
1.2	-0.232340131219567\\
};
\addlegendentry{$\eSystemRed$ IRKA};

\addplot [color=mycolor4,forget plot,solid,line width=2,mark=diamond*,mark repeat=8]
  table[row sep=crcr]{%
0	0\\
0.00615260141840658	0.0162206408897668\\
0.00924125674893198	0.0242613616684371\\
0.0154185674099828	0.0401349968651363\\
0.0254185674099828	0.0652340773891419\\
0.0354185674099828	0.0895769073925738\\
0.0454185674099828	0.113143646362703\\
0.0554185674099828	0.135914772699067\\
0.0654185674099828	0.157871065283808\\
0.0754185674099828	0.178993527686871\\
0.0854185674099828	0.199263732682798\\
0.0954185674099828	0.218663943924939\\
0.105418567409983	0.237177096764497\\
0.115418567409983	0.254786732314038\\
0.125418567409983	0.271477275885103\\
0.135418567409983	0.287234132598627\\
0.145418567409983	0.302043667303869\\
0.155418567409983	0.315893147598784\\
0.165418567409983	0.328770961323158\\
0.175418567409983	0.340666687968462\\
0.185418567409983	0.351571077698072\\
0.195418567409983	0.361476002651463\\
0.205418567409983	0.370374618058869\\
0.215418567409983	0.378261411198319\\
0.225418567409983	0.385132179640768\\
0.235418567409983	0.390983990321631\\
0.245418567409983	0.395815288552635\\
0.255418567409983	0.39962592624704\\
0.265418567409983	0.402417139609693\\
0.275418567409983	0.404191515558049\\
0.285418567409983	0.404953052821339\\
0.295418567409983	0.404707171190016\\
0.305418567409983	0.403460688943273\\
0.315418567409983	0.401221796254296\\
0.325418567409983	0.398000072623545\\
0.335418567409983	0.393806478982282\\
0.345418567409983	0.388653335200548\\
0.355418567409983	0.382554300128403\\
0.365418567409983	0.37552434975571\\
0.375418567409983	0.367579754092966\\
0.381563925557487	0.362251299798387\\
0.387709283704991	0.356588335147387\\
0.393854641852496	0.350595341013112\\
0.4	0.344276968590795\\
};
\addplot [color=mycolor4,forget plot,solid,line width=2,mark=diamond*,mark repeat=8]
  table[row sep=crcr]{%
0.4	0.335883374228174\\
0.405630715083659	0.335412035146201\\
0.411261430167318	0.335145894734036\\
0.416892145250976	0.335056520606712\\
0.422522860334635	0.335115431544246\\
0.430022860334635	0.335375101615274\\
0.437522860334635	0.33577982442807\\
0.445022860334635	0.336262787797911\\
0.452522860334635	0.336758420107654\\
0.460022860334635	0.337202432532151\\
0.467522860334635	0.337531670189286\\
0.475022860334635	0.337685143118274\\
0.482522860334635	0.337604461779815\\
0.490022860334635	0.337233826068745\\
0.497522860334635	0.336519841534134\\
0.505022860334635	0.335412441540447\\
0.512522860334635	0.33386524492721\\
0.520022860334635	0.331835494782175\\
0.527522860334635	0.329283856741341\\
0.535022860334635	0.326175158347096\\
0.542522860334635	0.322478641507417\\
0.550022860334635	0.318167858111417\\
0.557522860334635	0.313220468292348\\
0.565022860334635	0.30761874110073\\
0.572522860334635	0.301349684256619\\
0.580022860334635	0.294404907055891\\
0.587522860334635	0.286780435819293\\
0.595022860334635	0.278476942135479\\
0.602522860334635	0.269499743365291\\
0.610022860334635	0.259858645654672\\
0.617522860334635	0.249567791711491\\
0.625022860334635	0.238645607039122\\
0.632522860334635	0.227114675768295\\
0.640022860334635	0.215001577798491\\
0.647522860334635	0.202336780767453\\
0.655022860334635	0.189154318788509\\
0.662522860334635	0.175491558514581\\
0.670022860334635	0.161389044414875\\
0.677522860334635	0.146890442648086\\
0.685022860334635	0.132041988815896\\
0.692522860334635	0.116892167866428\\
0.694392145250976	0.113075213302746\\
0.696261430167318	0.109243484286474\\
0.698130715083659	0.105397791340392\\
0.7	0.101538951211925\\
};
\addplot [color=mycolor4,forget plot,solid,line width=2,mark=diamond*,mark repeat=8]
  table[row sep=crcr]{%
0.7	0.679810292466387\\
0.703817032901321	0.677655515138679\\
0.707634065802641	0.67535038830774\\
0.711451098703962	0.672896187983387\\
0.715268131605282	0.670294209087594\\
0.720268131605282	0.666664451045593\\
0.725268131605282	0.662786379693968\\
0.730268131605282	0.658663051356706\\
0.735268131605282	0.654297576773549\\
0.740268131605282	0.64969312070075\\
0.745268131605282	0.644852901640859\\
0.750268131605282	0.639780190875288\\
0.755268131605282	0.634478311832651\\
0.760268131605282	0.628950639685987\\
0.765268131605282	0.623200601172781\\
0.770268131605282	0.61723167305614\\
0.775268131605282	0.611047381269073\\
0.780268131605282	0.604651300510683\\
0.785268131605282	0.598047054150185\\
0.790268131605282	0.591238312170851\\
0.795268131605282	0.584228790112842\\
0.800268131605282	0.577022248670857\\
0.805268131605282	0.569622493676054\\
0.810268131605282	0.562033373575801\\
0.815268131605282	0.554258778197257\\
0.820268131605282	0.546302638348902\\
0.825268131605282	0.538168925874393\\
0.830268131605282	0.529861650720101\\
0.835268131605282	0.521384859541186\\
0.840268131605282	0.512742635309383\\
0.845268131605282	0.503939097432928\\
0.850268131605282	0.494978398462796\\
0.855268131605282	0.485864722562584\\
0.860268131605282	0.476602285124842\\
0.865268131605282	0.467195332951292\\
0.870268131605282	0.457648140647626\\
0.875268131605282	0.447965008978024\\
0.880268131605283	0.438150264492245\\
0.885268131605283	0.428208259760458\\
0.890268131605283	0.418143369505312\\
0.895268131605283	0.407959988861453\\
0.896451098703962	0.405533815802182\\
0.897634065802641	0.403101315821078\\
0.898817032901321	0.400662547613992\\
0.9	0.398217569931787\\
};
\addplot [color=mycolor4,solid,line width=2,mark=diamond*,mark repeat=8]
  table[row sep=crcr]{%
0.9	0.122254833739653\\
0.905146364855839	0.108173520155801\\
0.910507161580671	0.0936810094456703\\
0.915867958305503	0.0793884146825076\\
0.921228755030336	0.0653163281601307\\
0.926589551755168	0.051484758108555\\
0.934089551755168	0.0325741378337732\\
0.941589551755168	0.014223493775401\\
0.949089551755168	-0.00351858836119193\\
0.956589551755168	-0.0206066701921707\\
0.964089551755168	-0.0369984153673549\\
0.971589551755168	-0.0526544149600905\\
0.979089551755168	-0.0675388611007154\\
0.986589551755168	-0.0816197844365421\\
0.994089551755168	-0.0948689705400078\\
1.00158955175517	-0.107261782166136\\
1.00908955175517	-0.118777637589767\\
1.01658955175517	-0.129400145706969\\
1.02408955175517	-0.139116992438741\\
1.03158955175517	-0.147919774714921\\
1.03908955175517	-0.155804250457466\\
1.04658955175517	-0.162770363650636\\
1.05408955175517	-0.168822111261087\\
1.06158955175517	-0.173967402300869\\
1.06908955175517	-0.178218066698497\\
1.07658955175517	-0.181589772155074\\
1.08408955175517	-0.184101883359981\\
1.09158955175517	-0.185777356912059\\
1.09908955175517	-0.186642516965171\\
1.10658955175517	-0.186726874685344\\
1.11408955175517	-0.186062991765716\\
1.12158955175517	-0.184686418598557\\
1.12908955175517	-0.182635263827028\\
1.13658955175517	-0.179949934986709\\
1.14408955175517	-0.176673017532463\\
1.15158955175517	-0.172849259768172\\
1.15908955175517	-0.168524979716288\\
1.16658955175517	-0.163747751457965\\
1.17408955175517	-0.158566309180985\\
1.18158955175517	-0.153030578333195\\
1.18908955175517	-0.14719097527632\\
1.19658955175517	-0.141098067500674\\
1.19744216381638	-0.140391642445233\\
1.19829477587758	-0.139682675196306\\
1.19914738793879	-0.138971240280764\\
1.2	-0.138257412180633\\
};
\addlegendentry{nice selection};

\end{axis}
\end{tikzpicture}%

%% file: TwoRoomsStatic-HankelSingularValues.tex
%
\definecolor{mycolor1}{rgb}{0.00000,0.44700,0.74100}%
\definecolor{mycolor2}{rgb}{0.85000,0.32500,0.09800}%
\definecolor{mycolor3}{rgb}{0.92900,0.69400,0.12500}%
\begin{tikzpicture}

\begin{axis}[%
width=2.521in,
height=1.866in,
at={(0.758in,0.481in)},
scale only axis,
xmin=1,
xmax=103,
ymode=log,
ymin=1e-17,
ymax=1,
yminorticks=true,
axis background/.style={fill=white},
legend style={legend cell align=left,align=left,draw=white!15!black}
]
\addplot [color=mycolor1,line width=2,mark=square*]
  table[row sep=crcr]{%
1	1\\
2	0.130285910466108\\
3	0.00767432353565992\\
4	6.06446369166999e-07\\
5	7.32916642676969e-11\\
6	1.15091616310612e-13\\
7	3.20258816156763e-16\\
8	2.21244180231175e-16\\
9	9.69217018132452e-17\\
10	9.69217018132452e-17\\
11	9.69217018132452e-17\\
12	9.69217018132452e-17\\
13	9.69217018132452e-17\\
14	9.69217018132452e-17\\
15	9.69217018132452e-17\\
16	9.69217018132452e-17\\
17	9.69217018132452e-17\\
18	9.69217018132452e-17\\
19	9.69217018132452e-17\\
20	9.69217018132452e-17\\
21	9.69217018132452e-17\\
22	9.69217018132452e-17\\
23	9.69217018132452e-17\\
24	9.69217018132452e-17\\
25	9.69217018132452e-17\\
26	9.69217018132452e-17\\
27	9.69217018132452e-17\\
28	9.69217018132452e-17\\
29	9.69217018132452e-17\\
30	9.69217018132452e-17\\
31	9.69217018132452e-17\\
32	9.69217018132452e-17\\
33	9.69217018132452e-17\\
34	9.69217018132452e-17\\
35	9.69217018132452e-17\\
36	9.69217018132452e-17\\
37	9.69217018132452e-17\\
38	9.69217018132452e-17\\
39	9.69217018132452e-17\\
40	9.69217018132452e-17\\
41	9.69217018132452e-17\\
42	9.69217018132452e-17\\
43	9.69217018132452e-17\\
44	9.69217018132452e-17\\
45	9.69217018132452e-17\\
46	9.69217018132452e-17\\
47	9.69217018132452e-17\\
48	9.69217018132452e-17\\
49	9.69217018132452e-17\\
50	9.69217018132452e-17\\
51	9.69217018132452e-17\\
52	9.69217018132452e-17\\
53	9.69217018132452e-17\\
54	9.69217018132452e-17\\
55	9.69217018132452e-17\\
56	9.69217018132452e-17\\
57	9.69217018132452e-17\\
58	9.69217018132452e-17\\
59	9.69217018132452e-17\\
60	9.69217018132452e-17\\
61	9.69217018132452e-17\\
62	9.69217018132452e-17\\
63	9.69217018132452e-17\\
64	9.69217018132452e-17\\
65	9.69217018132452e-17\\
66	9.69217018132452e-17\\
67	9.69217018132452e-17\\
68	9.69217018132452e-17\\
69	9.69217018132452e-17\\
70	9.69217018132452e-17\\
71	9.69217018132452e-17\\
72	9.69217018132452e-17\\
73	9.69217018132452e-17\\
74	9.69217018132452e-17\\
75	9.69217018132452e-17\\
76	9.69217018132452e-17\\
77	9.69217018132452e-17\\
78	9.69217018132452e-17\\
79	9.69217018132452e-17\\
80	9.69217018132452e-17\\
81	9.69217018132452e-17\\
82	9.69217018132452e-17\\
83	9.69217018132452e-17\\
84	9.69217018132452e-17\\
85	9.69217018132452e-17\\
86	9.69217018132452e-17\\
87	9.69217018132452e-17\\
88	9.69217018132452e-17\\
89	9.69217018132452e-17\\
90	9.69217018132452e-17\\
91	9.69217018132452e-17\\
92	9.69217018132452e-17\\
93	9.69217018132452e-17\\
94	9.69217018132452e-17\\
95	9.69217018132452e-17\\
96	9.69217018132452e-17\\
97	9.69217018132452e-17\\
98	9.69217018132452e-17\\
99	9.69217018132452e-17\\
100	9.69217018132452e-17\\
101	9.69217018132452e-17\\
102	9.69217018132452e-17\\
103	3.46998880491551e-17\\
};
\addlegendentry{$\system_1$};

\addplot [color=mycolor2,line width=2,mark=triangle*]
  table[row sep=crcr]{%
1	1\\
2	0.128368133072725\\
3	0.0104336757152911\\
4	0.000550339746218776\\
5	2.87805504531662e-05\\
6	6.06374654285586e-06\\
7	9.91506462831602e-07\\
8	1.70891607121309e-07\\
9	1.59021400688505e-08\\
10	3.90334954303179e-09\\
11	1.10515179656523e-09\\
12	1.57435489306885e-10\\
13	1.64571215404177e-11\\
14	2.23411361179628e-12\\
15	3.37077696898614e-13\\
16	1.52668532641179e-13\\
17	1.50031195887263e-14\\
18	1.76264906069736e-15\\
19	1.74041537067918e-16\\
20	9.23783312447503e-17\\
21	9.23783312447503e-17\\
22	9.23783312447503e-17\\
23	9.23783312447503e-17\\
24	9.23783312447503e-17\\
25	9.23783312447503e-17\\
26	9.23783312447503e-17\\
27	9.23783312447503e-17\\
28	9.23783312447503e-17\\
29	9.23783312447503e-17\\
30	9.23783312447503e-17\\
31	9.23783312447503e-17\\
32	9.23783312447503e-17\\
33	9.23783312447503e-17\\
34	9.23783312447503e-17\\
35	9.23783312447503e-17\\
36	9.23783312447503e-17\\
37	9.23783312447503e-17\\
38	9.23783312447503e-17\\
39	9.23783312447503e-17\\
40	9.23783312447503e-17\\
41	9.23783312447503e-17\\
42	9.23783312447503e-17\\
43	9.23783312447503e-17\\
44	9.23783312447503e-17\\
45	9.23783312447503e-17\\
46	9.23783312447503e-17\\
47	9.23783312447503e-17\\
48	9.23783312447503e-17\\
49	9.23783312447503e-17\\
50	9.23783312447503e-17\\
51	9.23783312447503e-17\\
52	9.23783312447503e-17\\
53	9.23783312447503e-17\\
54	9.23783312447503e-17\\
55	9.23783312447503e-17\\
56	9.23783312447503e-17\\
57	9.23783312447503e-17\\
58	9.23783312447503e-17\\
59	9.23783312447503e-17\\
60	9.23783312447503e-17\\
61	9.23783312447503e-17\\
62	9.23783312447503e-17\\
63	9.23783312447503e-17\\
64	9.23783312447503e-17\\
65	9.23783312447503e-17\\
66	9.23783312447503e-17\\
67	9.23783312447503e-17\\
68	9.23783312447503e-17\\
69	9.23783312447503e-17\\
70	9.23783312447503e-17\\
71	9.23783312447503e-17\\
72	9.23783312447503e-17\\
73	9.23783312447503e-17\\
74	9.23783312447503e-17\\
75	9.23783312447503e-17\\
76	9.23783312447503e-17\\
77	9.23783312447503e-17\\
78	9.23783312447503e-17\\
79	9.23783312447503e-17\\
80	9.23783312447503e-17\\
81	9.23783312447503e-17\\
82	9.23783312447503e-17\\
83	9.23783312447503e-17\\
84	9.23783312447503e-17\\
85	9.23783312447503e-17\\
86	9.23783312447503e-17\\
87	9.23783312447503e-17\\
88	9.23783312447503e-17\\
89	9.23783312447503e-17\\
90	9.23783312447503e-17\\
91	9.23783312447503e-17\\
92	9.23783312447503e-17\\
93	9.23783312447503e-17\\
94	9.23783312447503e-17\\
95	9.23783312447503e-17\\
96	9.23783312447503e-17\\
97	9.23783312447503e-17\\
98	9.23783312447503e-17\\
99	9.23783312447503e-17\\
100	9.23783312447503e-17\\
101	9.23783312447503e-17\\
102	9.23783312447503e-17\\
103	6.19448031270253e-17\\
};
\addlegendentry{$\system_2$};

\addplot [color=mycolor3,line width=2,mark=*]
  table[row sep=crcr]{%
1	1\\
2	0.957547551683315\\
3	0.657886079228718\\
4	0.00830866424966674\\
5	0.00027775491460167\\
6	2.22399369115331e-05\\
7	8.36819806873606e-06\\
8	3.21103134832412e-06\\
9	1.7217042628827e-06\\
10	7.83565071836298e-07\\
11	4.62995825203506e-07\\
12	2.05528613577479e-07\\
13	1.27974678713286e-07\\
14	5.17971631458545e-08\\
15	3.44309249009879e-08\\
16	1.23489269646231e-08\\
17	8.79606872825697e-09\\
18	2.80039230658158e-09\\
19	2.11720574434769e-09\\
20	6.20173004512541e-10\\
21	4.90890885567971e-10\\
22	1.41642855929513e-10\\
23	1.13875030656879e-10\\
24	3.34622256603302e-11\\
25	2.53788166638354e-11\\
26	6.97059803288528e-12\\
27	5.25215048853118e-12\\
28	1.3272729821191e-12\\
29	1.04035699367767e-12\\
30	2.65562524160872e-13\\
31	1.98953936161739e-13\\
32	5.12240083205708e-14\\
33	3.60938312208108e-14\\
34	8.76368133478673e-15\\
35	6.27511198011802e-15\\
36	1.49068725952589e-15\\
37	1.04687628279028e-15\\
38	2.90105140329413e-16\\
39	1.54090086098437e-16\\
40	1.23653636447867e-16\\
41	9.82979681854495e-17\\
42	9.82979681854495e-17\\
43	9.82979681854495e-17\\
44	9.82979681854495e-17\\
45	9.82979681854495e-17\\
46	9.82979681854495e-17\\
47	9.82979681854495e-17\\
48	9.82979681854495e-17\\
49	9.82979681854495e-17\\
50	9.82979681854495e-17\\
51	9.82979681854495e-17\\
52	9.82979681854495e-17\\
53	9.82979681854495e-17\\
54	9.82979681854495e-17\\
55	9.82979681854495e-17\\
56	9.82979681854495e-17\\
57	9.82979681854495e-17\\
58	9.82979681854495e-17\\
59	9.82979681854495e-17\\
60	9.82979681854495e-17\\
61	9.82979681854495e-17\\
62	9.82979681854495e-17\\
63	9.82979681854495e-17\\
64	9.82979681854495e-17\\
65	9.82979681854495e-17\\
66	9.82979681854495e-17\\
67	9.82979681854495e-17\\
68	9.82979681854495e-17\\
69	9.82979681854495e-17\\
70	9.82979681854495e-17\\
71	9.82979681854495e-17\\
72	9.82979681854495e-17\\
73	9.82979681854495e-17\\
74	9.82979681854495e-17\\
75	9.82979681854495e-17\\
76	9.82979681854495e-17\\
77	9.82979681854495e-17\\
78	9.82979681854495e-17\\
79	9.82979681854495e-17\\
80	9.82979681854495e-17\\
81	9.82979681854495e-17\\
82	9.82979681854495e-17\\
83	9.82979681854495e-17\\
84	9.82979681854495e-17\\
85	9.82979681854495e-17\\
86	9.82979681854495e-17\\
87	9.82979681854495e-17\\
88	9.82979681854495e-17\\
89	9.82979681854495e-17\\
90	9.82979681854495e-17\\
91	9.82979681854495e-17\\
92	9.82979681854495e-17\\
93	9.82979681854495e-17\\
94	9.82979681854495e-17\\
95	9.82979681854495e-17\\
96	9.82979681854495e-17\\
97	9.82979681854495e-17\\
98	9.82979681854495e-17\\
99	9.82979681854495e-17\\
100	9.82979681854495e-17\\
101	9.82979681854495e-17\\
102	9.82979681854495e-17\\
103	4.89588340981849e-17\\
};
\addlegendentry{$\eSystem$};

\end{axis}
\end{tikzpicture}%

%% file: SchU18-ppt.bbl
\begin{thebibliography}{10}

\bibitem{Ant05}
A.~C. Antoulas.
\newblock {\em Approximation of Large-Scale Dynamical Systems}.
\newblock SIAM, Philadelphia, PA, USA, 2005.

\bibitem{AntBG10}
A.~C. Antoulas, C.~A. Beattie, and S.~Gugercin.
\newblock Interpolatory model reduction of large-scale dynamical systems.
\newblock In J.~Mohammadpour and K.~M. Grigoriadis, editors, {\em Efficient
  Modeling and Control of Large-Scale Systems}, pages 3--58. Springer, New
  York, NY, USA, 2010.

\bibitem{Ast10}
A.~Astolfi.
\newblock Model reduction by moment matching for linear and nonlinear systems.
\newblock {\em IEEE Trans. Autom. Control}, 55(10):2321--2336, 2010.

\bibitem{BasPWL14b}
M.~Ba\c{s}tu\u{g}, M.~Petreczky, R.~Wisniewski, and J.~Leth.
\newblock Model reduction by moment matching for linear switched systems.
\newblock In {\em 2014 Am. Control Conf.}, pages 3942--3947, Los Angeles, CA,
  USA, 2014.

\bibitem{BasPWL16}
M.~Ba\c{s}tu\u{g}, M.~Petreczky, R.~Wisniewski, and J.~Leth.
\newblock Model reduction by nice selections for linear switched systems.
\newblock {\em IEEE Trans. Automat. Contr.}, 61(11):3422--3437, 2016.

\bibitem{BalBBPS00}
A.~Balluchi, L.~Benvenuti, M.~D. di~Benedetto, C.~Pinello, and A.~L.
  Sangiovanni-Vincentelli.
\newblock Automotive engine control and hybrid systems: {c}hallenges and
  opportunities.
\newblock {\em Proc. IEEE}, 88(7):888--912, 2000.

\bibitem{BauBB14}
U.~Baur, C.~Beattie, and P.~Benner.
\newblock {Mapping parameters across system boundaries: parameterized model
  reduction with low rank variability in dynamics}.
\newblock {\em Proc. Appl. Math. and Mech.}, 14(1):19--22, 2014.

\bibitem{BauBF14}
U.~Baur, P.~Benner, and L.~Feng.
\newblock Model order reduction for linear and nonlinear systems: {a}
  system-theoretic perspective.
\newblock {\em Arch. Comput. Methods Eng.}, 21(4):331--358, 2014.

\bibitem{BeaG09}
C.~Beattie and S.~Gugercin.
\newblock Interpolatory projection methods for structure-preserving model
  reduction.
\newblock {\em Syst. Control Lett.}, 58(3):225--232, 2009.

\bibitem{BeaGM17}
C.~Beattie, S.~Gugercin, and V.~Mehrmann.
\newblock {Model reduction for systems with inhomogeneous initial conditions}.
\newblock {\em Syst. Control Lett.}, 99:99--106, 2017.

\bibitem{BeaMXZ17_ppt}
C.~{Beattie}, V.~{Mehrmann}, H.~{Xu}, and H.~{Zwart}.
\newblock Port-{H}amiltonian descriptor systems.
\newblock {\em ArXiv e-prints, 1705.09081}, 2017.

\bibitem{BenCOW17}
P.~Benner, A.~Cohen, M.~Ohlberger, and K.~Willcox.
\newblock {\em Model Reduction and Approximation}.
\newblock SIAM, Philadelphia, PA, 2017.

\bibitem{Ber92}
Michael~W Berry.
\newblock Large scale sparse singular value computations.
\newblock {\em Int. J. Supercomput. Appl.}, 6:13--49, 1992.

\bibitem{ChaV02}
Y.~Chahlaoui and P.~{Van Dooren}.
\newblock {A collection of benchmark examples for model reduction of linear
  time invariant dynamical systems}.
\newblock Slicot working note 2002-2: February 2002.

\bibitem{ChaBG16}
S.~Chaturantabut, C.~Beattie, and S.~Gugercin.
\newblock Structure-preserving model reduction for nonlinear port-{H}amiltonian
  systems.
\newblock {\em SIAM J. Sci. Comput.}, 38(5):B837--B865, 2016.

\bibitem{Fre08}
R.~W. Freund.
\newblock Structure-preserving model order reduction of {RCL} circuit
  equations.
\newblock In W.~H.~A. Schilders, H.~A. van~der Vorst, and J.~Rommes, editors,
  {\em Model Order Reduction: Theory, Research Aspects and Applications}, pages
  49--73. Springer Berlin Heidelberg, Germany, 2008.

\bibitem{GilS17}
N.~Gillis and P.~Sharma.
\newblock On computing the distance to stability for matrices using linear
  dissipative {H}amiltonian systems.
\newblock {\em Automatica}, 85:113--121, 2017.

\bibitem{GosPAF17}
I.~V. {Gosea}, M.~{Petreczky}, A.~C. {Antoulas}, and C.~{Fiter}.
\newblock {Balanced truncation for linear switched systems}.
\newblock {\em ArXiv e-prints, 1712.02158}, 2017.

\bibitem{GugAB08}
S.~Gugercin, A.~C. Antoulas, and C.~Beattie.
\newblock {$\mathcal{H}_2$ model reduction for large-scale linear dynamical
  systems}.
\newblock {\em SIAM J. Matrix Anal. Appl.}, 30(2):609--638, 2008.

\bibitem{GugPBS12}
S.~Gugercin, R.~V. Polyuga, C.~Beattie, and A.~van~der Schaft.
\newblock Structure-preserving tangential interpolation for model reduction of
  port-{H}amiltonian systems.
\newblock {\em Automatica}, 48(9):1963--1974, 2012.

\bibitem{HeiRA11}
M.~Heinkenschloss, T.~Reis, and A.~C. Antoulas.
\newblock {Balanced truncation model reduction for systems with inhomogeneous
  initial conditions}.
\newblock {\em Automatica}, 47(3):559--564, 2011.

\bibitem{HinV05}
M.~Hinze and S.~Volkwein.
\newblock Proper orthogonal decomposition surrogate models for nonlinear
  dynamical systems: {e}rror estimates and suboptimal control.
\newblock In P.~Benner, V.~Mehrmann, and D.~C. Sorensen, editors, {\em
  Dimension reduction of large-scale systems}, pages 261--306. Springer
  Berlin/Heidelberg, Germany, 2005.

\bibitem{JarDM13}
E.~Jarlebring, T.~Damm, and W.~Michiels.
\newblock {Model reduction of time-delay systems using position balancing and
  delay Lyapunov equations}.
\newblock {\em Math. Control. Signals, Syst.}, 25:147--166, 2013.

\bibitem{KunM17ppt}
P.~{Kunkel} and V.~{Mehrmann}.
\newblock Regular solutions of {DAE} hybrid systems and regularization
  techniques.
\newblock Preprint 05-2017, Institut f\"ur Mathematik, TU Berlin, 2017.

\bibitem{LalKM03}
S.~Lall, P.~Krysl, and J.~E. Marsden.
\newblock Structure-preserving model reduction for mechanical systems.
\newblock {\em Phys. D Nonlinear Phenom.}, 184(1-4):304--318, 2003.

\bibitem{LeeS03}
J.~Lee and S.~Song.
\newblock Modeling urban transportation systems with hybrid systems: {a}n
  overview.
\newblock In {\em Proceedings of the 2003 IEEE International Conference on
  Intelligent Transportation Systems}, pages 1269--1272, Shanghai, China, 2003.

\bibitem{Lib03}
D.~Liberzon.
\newblock {\em {Switching in Systems and Control}}.
\newblock Springer Science \& Business Media, New York, NY, USA, 2003.

\bibitem{MazVBB08}
E.~Mazzi, A.~S. Vincentelli, A.~Balluchi, and A.~Bicchi.
\newblock {Hybrid system reduction}.
\newblock In {\em Proc. IEEE Conf. Decis. Control}, pages 227--232, Cancun,
  Mexico, 2008.

\bibitem{MehMW18}
C.~{Mehl}, V.~{Mehrmann}, and M.~{Wojtylak}.
\newblock {Linear algebra properties of dissipative Hamiltonian descriptor
  systems}.
\newblock {\em ArXiv e-prints, 1801.02214}, January 2018.

\bibitem{Moo81}
B.~Moore.
\newblock Principal component analysis in linear systems: {c}ontrollability,
  observability, and model reduction.
\newblock {\em IEEE Trans. Automat. Control}, 26(1):17--32, 1981.

\bibitem{MulR76}
C.~Mullis and R.~Roberts.
\newblock Synthesis of minimum roundoff noise fixed point digital filters.
\newblock {\em IEEE Trans. Circuits Syst.}, 23(9):551--562, 1976.

\bibitem{OhlR16}
M.~Ohlberger and S.~Rave.
\newblock {Reduced basis methods: success, limitations and future challenges}.
\newblock In {\em Algoritmy 2016, Conference on Scientific Computing}, pages
  1--12, Podbansk{\'e}, Slovakia, 2016.

\bibitem{Oez68}
M.~N. {\"O}zisik.
\newblock {\em Boundary Value Problems of Heat Conduction}.
\newblock Dover Publications, Mineola, NY, USA, 1968.

\bibitem{PapP14}
A.~V. Papadopoulos and M.~Prandini.
\newblock Model reduction of switched affine systems: {a} method based on
  balanced truncation and randomized optimization.
\newblock In {\em Proc. 17th Int. Conf. Hybrid Syst. Comput. Control}, pages
  113--122, Berlin, Germany, 2014.

\bibitem{PapP16}
A.~V. Papadopoulos and M.~Prandini.
\newblock Model reduction of switched affine systems.
\newblock {\em Automatica}, 70:57--65, 2016.

\bibitem{Pat80}
S.~V. Patankar.
\newblock {\em Numerical Heat Transfer and Fluid Flow}.
\newblock Taylor \& Francis, Abingdon-on-Thames, UK, 1980.

\bibitem{PepC00}
D.~L. Pepyne and C.~G. Cassandras.
\newblock Optimal control of hybrid systems in manufacturing.
\newblock {\em Proc. IEEE}, 88(7):1108--1123, 2000.

\bibitem{Pet11}
M.~Petreczky.
\newblock Realization theory for linear and bilinear switched systems: {f}ormal
  power series approach -- {P}art {I}: {r}ealization theory of linear switched
  systems.
\newblock {\em ESAIM Control. Optim. Calc. Var.}, 17(2):410--445, 2011.

\bibitem{PetWL13}
M.~Petreczky, R.~Wisniewski, and J.~Leth.
\newblock {Balanced truncation for linear switched systems}.
\newblock {\em Nonlinear Anal. Hybrid Syst.}, 10(1):4--20, 2013.

\bibitem{ScaA16}
G.~Scarciotti and A.~Astolfi.
\newblock {Model reduction for hybrid systems with state-dependent jumps}.
\newblock {\em IFAC-PapersOnLine}, 49(18):850--855, 2016.

\bibitem{SchU18_code}
P.~Schulze and B.~Unger.
\newblock Code for the paper '{M}odel reduction for linear systems with
  low-rank switching', January 2018.
\newblock Available online from \url{https://doi.org/10.5281/zenodo.1161994}.

\bibitem{SchUBG18}
P.~Schulze, B.~Unger, C.~Beattie, and S.~Gugercin.
\newblock Data-driven structured realization.
\newblock {\em Linear Algebra Appl.}, 537:250--286, 2018.

\bibitem{SelBA13_ppt}
K.~Selvaraj, M.~N. Belur, and R.~Abdulrazak.
\newblock Dissipative systems: {u}ncontrollability, observability and {RLC}
  realizability.
\newblock {\em ArXiv e-prints, 1110.2102v1}, 2013.

\bibitem{ShaW09}
H.~R. Shaker and R.~Wisniewski.
\newblock Generalized {G}ramian framework for model reduction of switched
  systems.
\newblock In {\em European Control Conference (ECC)}, pages 1029--1034,
  Budapest, Hungary, 2009.

\bibitem{ShaW12}
H.~R. Shaker and R.~Wisniewski.
\newblock Model reduction of switched systems based on switching generalized
  {G}ramians.
\newblock {\em Int. J. Innov. Comput. I.}, 8(7(B)):5025--5044, 2012.

\bibitem{SunGL02}
Z.~Sun, S.~S. Ge, and T.~H. Lee.
\newblock {Controllability and reachability criteria for switched linear
  systems}.
\newblock {\em Automatica}, 38(5):775--786, 2002.

\bibitem{WouMB15}
N.~van~de Wouw, W.~Michiels, and B.~Besselink.
\newblock {Model reduction for delay differential equations with guaranteed
  stability and error bound}.
\newblock {\em Automatica}, 55:132--139, 2015.

\bibitem{SchJ14}
A.~van~der Schaft and D.~Jeltsema.
\newblock {\em Port-{H}amiltonian Systems Theory: {A}n Introductory Overview}.
\newblock now Publishers Inc., Hanover, MA, USA, 2014.

\bibitem{Vol01}
S.~Volkwein.
\newblock Optimal control of a phase-field model using proper orthogonal
  decomposition.
\newblock {\em ZAMM Z. Angew. Math. Mech.}, 81(2):83--97, 2001.

\bibitem{WolLEK10}
T.~Wolf, B.~Lohmann, R.~Eid, and P.~Kotyczka.
\newblock Passivity and structure preserving order reduction of linear
  port-{H}amiltonian systems using {K}rylov subspaces.
\newblock {\em Eur. J. Control}, 16(4):401--406, 2010.

\end{thebibliography}
